\documentclass{aa}  

\usepackage{graphicx}
\usepackage{natbib}
\usepackage{txfonts}
\usepackage{pdflscape}
\begin{document}

   \title{The ATESP 5 GHz radio survey}

   \subtitle{IV. 19, 38, and 94 GHz observations and radio spectral energy distributions}

   \author{R. Ricci\inst{1}, I. Prandoni\inst{1}, H.R. De Ruiter\inst{1}
          \and
          P. Parma\inst{1}
          }

   \institute{INAF-Istituto di Radioastronomia, via Gobetti 101, 40129, 
              Bologna, Italy\\
              \email{r.ricci@ira.inaf.it}
             }

   \date{Received April 1, 2015; accepted May 1, 2015}

 
  \abstract
  {}  
   {It is now established that the faint radio population is a mixture of star-forming galaxies and faint active 
galactic nuclei (AGNs), with the former dominating below $S_{\rm 1.4~GHz} \sim 100 $ $\mu$Jy and the latter at larger flux densities. 
The faint radio AGN component can itself be separated into two main classes, mainly based on the host-galaxy properties: 
sources associated with red/early-type galaxies  (like radio galaxies) are the dominant class down to $\sim 100 $ $\mu$Jy; 
quasar/Seyfert--like sources  contribute an additional 10-20\%. One of the major open questions regarding faint radio AGNs is 
the physical process responsible for their radio emission. This work aims at investigating this issue, with particular respect 
to the AGN component associated with red/early-type galaxies.  Such AGNs show, on average, flatter radio spectra than radio galaxies 
and are mostly compact ($\leq 30$ kpc in size). Various scenarios have been proposed to explain their radio emission. For instance 
they could  be core/core-jet dominated radio galaxies, low-power BL LACs, or advection-dominated accretion flow  (ADAF) systems.}
   {We used the Australia Telescope Compact Array (ATCA) to extend a previous follow-up multi-frequency campaign to 38
and 94 GHz. This campaign focuses on  
a sample of 28 faint radio sources associated with early-type galaxies extracted from the ATESP 5 GHz survey. Such data, together with 
those already at hand, are used to perform  radio spectral and variability analyses. Both analyses can help us to disentangle 
between core- and jet-dominated sources, as well as to verify the presence of ADAF/ADAF+jet systems. Additional high-resolution 
observations at 38 GHz were carried out to characterise the radio morphology of these sources on kiloparsec scales.}
   {Most of the sources (25/28) were detected at 38 GHz, while only one (ATESP5J224547-400324) of the twelve sources 
observed at 94 GHz was detected. From the analysis of the radio spectra we confirmed our previous findings that pure ADAF models  
can be ruled out. Only eight out of the 28 sources were detected in the 38-GHz high-resolution (0.6 arcsec) radio images and of those 
eight only one showed a tentative core-jet structure. 
Putting together spectral, variability, luminosity, and linear size information we conclude that different kinds of sources compose our 
AGN sample: (a) luminous and large ($\geq 100$ kpc) classical radio galaxies ($\sim 18\%$  of the sample); (b) compact (confined within 
their host galaxies), low-luminosity, power-law (jet-dominated) sources ($\sim 46\%$  of the sample);  and (c)  compact, 
flat (or peaked) spectrum, presumably core-dominated, radio sources ($\sim 36\%$  of the sample). 
Variability is indeed preferentially associated with the latter. }
  {}


   \maketitle
%

\section{Introduction}

Multi-wavelength studies of deep radio fields show that the sub-mJy population
responsible for the steepening of the 1.4-GHz source counts (\citealt{Windhorst90})
has a composite nature. Star-forming galaxies dominate 
at $S < 100-200 \mu$Jy, but represent no more than $20-30\%$ of the entire 
population above this threshold (\citealt{Seymour08}), where radio sources associated with early-type
galaxies (plausably triggered by nuclear activity) are the dominant population (they account for 64\% of the 
total at $S > 400 \mu$Jy; \citealt{Mignano08}).  
Radio sources associated to quasars and Seyfert galaxies provide another 10--20\% contribution at 
sub-mJy flux densities (see e.g. \citealt{Bonzini13}). 

Faint radio sources associated with early-type galaxies could represent the low-power tail of 
the radio galaxy parent population. However, they show properties that are not entirely consistent 
with those of classical radio galaxies. \cite{Mignano08} showed  that they tend to have 
small linear  sizes ($<10-30$ kpc) and flat ($\alpha>-0.5$, where $S\sim \nu^{\alpha}$) or 
even inverted ($\alpha>-0.5$) radio spectra between 1.4 and 5 GHz. Both compactness and 
spectral shape suggest a core emission with strong synchrotron or free-free self-absorption. 
Such sources are known to exist among low-power FR I (\citealt{Fanaroff74}) radio galaxies, 
but they are relatively rare. 

These sources may represent a composite class of objects very similar to the so-called low-power 
($P_{\rm 408 MHz} < 10^{25.5}$ W/Hz) compact ($<10$ kpc) LPC radio sources studied by \cite{Giroletti05}. 
Multiple causes are invoked to produce LPC sources: geometrical-relativistic effects 
(low power BL-Lacertae objects), youth (GHz Peaked Sources(GPS)-like sources), instabilities in the jets, 
frustration by a denser-than-average interstellar medium (ISM) and a premature end of nuclear activity (sources characterised 
by low accretion/radiative efficiency, i.e. ADAF/ADIOS systems).

\cite{Whittam13} analysed a sample of 296 faint (S$_{\rm 15 GHz} > 0.4$~mJy) radio sources
selected from the Tenth Cambridge (10C, \citealt{Franzen11}; \citealt{Davies11})
survey in the Lockman Hole region. By matching this 15-GHz selected catalogue with lower-frequency 
surveys (with Giant Metre-wavelength Radio Telescope, GMRT at 610~MHz, Westerbork Synthesis Radio Telescope, WSRT
at 1.4~GHz plus NRAO VLA Sky Survey NVSS, Faint Images of the Radio Sky at Twenty-Centimeters, FIRST
and Westerbork Northern Sky Survey, WENSS), they confirmed  that 
a population of faint flat-spectrum sources emerges at flux densities lower than 1~mJy.
In \cite{Whittam15} and \cite{Whittam16} the study of a complete sub-sample of 96 sources showed that 
this new population is dominated by cores of radio galaxies. \cite{Whittam17} carried out
a deep (rms noise $=18 \mu$Jy beam$^{-1}$) survey with GMRT in the AMI001 field and found that
the flat-spectrum population dominates down to S$_{\rm 15.7 GHz} =0.1$ mJy and that star-formation
galaxies make no significant contribution to this population. 

\cite{Sadler14} studied a sample of 202 sources in the local universe (median z=0.058)
extracted from the AT20G survey and matched with the 6dF Galaxy Redshift survey  
(\citealt{Jones09}). Most of the galaxies in the sample host compact radio active galactic nuclei (AGNs)
which appear to lack extended emission even at frequencies lower than 20~GHz. 
These compact sources show no evidence of relativistic beaming and present a mix 
of steep and peaked radio spectra.

The present paper is the fourth in a series based on the 5-GHz Australia Telescope ESO Slice Project 
(ATESP) survey, and the second devoted to the investigation of a sample of ATESP sources associated 
with red/early-type galaxies. The ATESP 5 GHz survey consists of a sample of 131 radio sources with 
limiting flux density S$_{\rm lim} \sim 0.4$~mJy, covering one square degree and imaged at both 1.4 
(see \citealt{Prandoni00a}, \citealt{Prandoni00b}) and 5~GHz. \cite{Prandoni06} (Paper I) present the surveys 
and the catalogue, as well as the spectral index properties of the radio sources.  
\cite{Mignano08} (Paper II) provide the optical identifications and redshifts for about 
80\% of the ATESP 5 GHz sample, and discuss the source radio-optical properties. 
As mentioned above, an unexpected class of flat/inverted spectrum compact radio sources 
with linear size d $< 10-20$~kpc associated to optically inactive early-type galaxies 
was found. Their redshift distribution extends to z $\sim 2$ and peaks at z=0.5,
indicating that such sources may undergo significant evolution. The compactness of the 
sources together with their flat/inverted spectra suggests core emission with strong
synchrotron or free-free self-absorption. This could be associated either to very early
phases of nuclear radio activity (GHz Peaked Spectrum - GPS - sources, \citealt{ODea98};
\citealt{Snellen00}) or late stages of evolution of AGNs, characterised by 
low-accretion/radiative efficiency (ADAF/ADIOS) sources. In the case of the ADAF mechanism 
the spectral index of the overall synchrotron emission is expected to range from   
0.2 to 1.1 (S$_{\nu}\propto \nu^{\alpha}$) up to millimetre (mm) wavelengths (\citealt{Nagar01}).
In the presence of outflows, however, the spectral index can flatten and shift the peak of 
the radio emission from mm to centimetre (cm) wavelengths (\citealt{Quataert99}). To verify 
this hypothesis \cite{Prandoni10} (Paper III) carried out a follow-up campaign 
in 2007-2008 at the frequencies of 5, 8, and 19~GHz of all the early-type galaxies
in the 5-GHz ATESP survey brighter than 0.6 mJy. The ADAF mechanism was 
ruled out, but the ADAF+jet scenario was still consistent with the overall spectral behaviour.
The AGN component in the ATESP sample was found to be mainly the low-power extrapolation
of the classical bright radio galaxy population, rather than the radio quiet counterpart 
of bright quasars. From the study of luminosity and linear size, Paper III
also showed that a significant fraction of ATESP sources has sizes consistent with 
the ones in B2 (\citealt{Colla75}) and 3CR (\citealt{Blundell99}) radio galaxies.
However, most of the ATESP sources appear point-like at the studied frequencies 
(angular resolution $\theta > 2$~arcsec). In Paper IV (the present work) we 
extend the spectral study carried out in Paper III to 38 and 94~GHz in order to further 
tighten down the spectral shapes over a much wider interval of frequencies. 
We also investigate the sub-arc angular scale in order to understand whether  
jet-dominated or compact-core sources dominate in the ATESP sample.                 

Many studies of high-frequency samples in different flux density regimes 
have been carried out in recent years. Studying radio spectral energy distributions (SEDs) of sources of this type
is important because different accreting regimes display different
spectral signatures in the radio domain.

\cite{Sajina11} observed 159 radio galaxies with the VLA in the frequency range 
4.8$-$43~GHz. The sample was selected from the Australia Telescope 20~GHz survey
(AT20G, \citealt{Murphy10}) sources with S$_{\rm 20 GHz} > $40 mJy in the equatorial 
field of the Atacama Cosmology Telescope (ATC, \citealt{Marriage11}) survey. 
The analysis of the sample showed that these sources have flatter spectra and are 
more compact than low-frequency-selected samples.

The bright Planck ATCA Co-eval Observations (PACO) sample (\citealt{Massardi11}) 
comprises 189 sources with AT20G flux densities S$_{\rm 20 GHz} > $500 mJy at $\delta < -30^{\circ}$.
The analysis of the radio spectra between 4.5 and 40~GHz showed that 14\% 
of the sources have SEDs described by a power law 
while the majority (66\%) of the sample is constituted by sources with 
down-turning spectra. The analysis of the faint (200 < S$_{\rm 20 GHz} < $500 mJy)
PACO sample (\citealt{Bonavera11}) showed an increase in the percentage
of steep-spectrum sources with respect to the bright sample. The radio spectra
of the 464 sources in the full PACO sample were studied by \cite{Massardi16}.
It was found that 91\% of the spectra are remarkably smooth and are fitted by a 
double power-law model. A spectral steepening above $\simeq 30$~GHz, consistent 
with synchrotron emission becoming optically thin, is present in most of the sources.
        
In Section 2 we describe how observations and data reduction were carried out. 
In Section 3 we describe the high-resolution analysis of the ATESP sample. In Section 4 
we analyse the source variability at 19~GHz. In Section 5 we model fit the radio
spectra. In Section 6 we present and analyse the colour-colour diagram and in Section 7 
we discuss the properties of the ATESP early-type sample. We summarise our conclusions in Section 8.  


\section{Radio observations and data reduction}

\subsection{Sample selection}

In Paper III a flux-limited (S$_{\rm 5 GHz}>0.6$ mJy) sample of 28 radio sources associated with 
early-type galaxies was selected from the ATESP 5-GHz survey and 26 of them were observed at 4.8, 8.6, and 19 GHz. 
In this paper we present follow-up observations of this sample (including this time the two missing 
sources of Paper III, J224628$-$401207 and J224719$-$401530) at 19, 38, and 94 GHz.

\begin{table*} [t]
 \centering
  \caption{Observation details: dates, number of sources, intermediate frequencies (IFs), array configurations, 
angular resolutions (ang. res.), primary calibrators used during the 2011-2012 observing campaigns. 
LR: low angular resolution; HR: high angular resolution.} 
\label{tab:list}
  \begin{tabular}{lcccccc}
\hline
Band      &    IFs        &   Date         & num. of sources & array config. & ang. res.        & Primary cal  \\
          &    MHz        &                &                 &               &   arcsec         &              \\
\hline
\hline         
19 GHz    & 18496/20544   & Sep 25th 2011  &   14            &     H75       &  31$\times$20    & 1934-638     \\ 
          & 18496/20544   & Jul 11th 2012  &   14+4          &     H168      &  14$\times$9     & 1934-638     \\
\hline
38 GHz LR & 37059/39104   & Sep 29th 2011  &   14            &     H75       &  14$\times$11    & uranus       \\         
          & 37059/39104   & Jul 20-21 2012 &   14+4          &     H75       &  16$\times$11    & uranus       \\
\hline
38 GHz HR & 37059/39104   & Sep 11th 2011  &   28            &     6B        &  0.7$\times$0.6  & uranus       \\
\hline
94 GHz    & 93504/95552   & Sep 26ht 2011  &   12            &     H75       &  6$\times$4      & uranus       \\           
\hline
\end{tabular}
\end{table*}

\subsection{Observations}

The observations at 19, 38, and 94 GHz with the Australia Telescope Compact Array (ATCA)
were carried out in two separate campaigns in 2011 and 2012. The total observing
time for each target was calculated based on the expected flux density at
38 and 94 GHz. To compute the expected flux density the spectral index between 
5 or 8 GHz and 19 GHz from the 2007 and 2008 observations was used. For the two new sources 
the 1.4-5~GHz spectral index from Paper I was used. In Table~\ref{tab:list} 
the relevant information on the 2011-2012 observing campaign is provided. 

\subsubsection{ Observations at 19 GHz}
The 15 mm receiver with simultaneous
intermediate frequencies (IFs) set at 18496 and 20544 MHz and a bandwidth of 2048 MHz for each IF
was used for these observations. On September 25, 2011, 
14 sources out of the list of 28 targets were observed. 
A hybrid and very short array configuration was 
chosen in order to map the sources with the same resolution as for the previous multi-frequency 
observations (Paper I and III, $\sim 10$ arcsec). 
The best array configuration for this purpose is H168 but, for scheduling constraints, 
we observed with the more compact array H75 which provides an angular resolution that is
a factor of two larger than H168. For bandpass calibration we took a 
single ten-minute scan on 1921-293. A five-minute scan on the primary ATCA calibrator 
1934-638 was used to bootstrap the absolute flux density scale and 2223-488 was 
used as pointing and phase calibrator and observed approximately every 20 min. 
On July 18, 2012, the remaining 14 sources in our sample, 
plus four objects which were not detected (signal to noise ratio (S/N) $<$ 5) in the 2011 run, were 
observed with the hybrid array configuration H168. The bandpass calibrator was
1921-293 which was observed in a single seven-minute scan. The absolute flux density 
calibrator was 1934-638 which was observed in a single five-minute scan. The pointing 
and phase calibrator was 2223-488.

\subsubsection{Low-resolution observations at 38 GHz }
The two IFs of the 7-mm receiver were set to frequencies   
37056 and 39104 MHz with a bandpass of 2048 MHz for each IF.
On September 29, 2011, 14 sources were observed  
with the array in its most compact configuration H75 in order to provide 
integrated flux density measurements for the radio spectra at similar 
resolution to the other frequencies. The bright calibrator 1921-293 was used for bandpass
corrections, 2223-488 as phase calibrator and the planet Uranus as absolute
flux density calibrator. On July 20 and 21, 2012, the remaining 14 sources were 
observed with the most compact array configuration H75. The four undetected objects 
mentioned above were also reobserved to obtain flux densities at 19 and 38 GHz at the same epoch. 
The same calibrator setting as for the 2011 observations was used.

\subsubsection{High-resolution observations at 38 GHz }
On September 10 and 11, 2011, the full sample of 28 sources was observed at 38 GHz with the array 
configuration 6B in order to achieve the highest angular resolution (0.6 arcsec). 
The frequency setting and bandwidth was the same as that in the low-resolution run. 
The Hour Angle (HA) coverage spanned the range
between -6$^{\rm h}$ and +6$^{\rm h}$ in order to maximize the $(u,v)$ coverage
of the linear array configuration. A single scan of 5 min on the strong 
source 1921-293 was used as band-pass calibration. For absolute flux density 
calibration the planet Uranus was observed in a single scan for 10 min. 
The bright calibrator 2223-488 was chosen as pointing and phase calibrator: the 
pointing was checked approximately every an hour during observations and a one-minute scan on the 
phase calibration was repeated every 20 minutes.

\subsubsection{Observations at 94 GHz}
On September 26, 2011, we observed a list of 12 
sources out of the full sample of 28 targets with the 3-mm receivers with two simultaneous 
IFs centred at 93504 and 95552 MHz and a bandwidth of 2048 MHz for each IF. The 12 sources 
showed an inverted- or flat spectrum between 8 and 19 GHz from previous ATCA
observations in 2007-2008 so they had the highest chance of being detected.
The most compact array configuration H75 was chosen in order to avoid losing flux density 
because of atmospheric phase instability. At 94 GHz this configuration provides 
a resolution of $\sim 5$ arcsec. A single five-minute scan on 1921-293 was used for bandpass 
calibration, a ten-minute scan on the planet Uranus provided absolute flux density 
calibration and three-minute scans every 20 minutes on 2223-488 provided phase calibration. 
A pointing check every hour during observations was also done on 2223-488 together 
with a paddling sequence for system temperature evaluation.

\subsection{Data reduction}

We used the astronomical software package MIRIAD (\citealt{Sault95}) 
to reduce all the raw visibility data. 
We imported the data into MIRIAD, then checked, flagged, and 
split them into single-source single-frequency data sets. To equalise 
the bandpass we used the standard ATCA calibrator 1921-293 for all the 
runs in 2011 and 2012. The absolute flux density scale was bootstrapped using the 
standard ATCA primary calibrator 1934-638 for the 19~GHz (K-band) observations.
The planet uranus was instead used for the 38 and 94~GHz observations,
because the model of 1934-638 at these bands was too poorly known at that time 
to perform absolute flux bootstrapping. The phase calibrator 2223-488 was 
consistently used for all 2011 and 2012 runs to solve for the amplitude and 
phase complex gains for each observing frequency. The calibrated visibilities 
were then imaged by Fourier inversion: for the data sets taken with the 
hybrid and compact arrays H75 and H168, the 6-km antenna data were split away
and not reduced because the baselines with the 6-km antenna provide a very poor 
$(u,v)$ coverage for such configurations. As the bandpass in 
continuum mode is very large (2 GHz bandwidth) an algorithm available in the package  
MIRIAD allows for bandpass slope correction over the bandwidth
of each IF. For mapping, the Multi Frequency Synthesis (MFS, \citealt{Sault94}) 
technique was used to merge the calibrated visibilities 
from the two IFs in each observing band (19, 38, and 94~GHz). If the two IFs are 
not too far apart, the MFS technique can be used to effectively increase 
the $(u,v)$ plane coverage, minimize the bandwidth smearing and improve 
the image quality. As we were carrying out a detection experiment the inverted
maps were inspected with the visualisation tool {\it kvis} of the KARMA software package
to search for at least a $3-\sigma$ detection in the map position 
where a source was expected to be found (the field centre). In case a detection
was found the inverted map was cleaned using the task CLEAN in MIRIAD which
implements one of the three standard cleaning algorithms (\citealt{Hogbom74}; 
\citealt{Clark80}; \citealt{Steer84}). When the restored I-Stokes images showed that 
more than one component was present, a Gaussian fit was performed on those components
with the MIRIAD task {\it imfit} to determine their flux densities, as well as 
the integrated flux density of the combined source. 

\section{Data analysis}

\subsection{Low-resolution data analysis}

In total 26 sources were found to be best fitted by 
a single-component point-like model at 19 GHz. One source (J225034-401936) appears 
extended and elongated in the southwest (SW) direction in agreement with
what was found in its previous 19~GHz observations (see Paper III) and with its lower-frequency 
morphology, where two components are clearly seen along the same direction (see Fig. 7 of Paper III).    
Finally, one source (J224719-401530) appeared as a double and was fitted with two Gaussian 
components with integrated flux densities of 0.422 and 0.390 mJy. This source is one of the two
not previously observed at 19 GHz. 

At 38 GHz 24 sources were best fitted by a single component point-like model,
three sources presented only upper limits, and one source
(J225004-402412) was fitted with two Gaussian components with integrated flux 
densities of 0.267 and 0.173 mJy. This source appears as compact at lower frequency, 
and even in the high-resolution (2~arcsec) 5-GHz images (see Fig. 7 of Paper III).
Error bars were computed by summing two terms in quadrature: a multiplicative term
related to the residual gain calibration and an additive term related to the 
root mean square (rms) noise in the V-Stokes maps. 

The integrated flux densities with their 1-$\sigma$ error bars are listed in
Table~\ref{tab:intflux} together with the original 1.4-GHz and 5-GHz flux densities from \cite{Prandoni00b}
and \cite{Prandoni06}. The radio spectra shown in Fig.~\ref{fig:spec1} include all observations 
available for each source (\citealt{Prandoni00b}, \citealt{Prandoni06}, \citealt{Prandoni10}). 
We often have detections and upper limits at the same frequency,
coming from different observing runs with apparently different quality. We note that there is only 
one observing run at 94 GHz resulting in only one detection out of 12 objects observed.


\begin{figure*}[h]
\centering
\includegraphics[width=8cm]{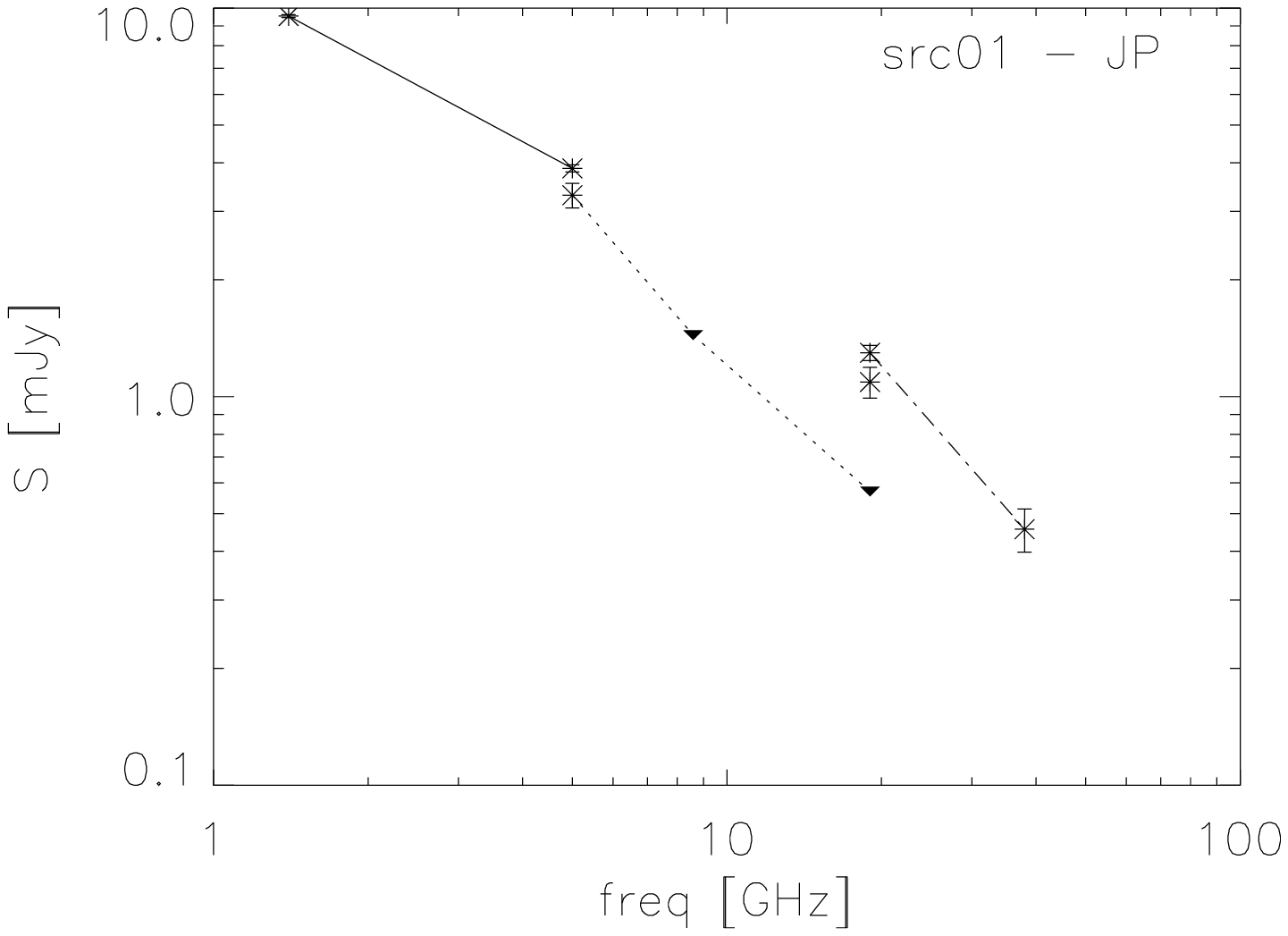}
\includegraphics[width=8cm]{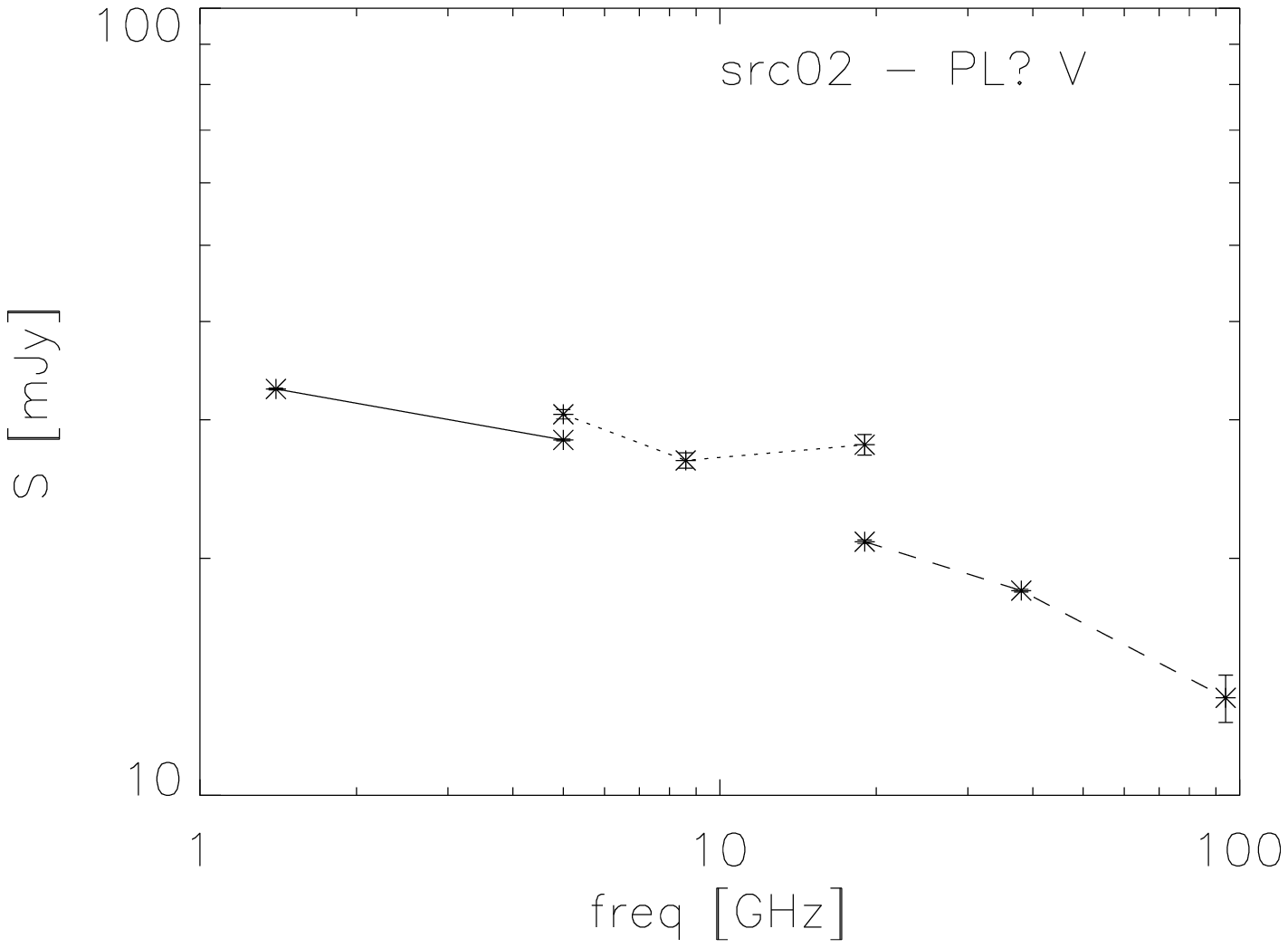}
\includegraphics[width=8cm]{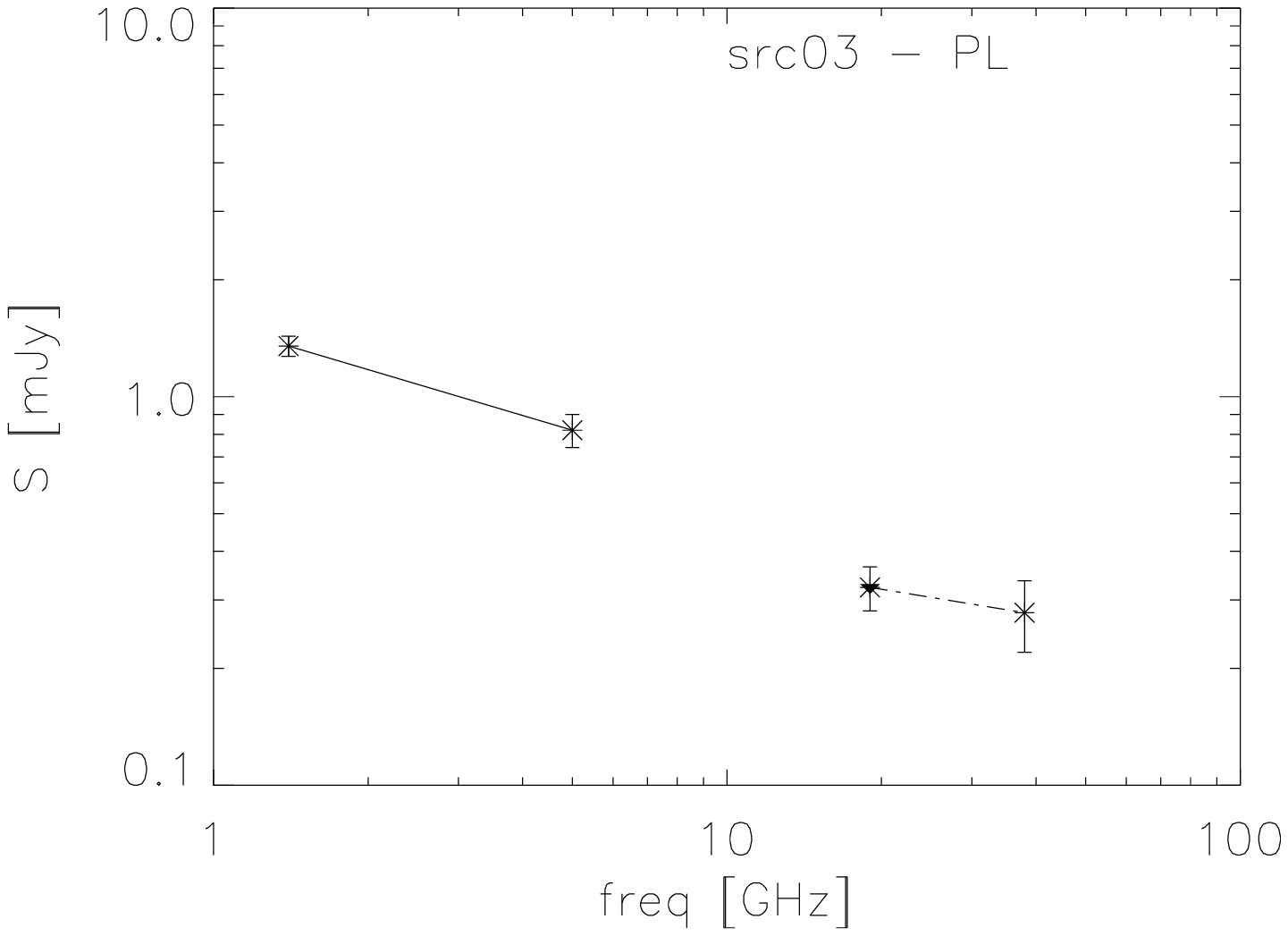}
\includegraphics[width=8cm]{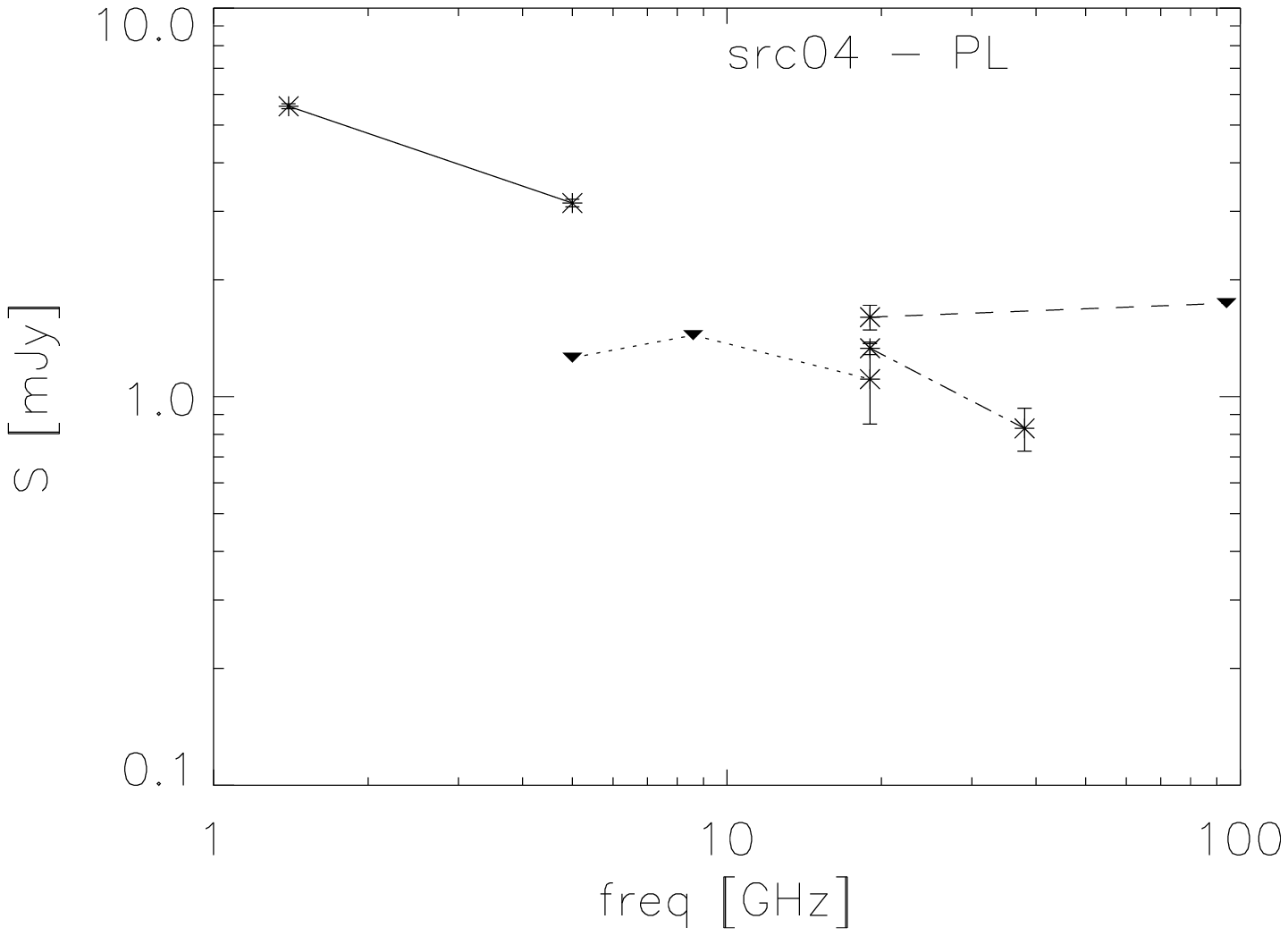}
\includegraphics[width=8cm]{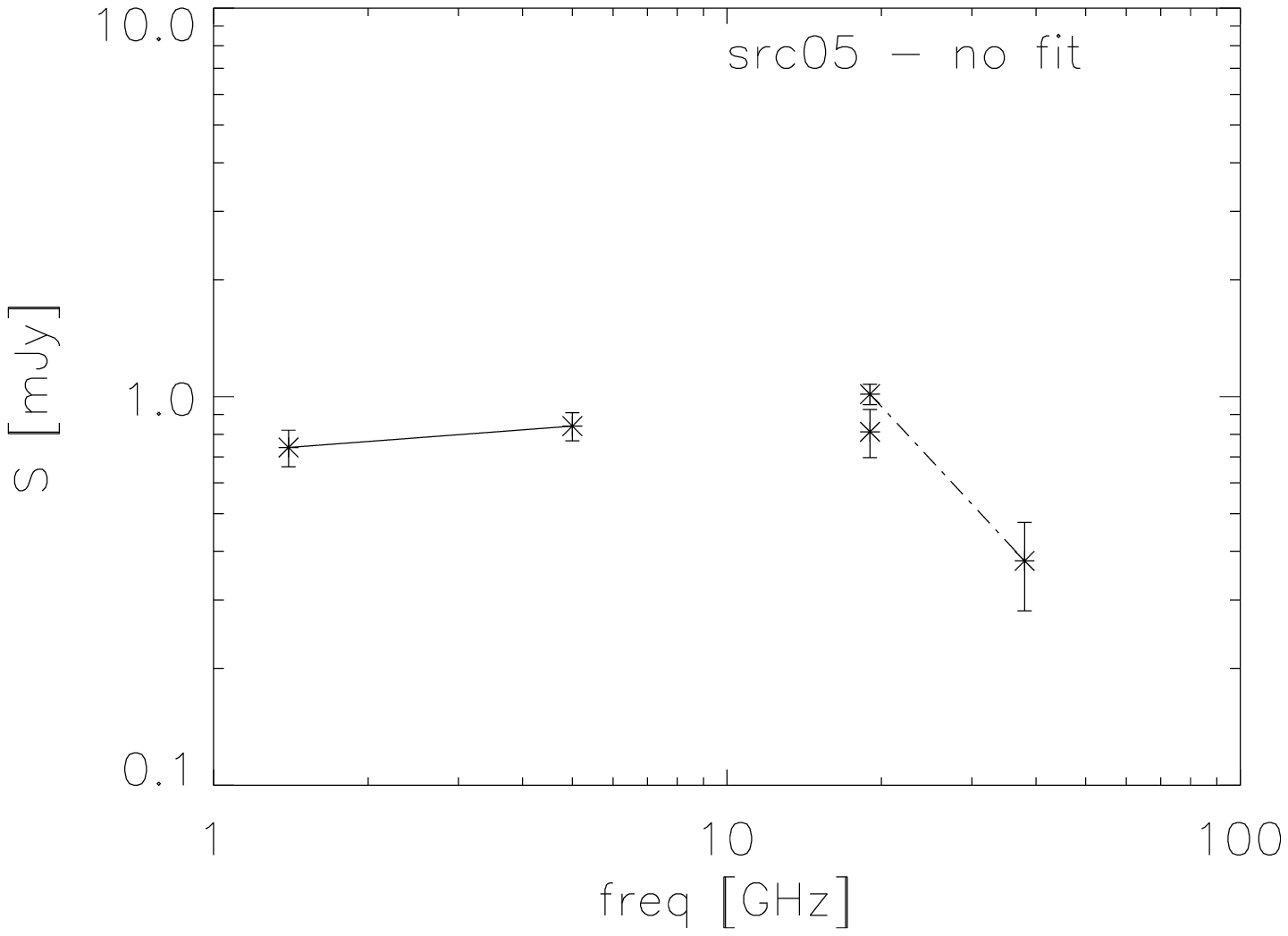}
\includegraphics[width=8cm]{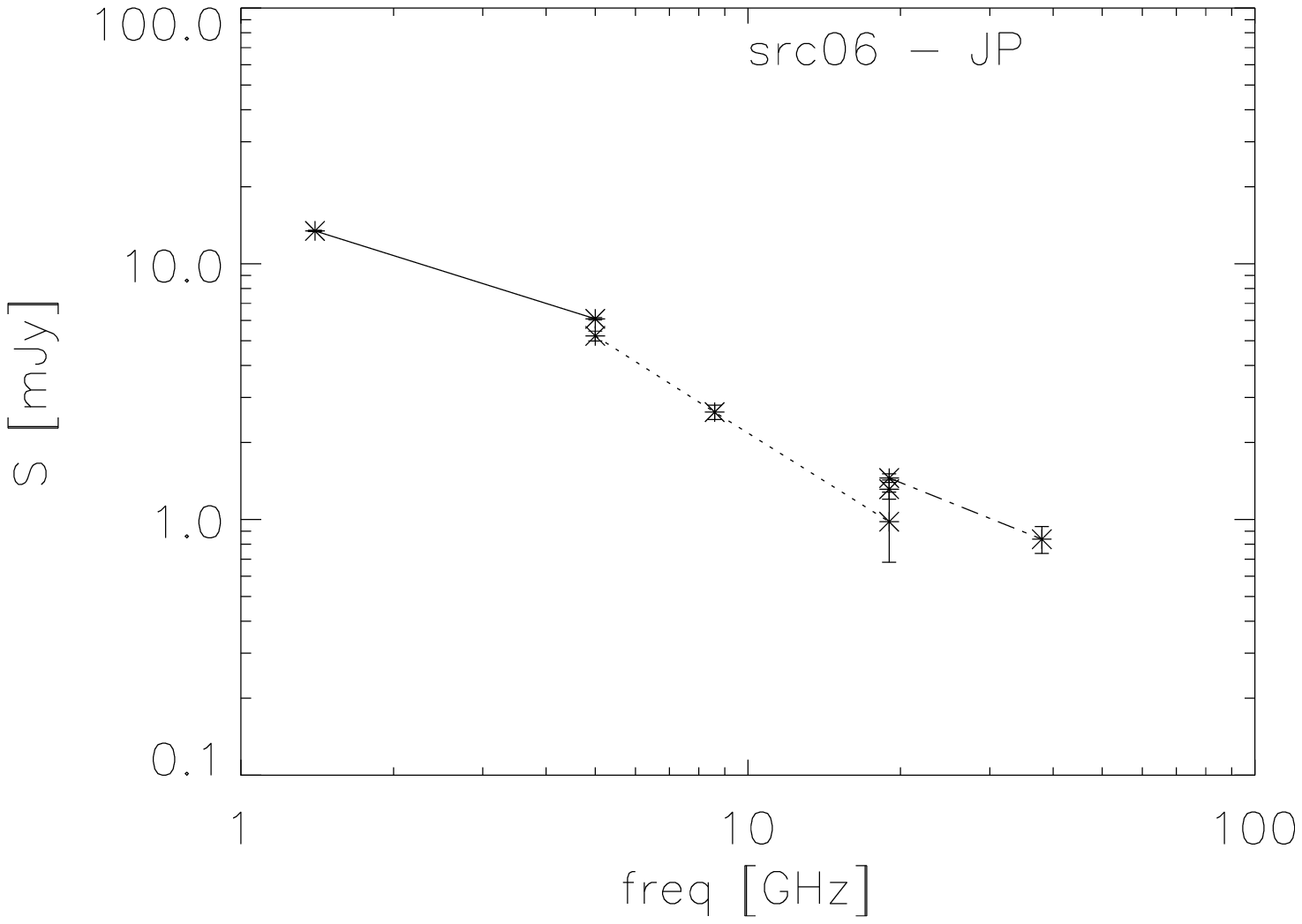}
\includegraphics[width=8cm]{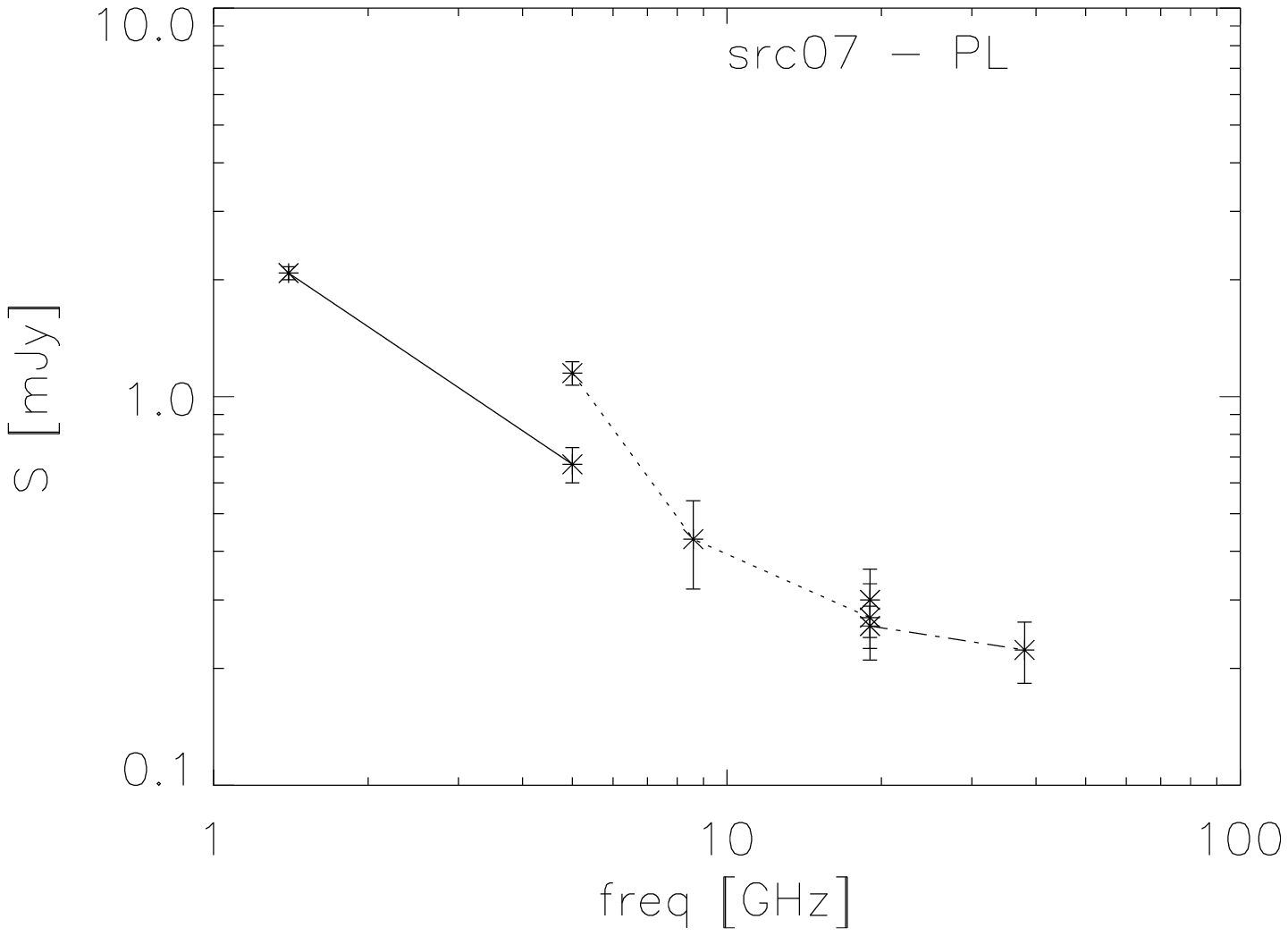}

\caption{ATESP sample radio spectra. Asterisks are detections, 
(with error bars) downward triangles are upper limits, solid lines join ATESP 
survey points, dotted lines join 2007-2008 follow-up points, dashed lines join 
2011 follow-up points, and dot-dashed lines join 2012 follow-up points.}
\label{fig:spec1}
\end{figure*}

\addtocounter{figure}{-1}
\begin{figure*}[h]
\centering
\includegraphics[width=8cm]{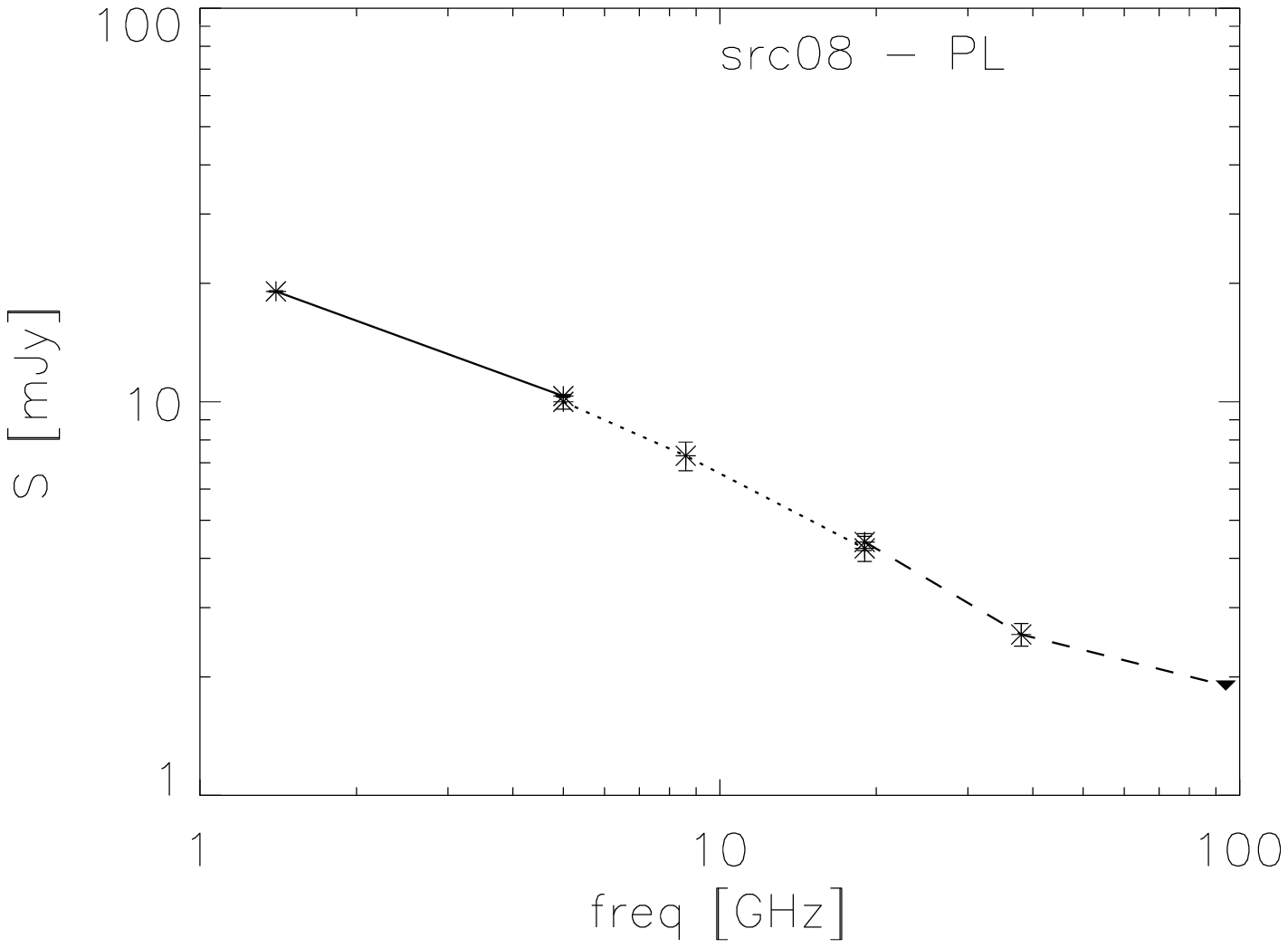}
\includegraphics[width=8cm]{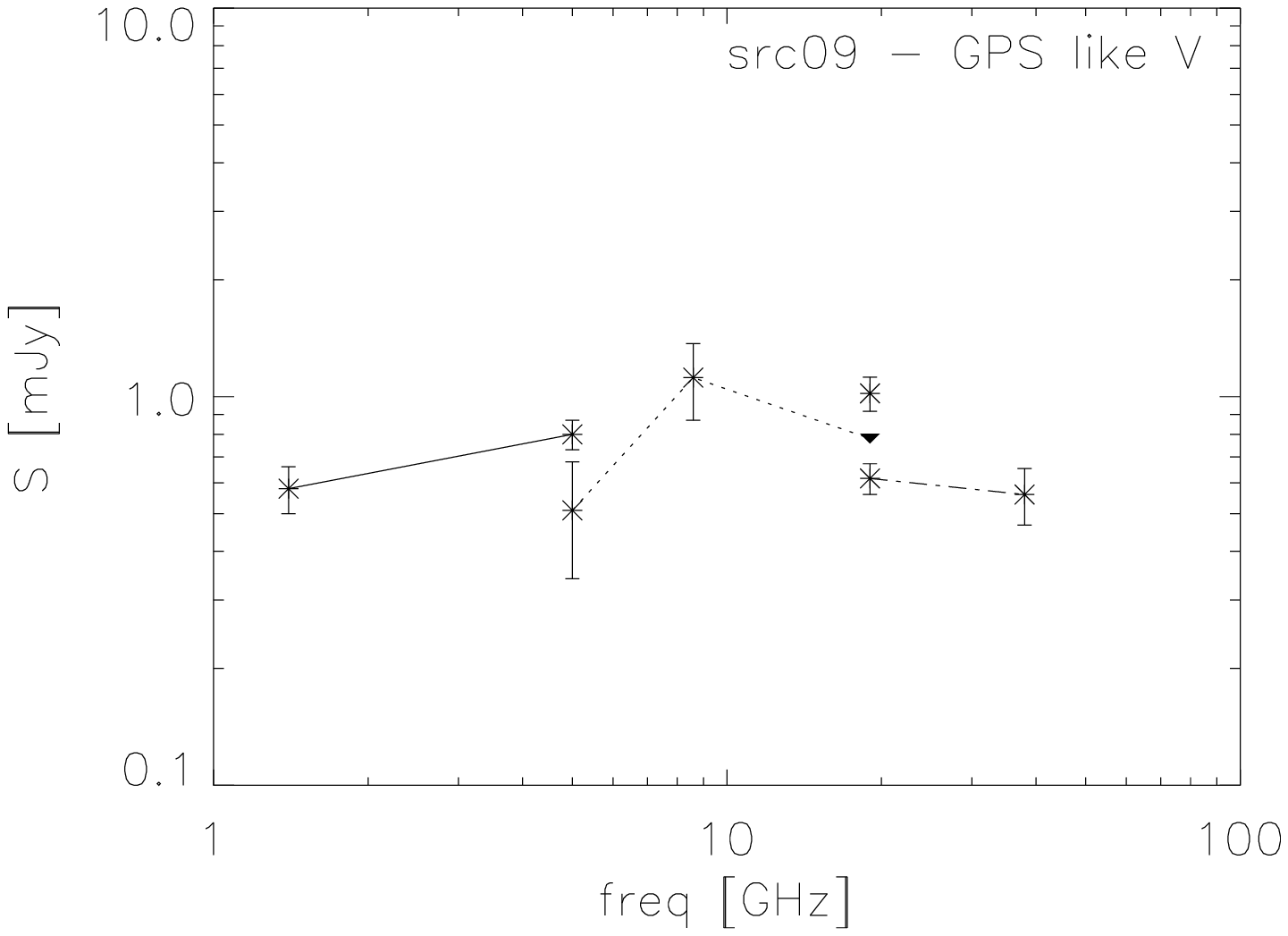}
\includegraphics[width=8cm]{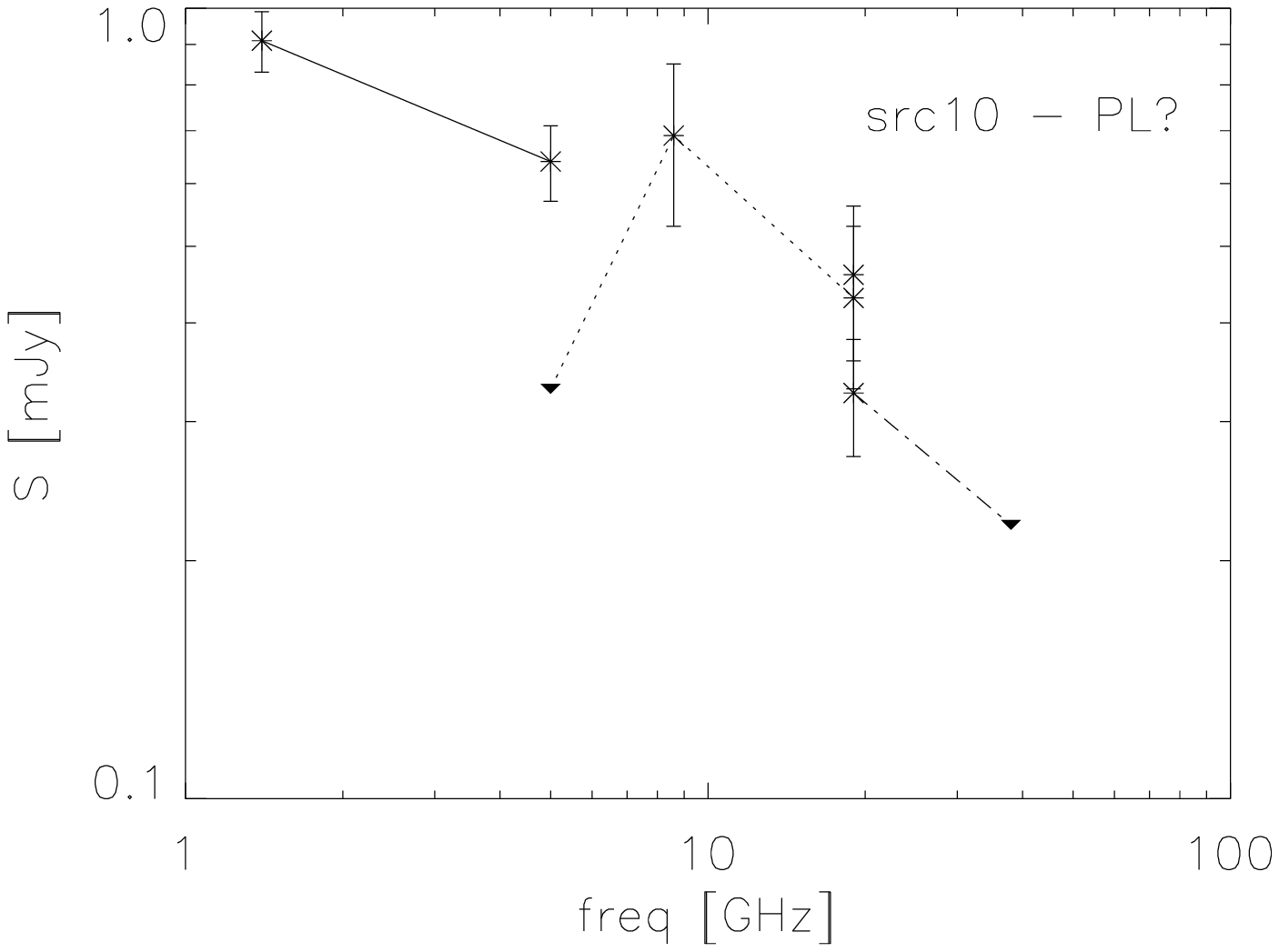}
\includegraphics[width=8cm]{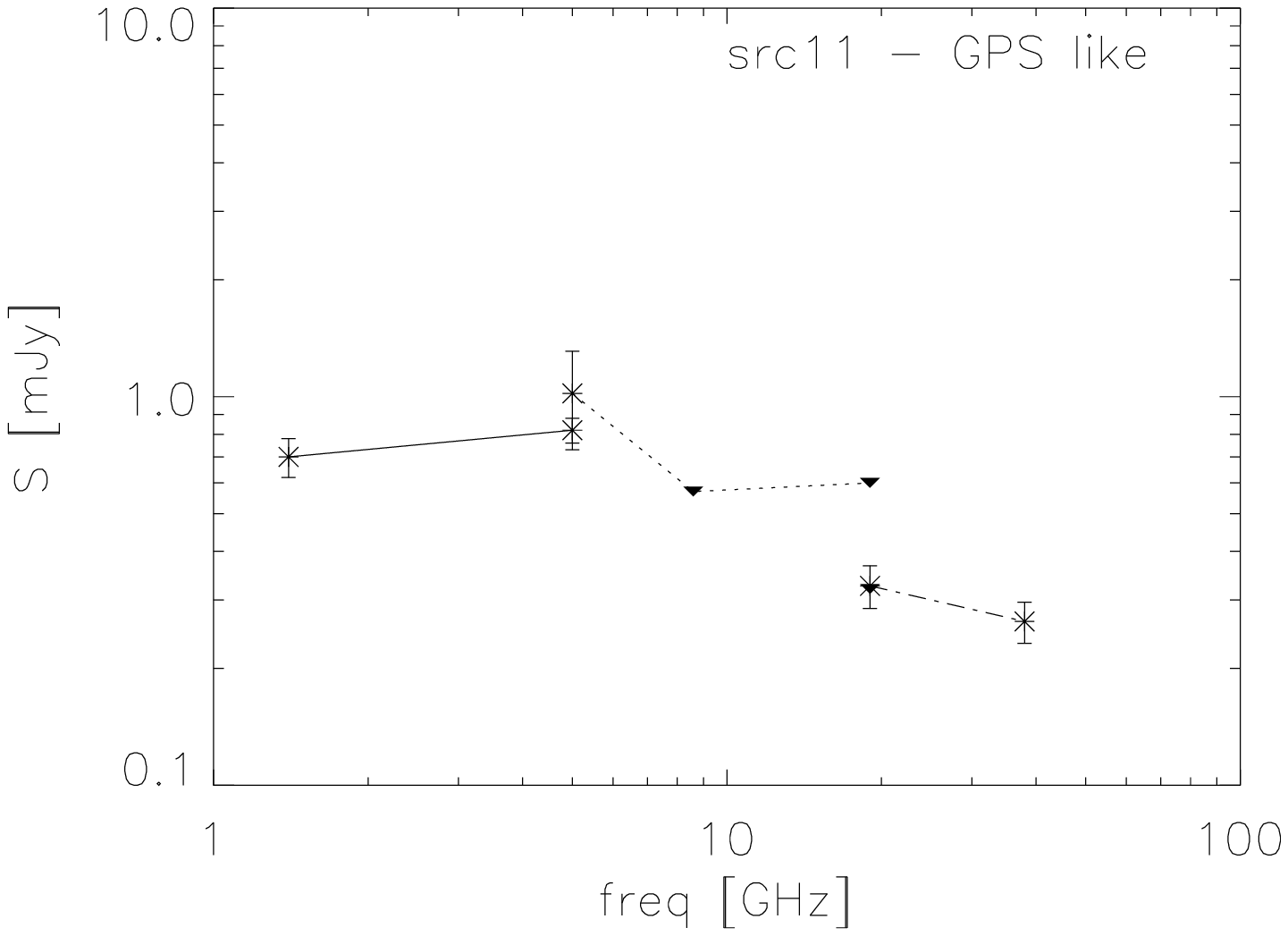}
\includegraphics[width=8cm]{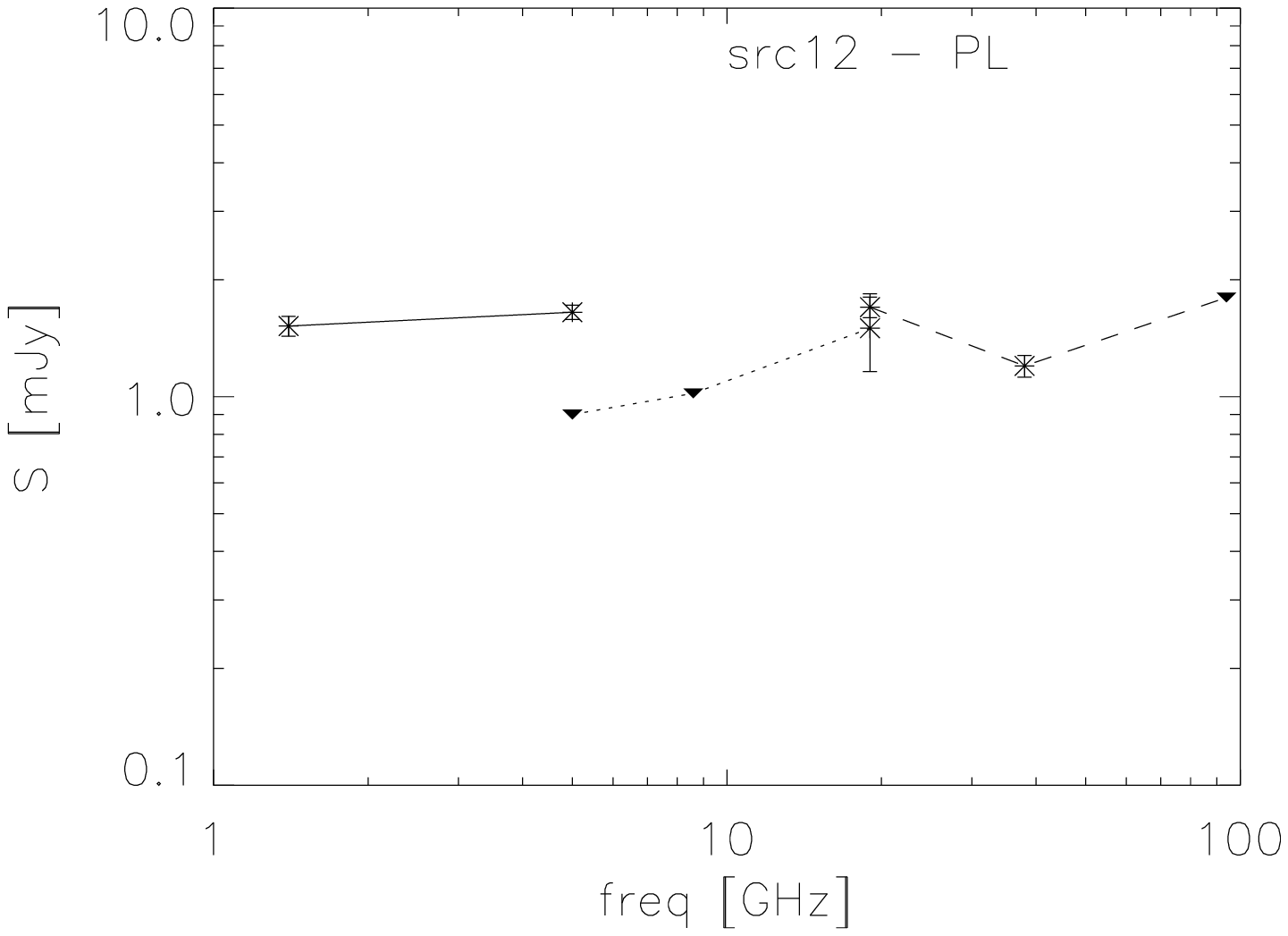}
\includegraphics[width=8cm]{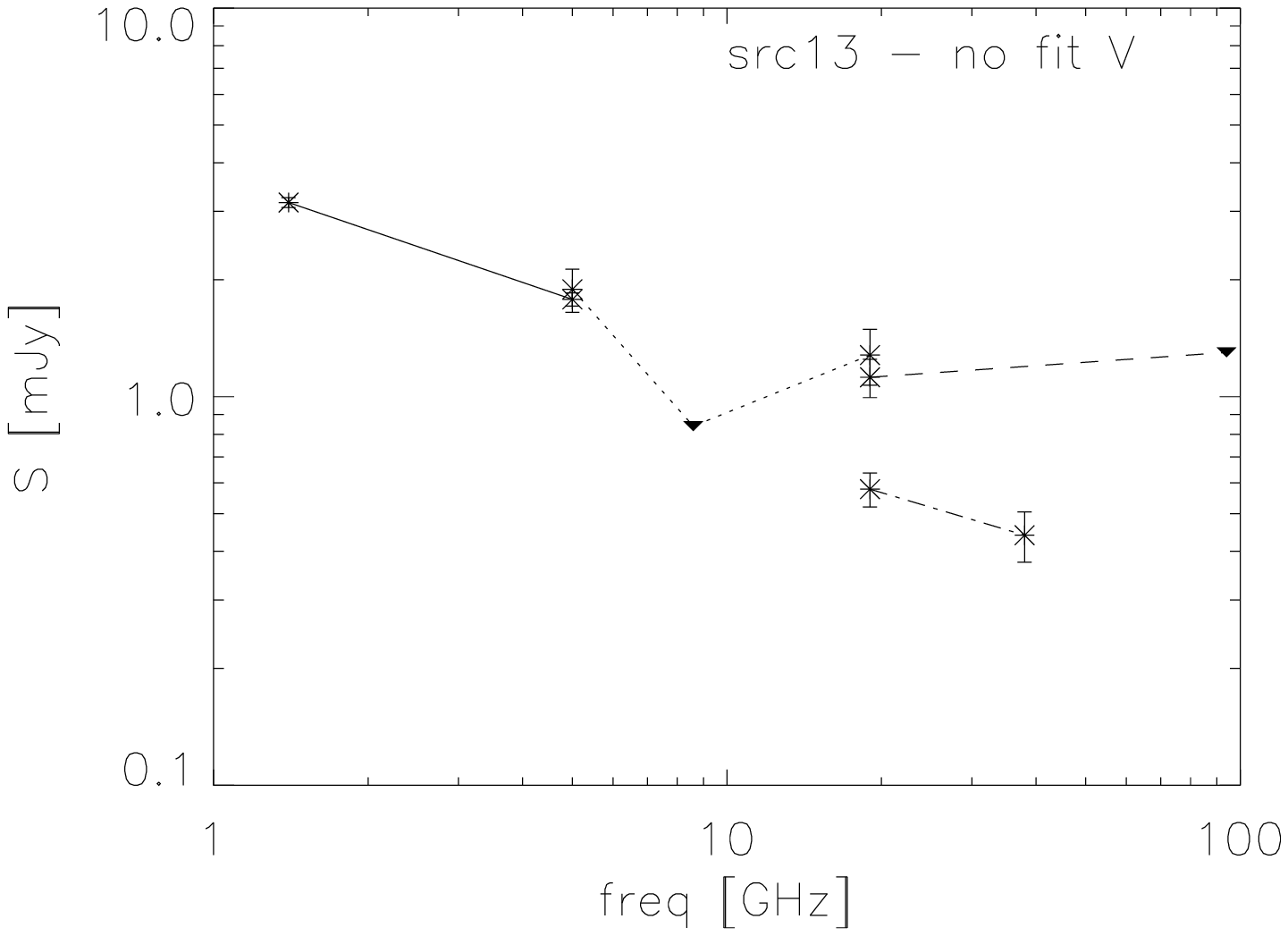}
\includegraphics[width=8cm]{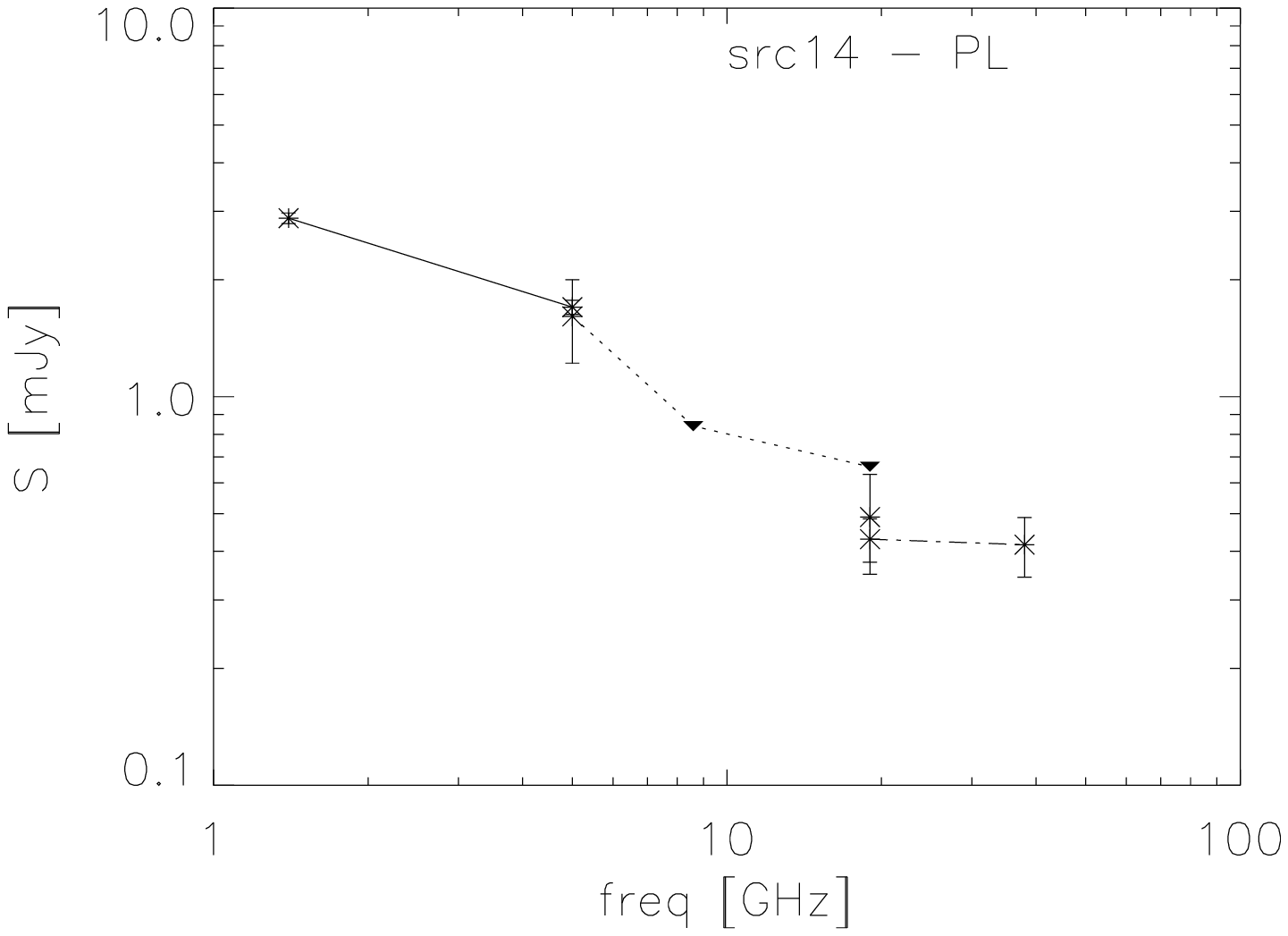}
\caption{Continued.}
\label{fig:spec2}
\end{figure*}

\addtocounter{figure}{-1}
\begin{figure*}[h]
\centering
\includegraphics[width=8cm]{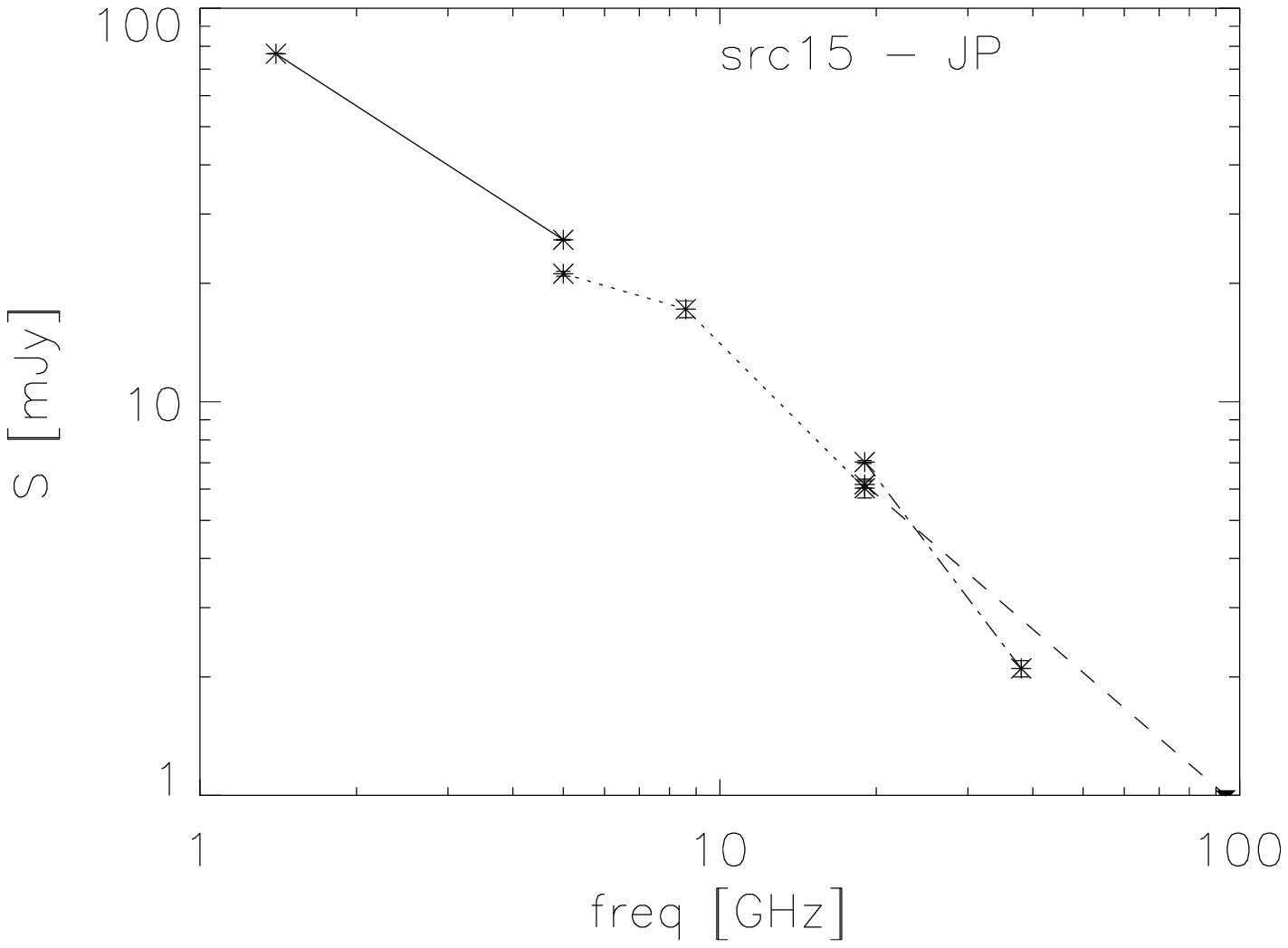}
\includegraphics[width=8cm]{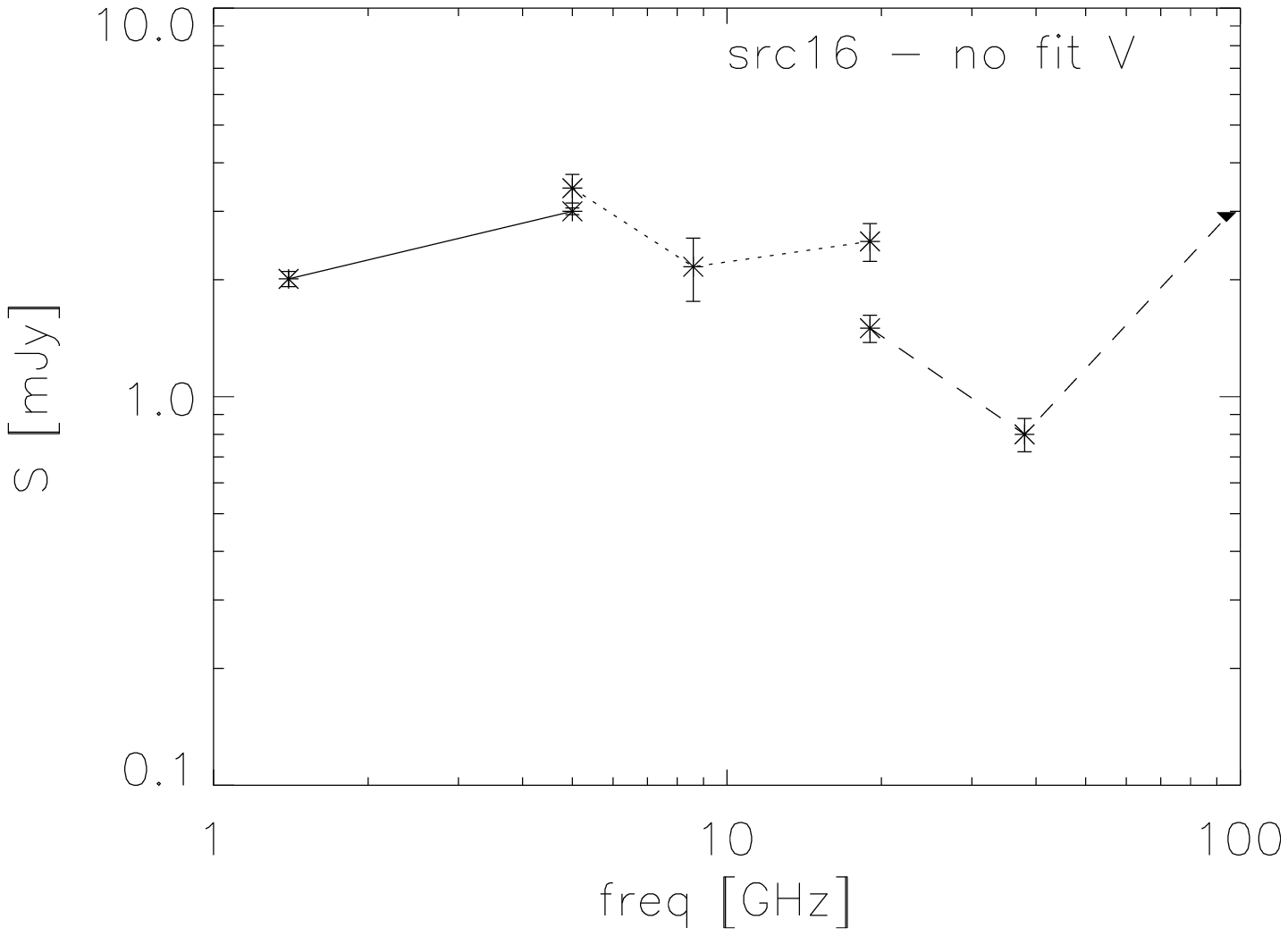}
\includegraphics[width=8cm]{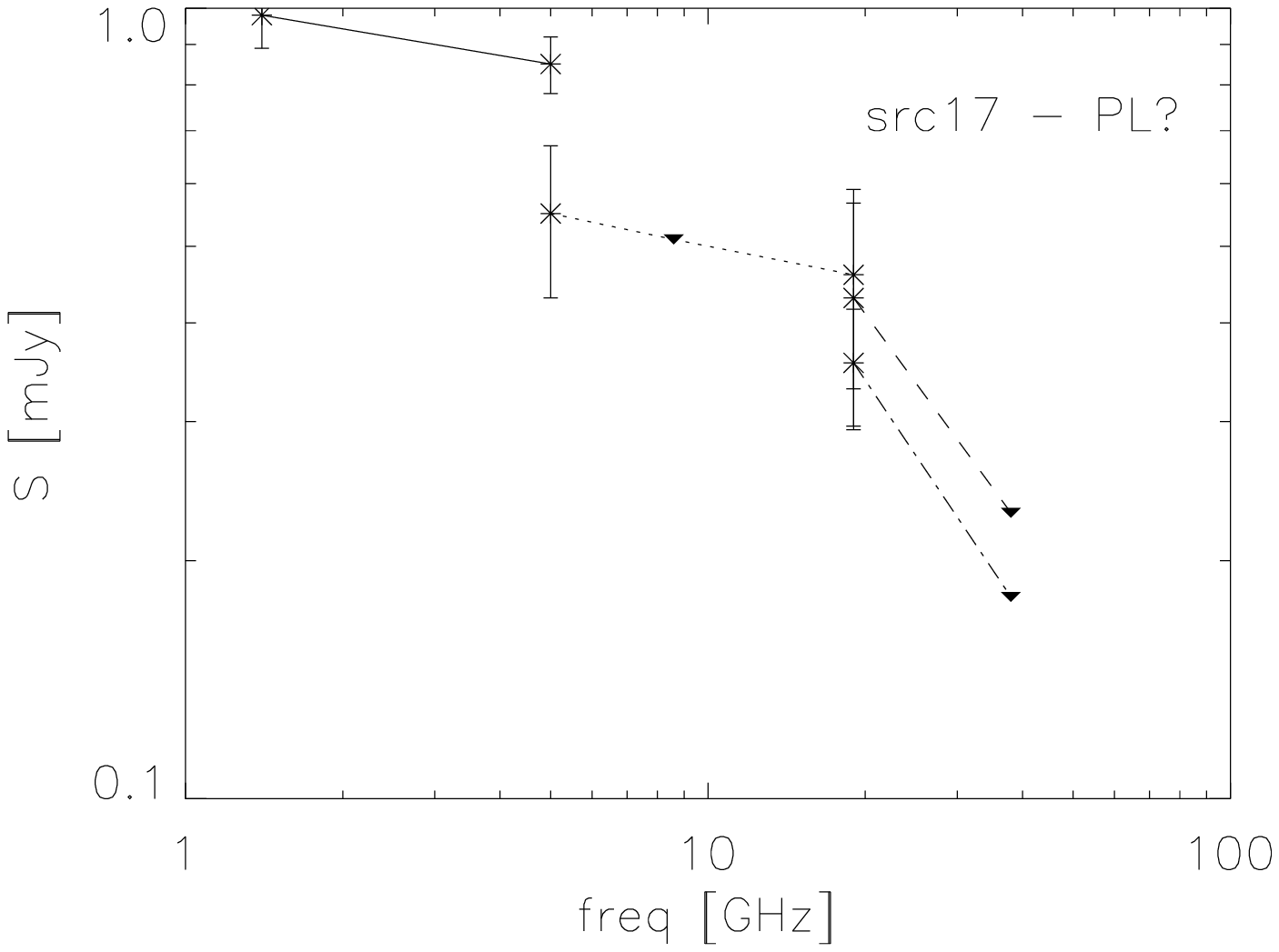}
\includegraphics[width=8cm]{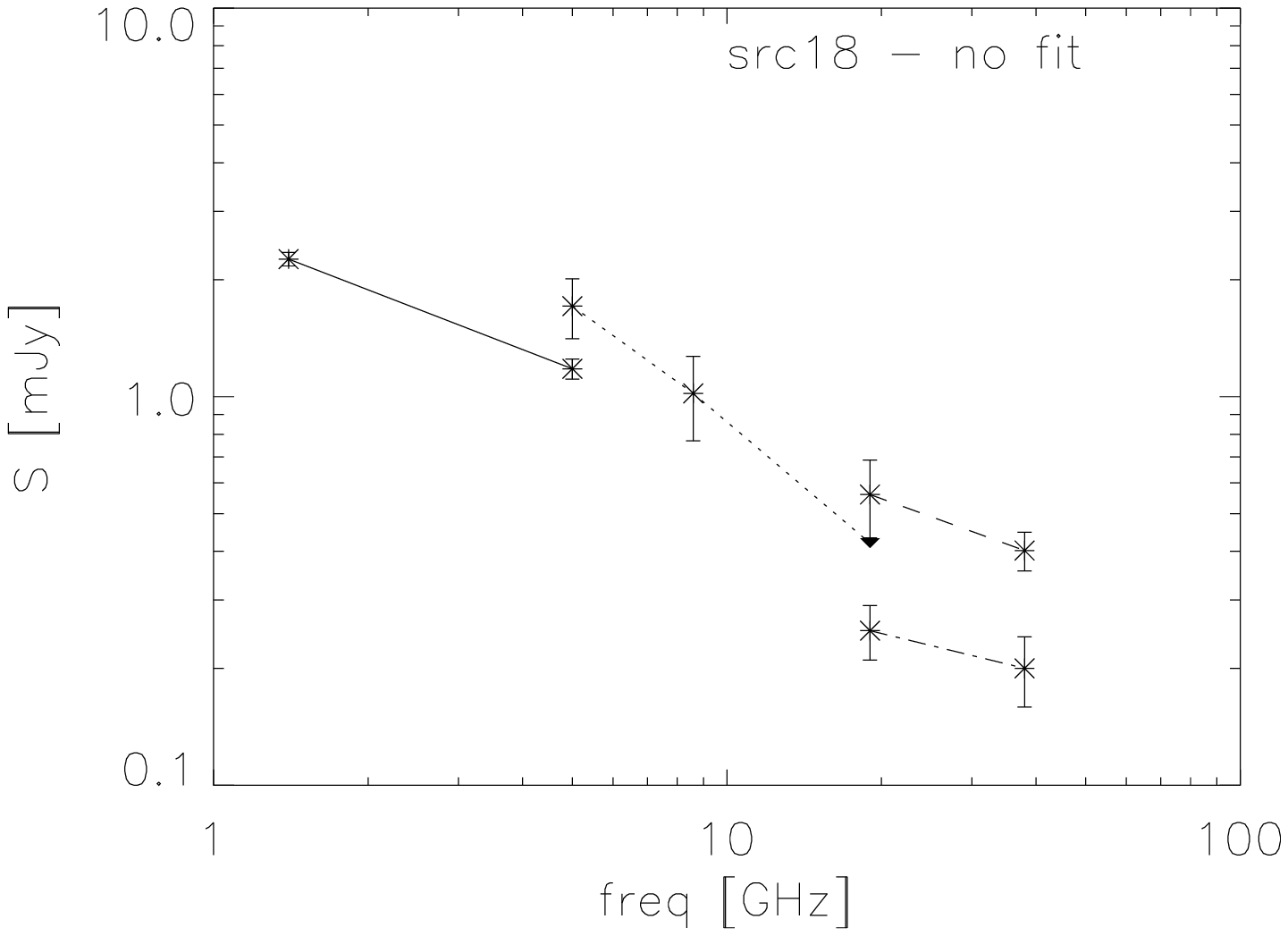}
\includegraphics[width=8cm]{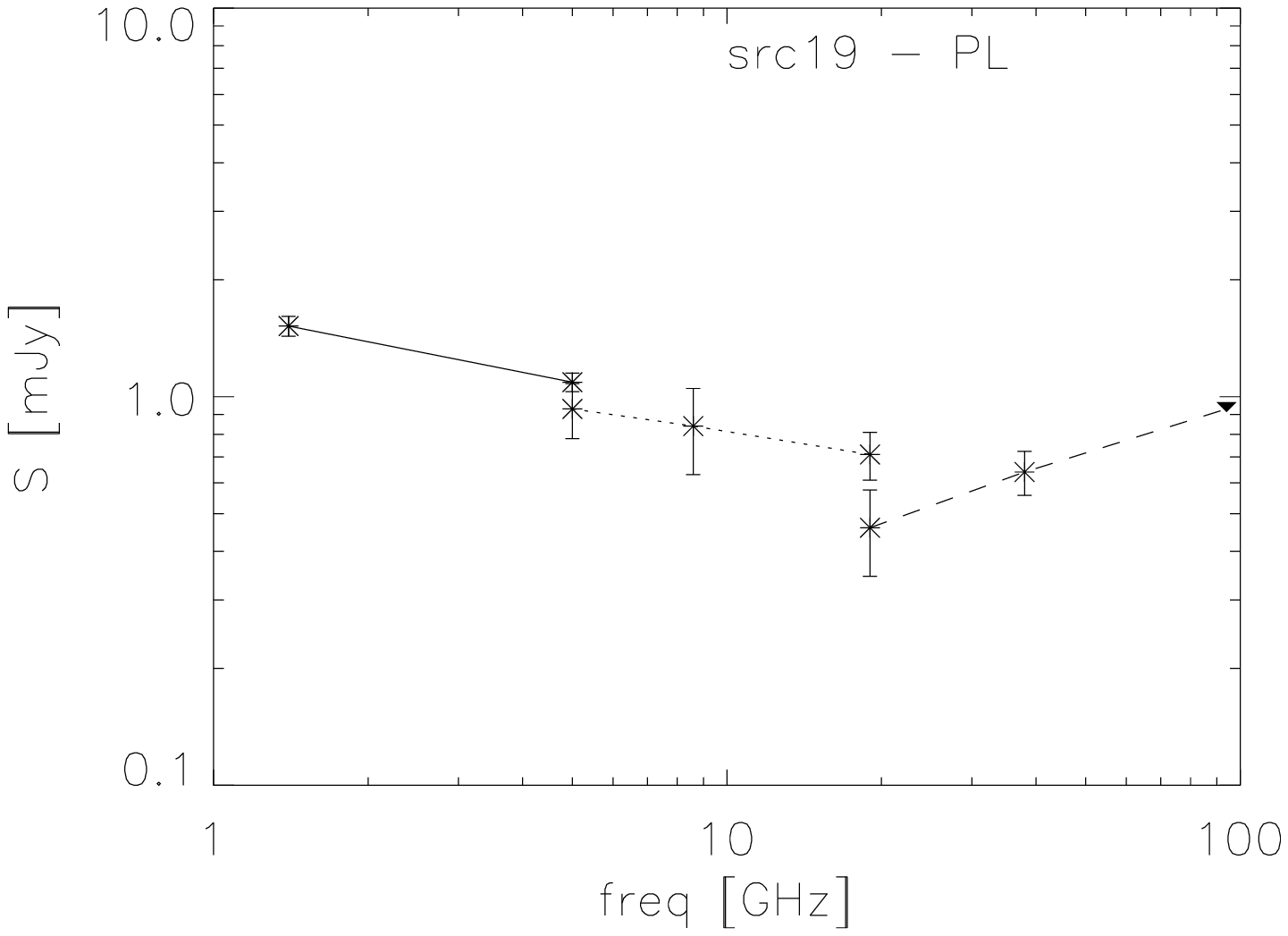}
\includegraphics[width=8cm]{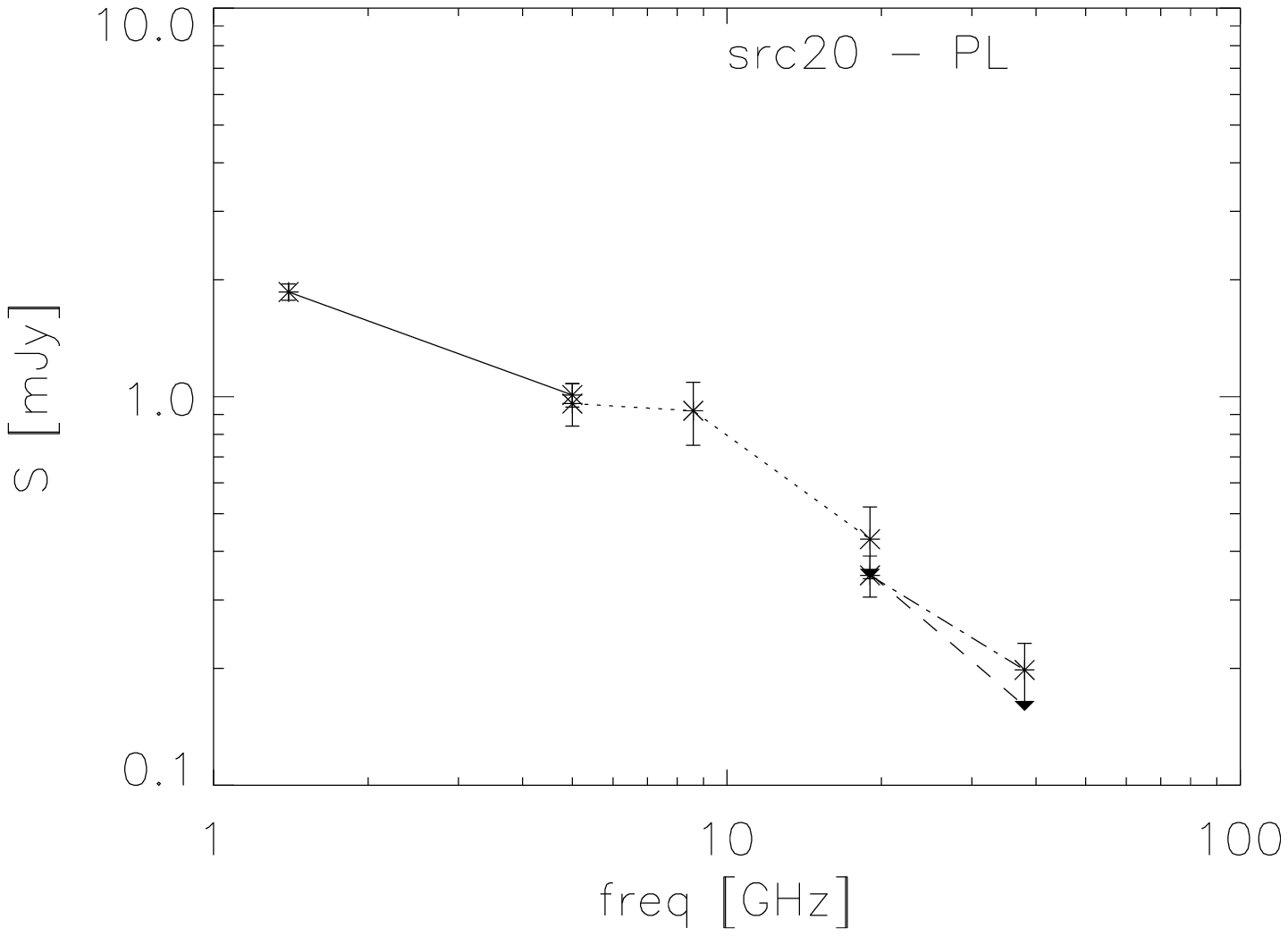}
\includegraphics[width=8cm]{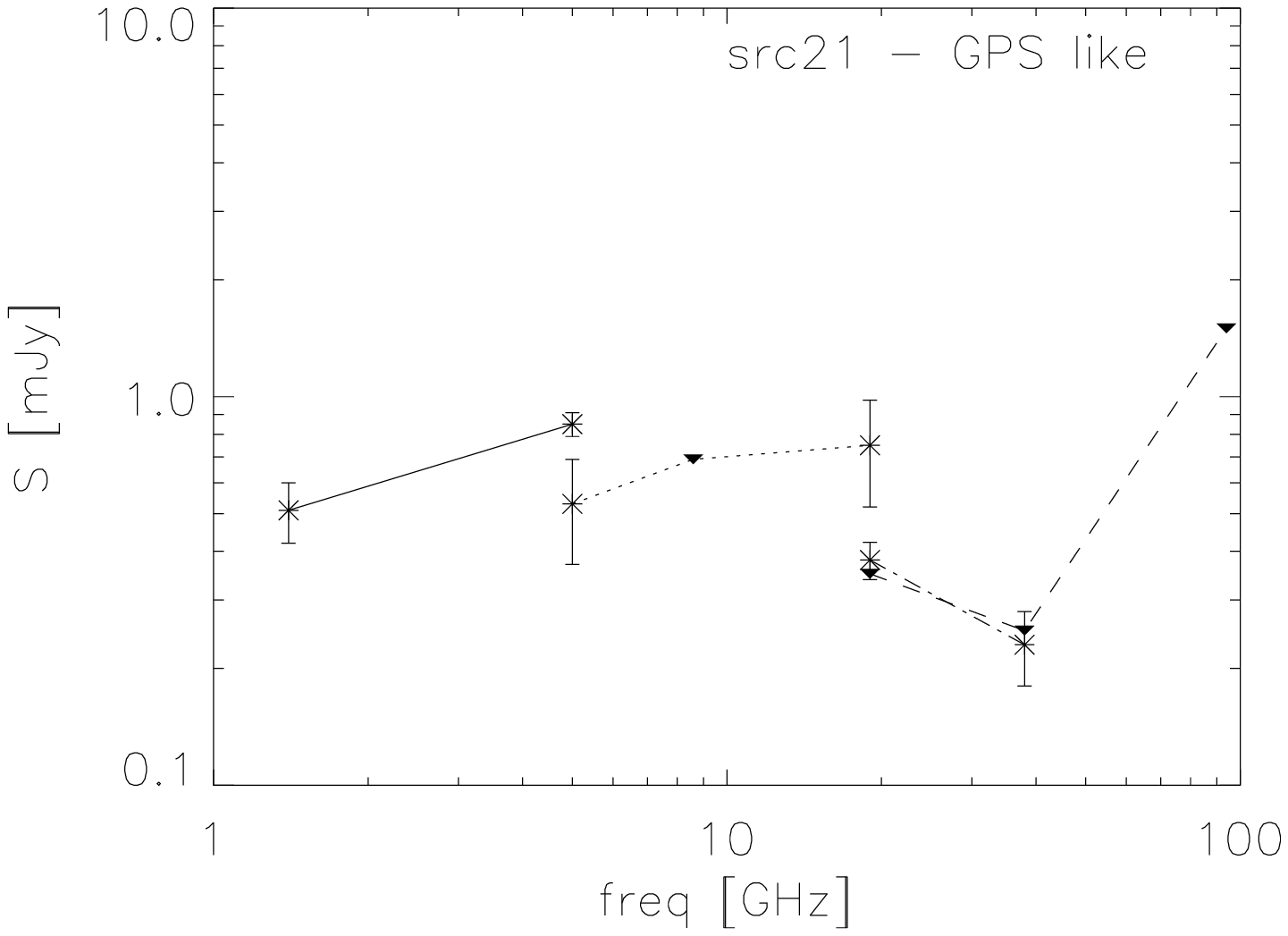}
\caption{Continued.}
\label{fig:spec3}
\end{figure*}

\addtocounter{figure}{-1}
\begin{figure*}[h]
\centering
\includegraphics[width=8cm]{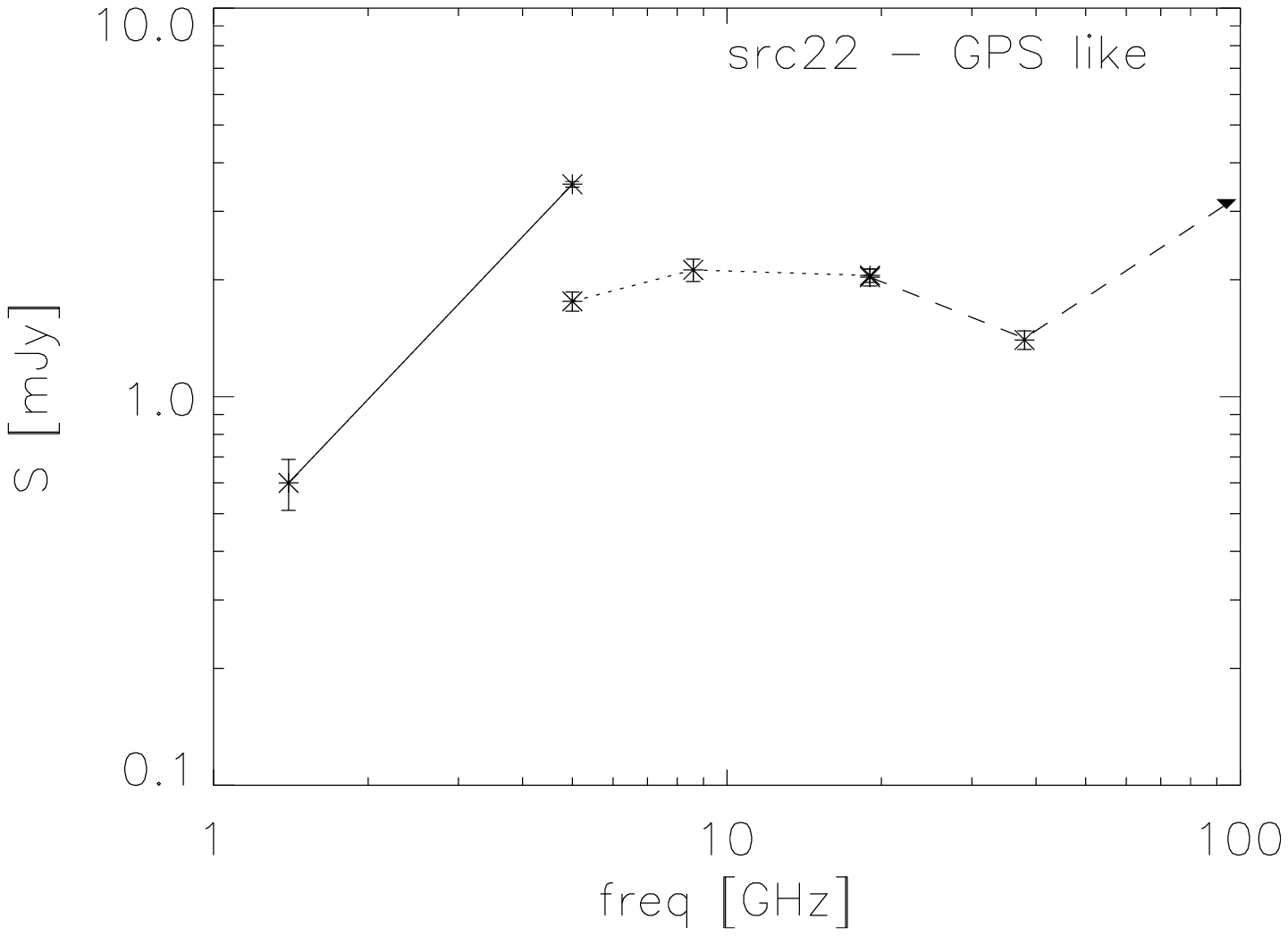}
\includegraphics[width=8cm]{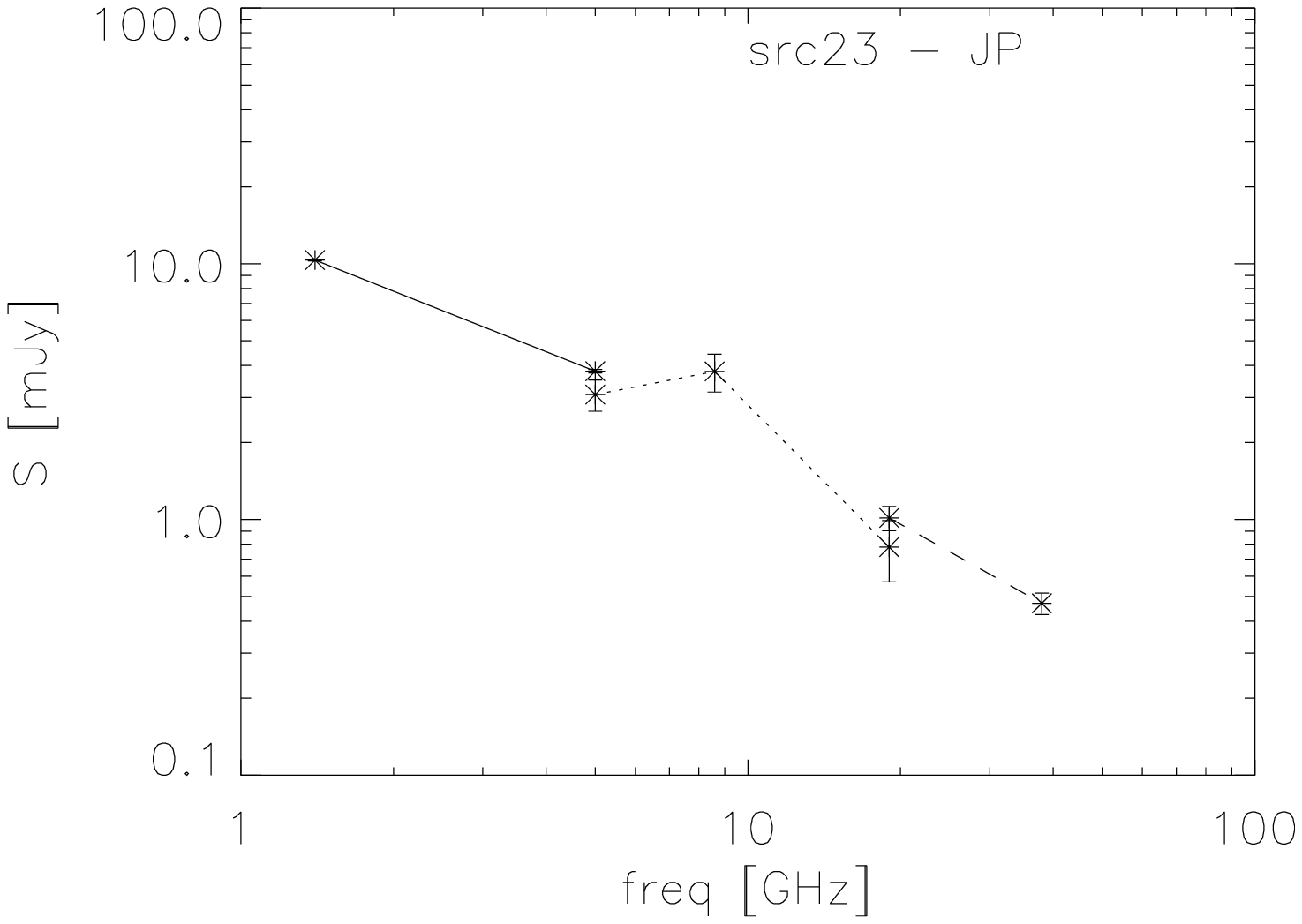}
\includegraphics[width=8cm]{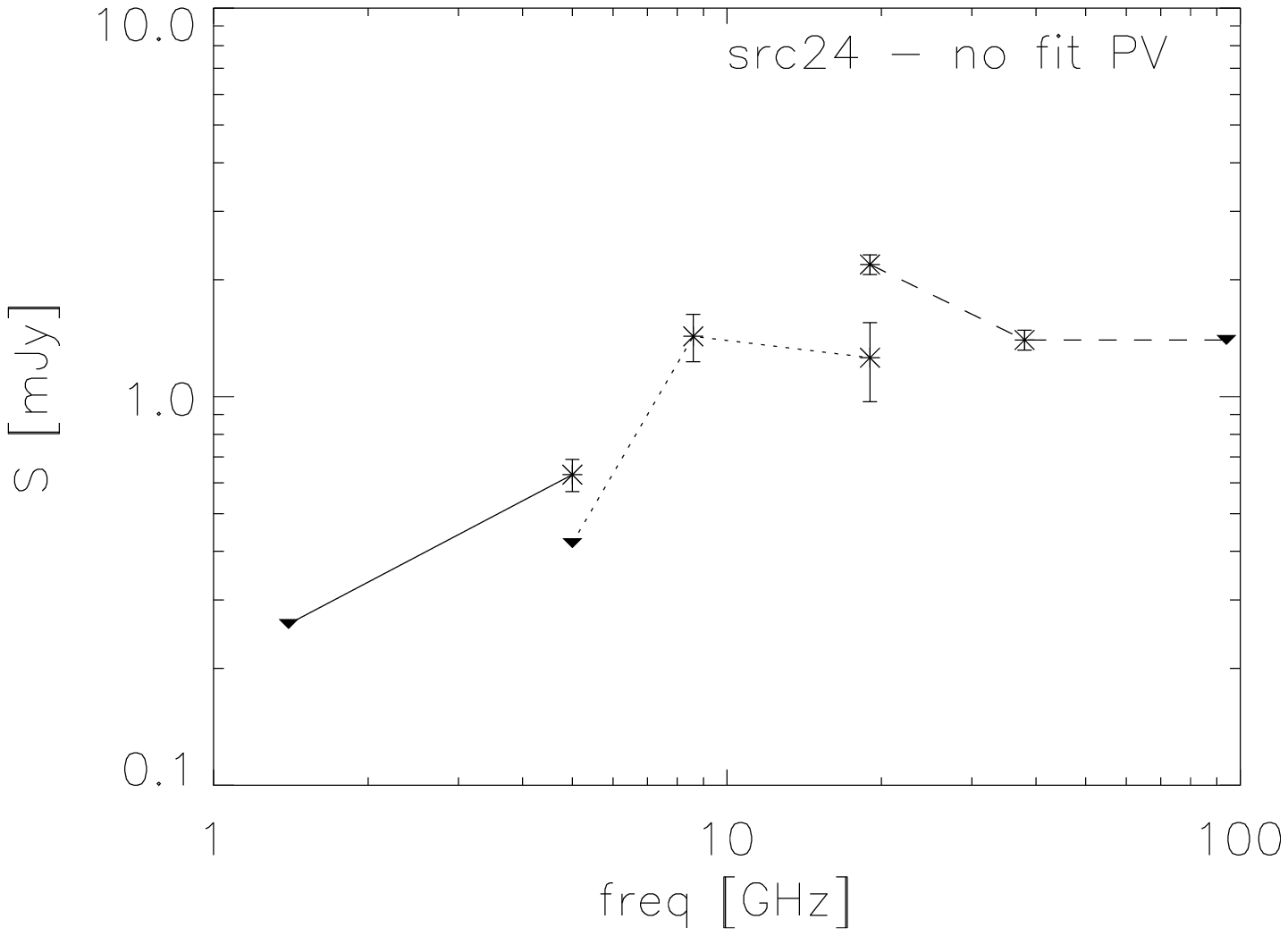}
\includegraphics[width=8cm]{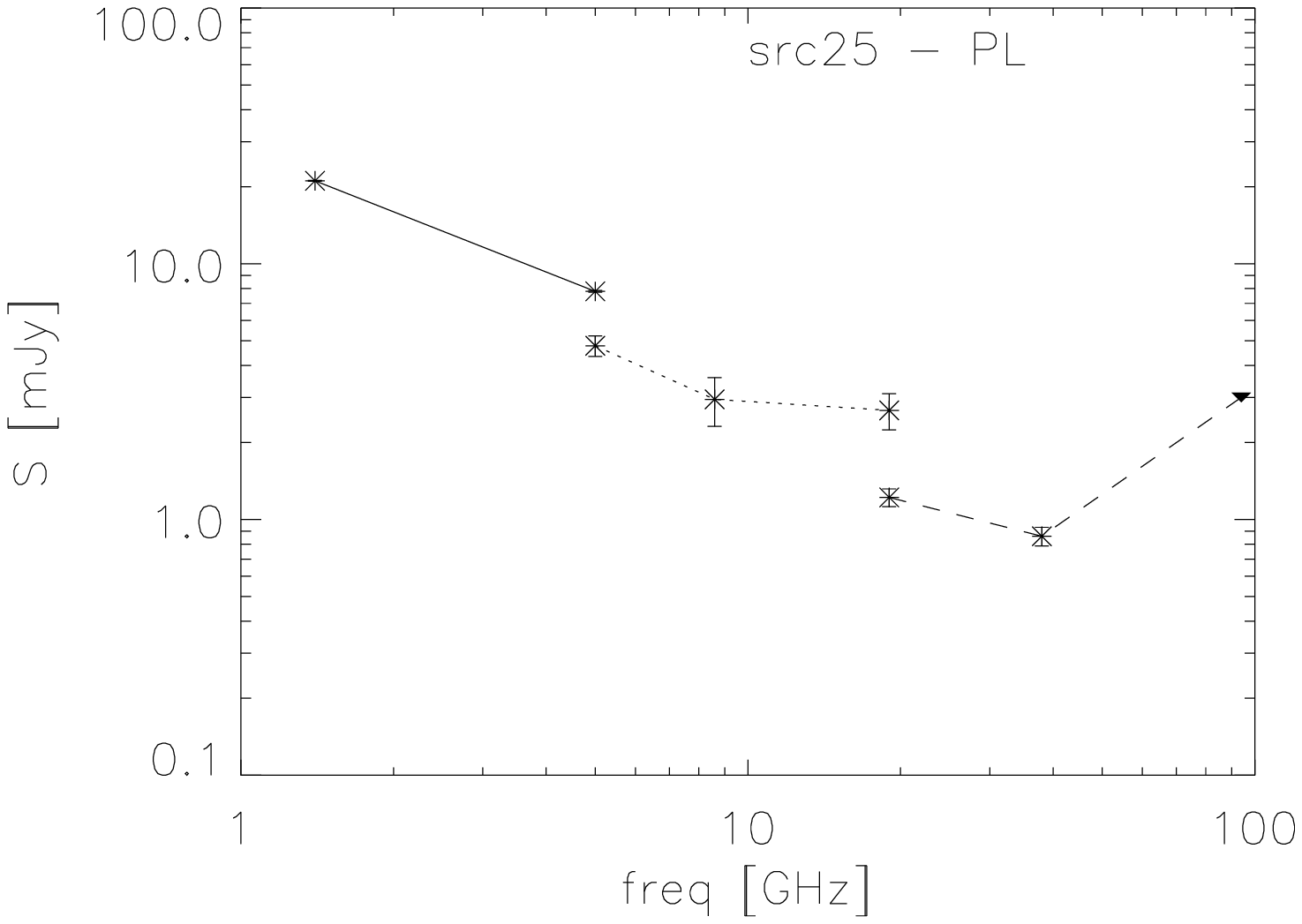}
\includegraphics[width=8cm]{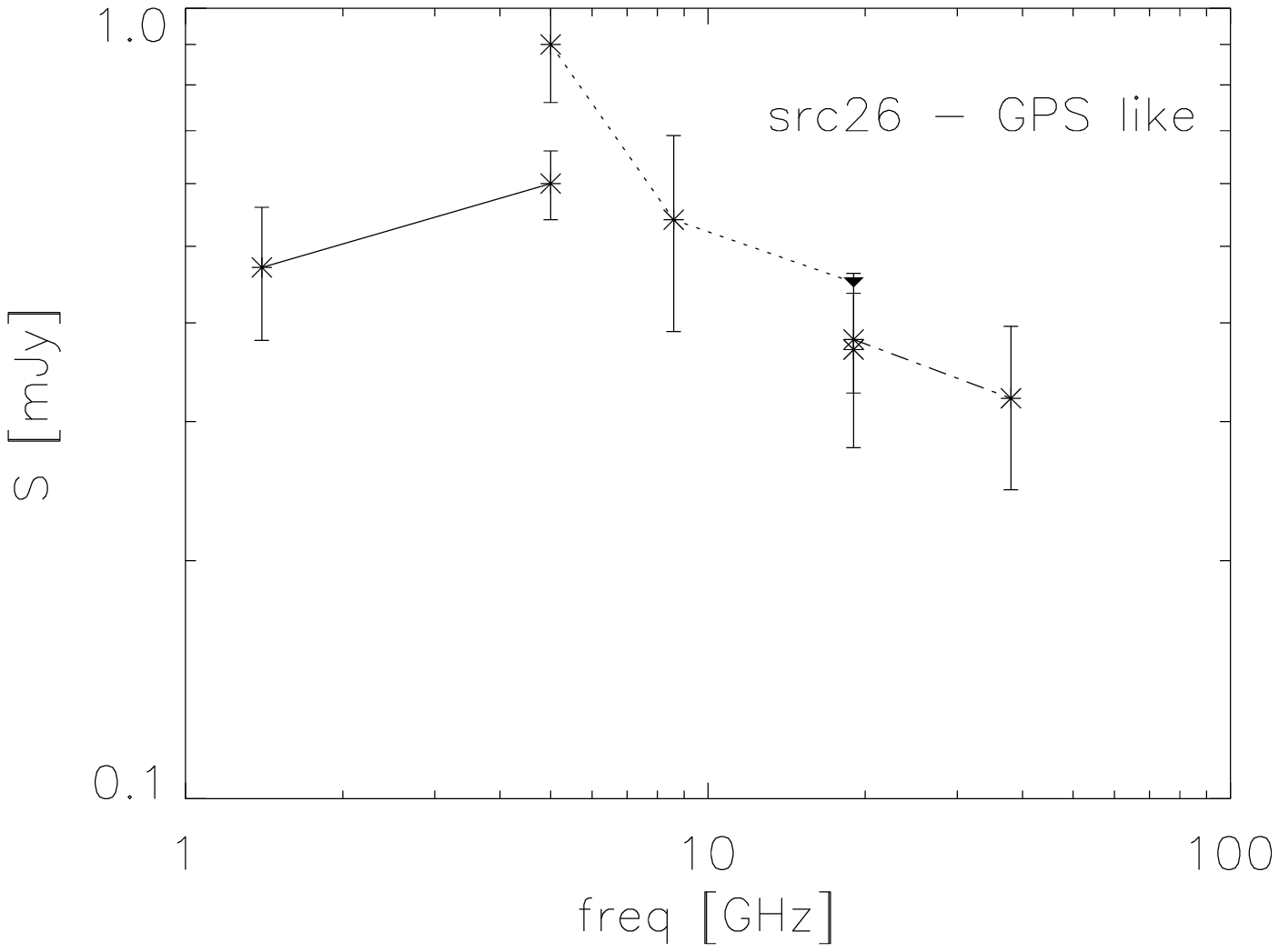}
\includegraphics[width=8cm]{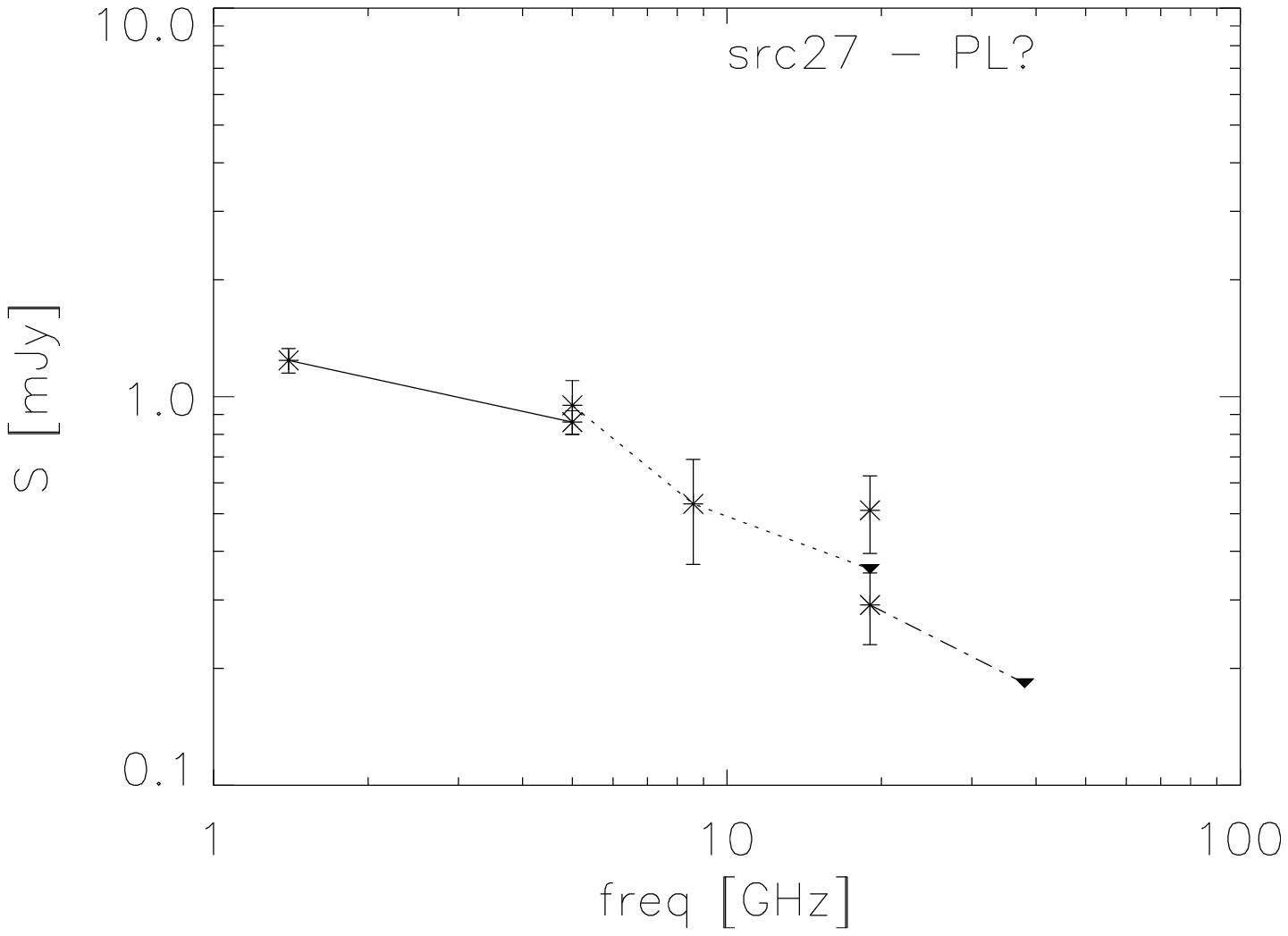}
\includegraphics[width=8cm]{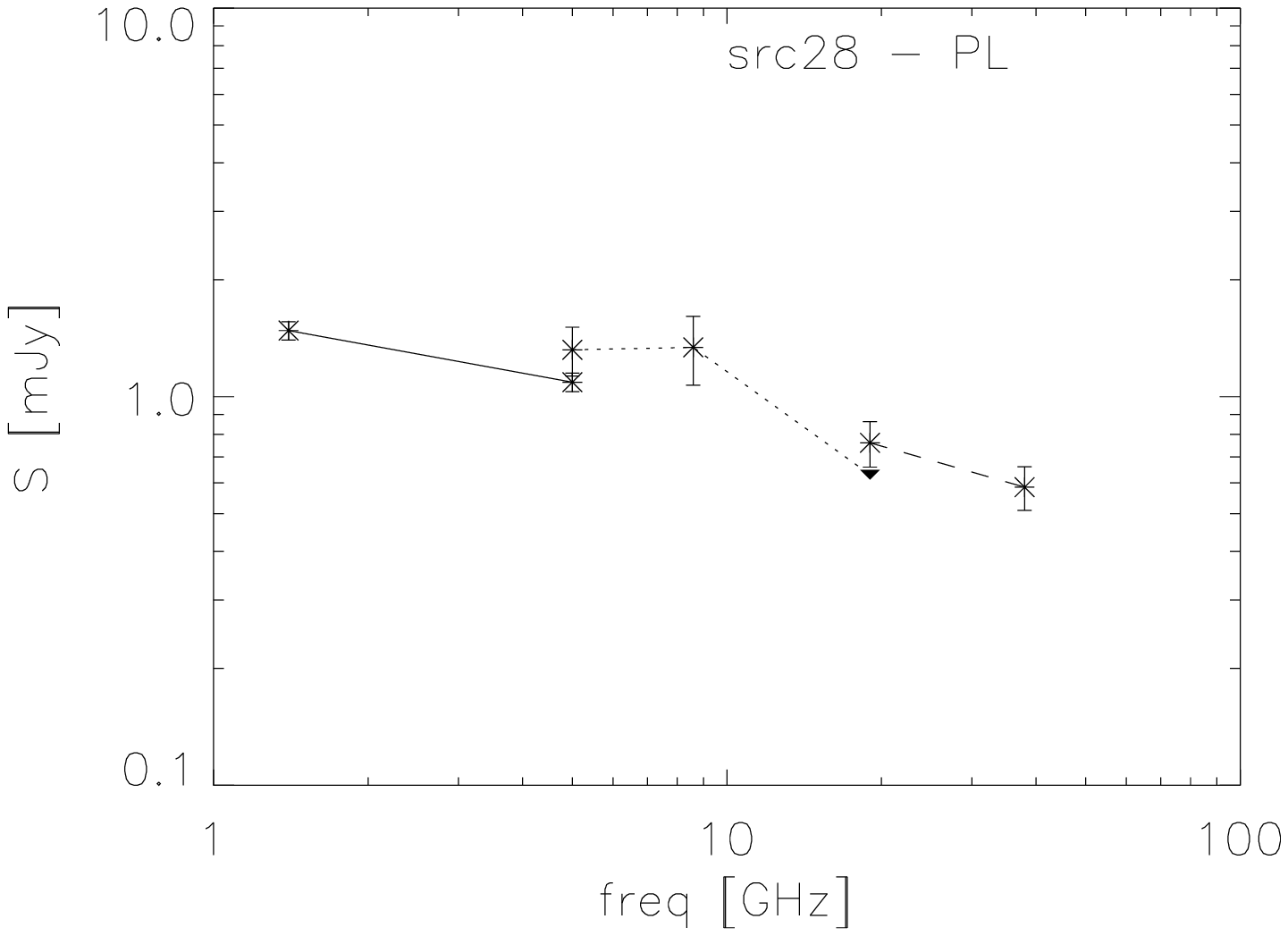}
\caption{Continued.}
\label{fig:spec4}
\end{figure*}


\subsection{High-resolution data analysis}

The 38-GHz I-Stokes images with an angular resolution of 0.6 arcsec 
were inspected with the KARMA software tool {\it kvis}. 
Only eight sources showed a detection above 3 $\sigma$ when the image peaks 
in the I-Stokes uncleaned maps were compared with 
the rms noise values in the V-Stokes maps. In 20 cases no bright peak was found 
in the centre of the map. Seven of the eight detected sources appeared point-like 
in these high-resolution maps.  
For the brightest source in the sample (J224547-400324) we self-calibrated the phases 
in the complex visibilities. This approach produced a structure slightly elongated 
to the northwest (NW). This source appeared as resolved also at low frequency 
(see Table~\ref{tab:lumlin} where the linear sizes of the sources are reported 
based on 5~GHz imaging with resolutions $\geq 2$ arcsec, see Paper I for more details). 
The 38-GHz flux densities (or upper limits) obtained from the analysis 
of the high-resolution images are also reported in Table~\ref{tab:intflux}. 

\begin{figure}[h]
\centering
\includegraphics[width=8cm]{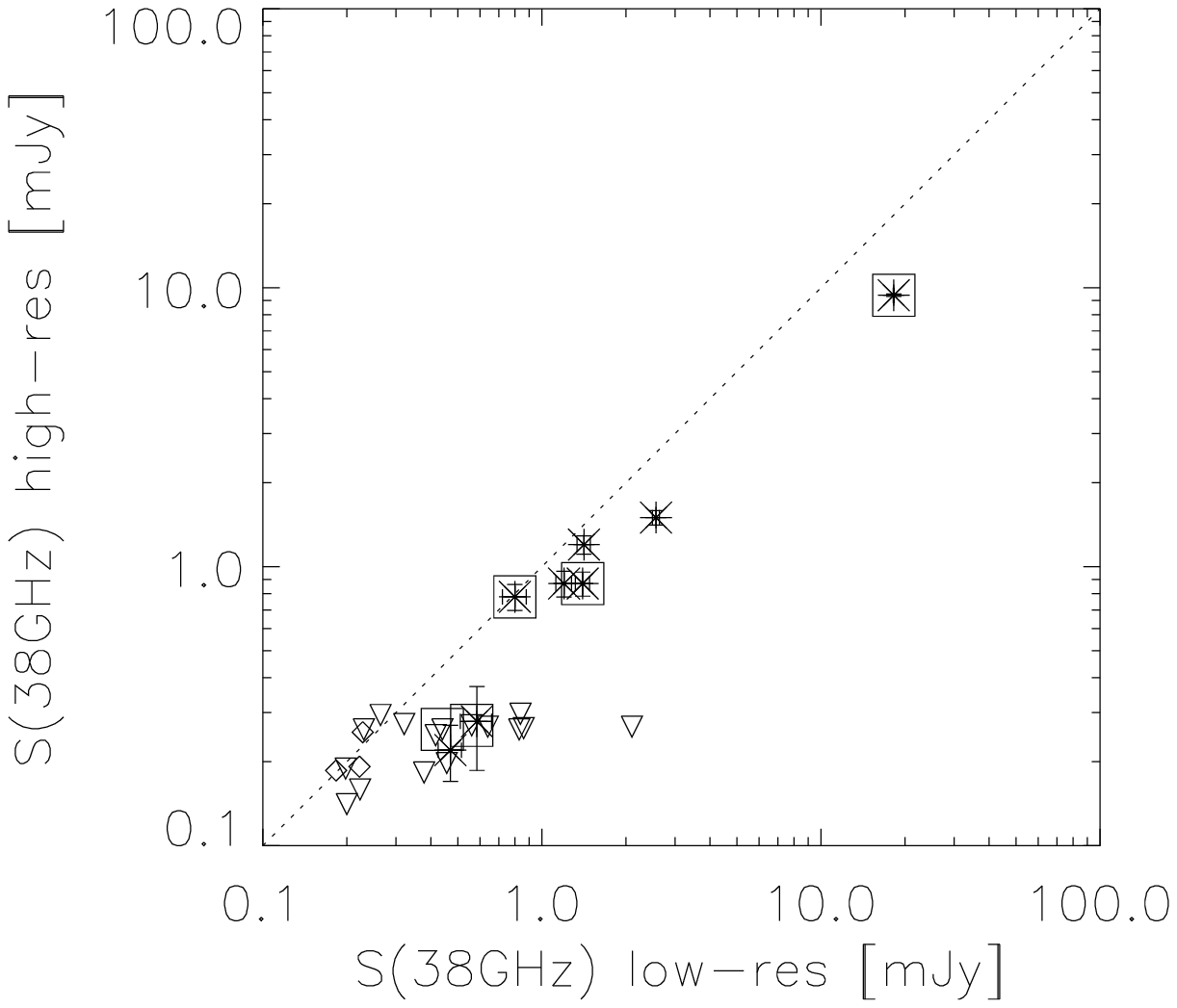}
\caption{Plot of high- vs. low-resolution 38-GHz flux density. Asterisks with error bars are 
detections at both high and low resolution. Downward triangles show sources with 38-GHz high-resolution 
upper limits and low-resolution detections. Diamonds are upper limits at both resolutions. Sources 
tagged as variable are also indicated by squares.}
\label{fig:hires}
\end{figure}

In Fig.~\ref{fig:hires} we show high- versus low-resolution 38-GHz 
flux densities. Apart from the 8 high-resolution detections, 17 sources present 
an upper limit in their high-resolution flux densities and 3 sources show an upper 
limit in both high- and low-resolution maps. The strongest 
sources are detected at both high- and low resolution. All sources are located   
below or on the one-to-one flux density line meaning that part of the total flux
density of the sources is resolved out when imaged at high resolution. The large majority of
sources with low-resolution 38-GHz flux density below 1 mJy lose up to 50\% 
of their total flux density when imaged at high resolution,
due to structure that is resolved out in high-resolution maps.  
The three sources with both high- and low-resolution upper limits    
(diamonds) are clustered on the one-to-one line. This is caused by 
a selection effect: both high- and low-resolution maps have similar detection limits. 
The seven sources resolved out in their 38-GHz high-resolution maps appeared as resolved
also at 5 GHz (see Table~\ref{tab:lumlin}), except for two sources 
(J225430-400334 and J225529-401101). For these latter sources we can infer 
an 0.6-arcsec lower limit on their 38-GHz angular size. 
       
An additional technique based on the amplitudes of the visibilities was investigated 
in order to extract source structure information from the data. This technique was first used by 
\cite{Chhetri13} to detect high-resolution structure in the visibility 
data of the sources in the AT20G catalogue. This technique consists of taking the 
ratio  of the averages of visibility amplitudes (called gamma factor) 
for all the baselines with and without antenna CA06 and computing it as a function of HA. 
The ratio of the averaged amplitudes between 
long and short baselines in these two sub-arrays provides a clear 
indication of the presence of structure along certain HAs. 
 
We applied this technique to the 19-GHz dataset only, because the 38-GHz data did 
not have high-enough S/N (i.e. the error bars in the amplitude ratio were too large)
to discriminate a significant offset from the threshold value, the point-like 
case. In addition, the 19-GHz visibilities are the most resistent to calibration instabilities. 
To test this technique, the gamma factor was computed for the phase 
calibrator 2223$-$488 for each scan and IF. The phase calibrator 2223$-$488 is known to be 
point-like at the angular resolution sampled by this technique (0.15 arcsec at 19~GHz). 
A gamma factor of 0.82 corresponds to a single-component Gaussian source with a
FWHM of 0.15 arcsec. Sources with gamma factor values higher than this threshold are 
therefore considered point-like for this array configuration.
Indeed the gamma factor values of 2223$-$488 are found to be consistent with the point-like case 
within the error bars. Having found that the gamma factor technique works with our 
test source (the phase calibrator), we computed the gamma factor for the ATESP sample sources 
and analysed their values versus HAs. 

In Table~\ref{tab:gfac} the gamma factor statistics is presented: for each source 
the mean value is obtained from the average of the gamma factor values over the full range 
of  HAs. For comparison the minimum gamma value ($\gamma_{\rm min}$) measured within that 
HA range is reported together with $\gamma_{\rm lim}$  (defined as 0.82-3$\sigma$, where $\sigma$ is the error 
associated with gamma mean), that is, the limit below which we consider a source to be significantly resolved. 
The analysis was performed on the 18 sources that appear point-like on the 
scatter plot in Fig.~\ref{fig:hires} based on the criterion $S_{\rm high} + 3\sigma \geq S_{low}$
where $S_{\rm low}$ is the 38-GHz low-resolution flux density, while $S_{\rm high}$ and $\sigma$ are 
the 38-GHz high-resolution flux density and its error bar. 

Sources with a mean gamma factor value $\gamma_{\rm mean} < \gamma_{\rm lim}$ at 19~GHz are likely to be extended 
on sub-arcsec angular scales along several directions. This never happens for our sources.
           
Comparing the minimum and limit gamma factor values in Table~\ref{tab:gfac} many sources seem to be resolved 
 in at least one direction (HA), having $\gamma_{\rm min} < \gamma_{\rm lim}$. Even if $\gamma_{\rm min}$ estimates are 
less robust to noise fluctuations and calibration errors, we note that this generally happens 
for sources known to be resolved from lower frequency - 5 GHz - observations (see Table~\ref{tab:lumlin}).
In three cases we measure $\gamma_{\rm min} < \gamma_{\rm lim}$ but sources appear unresolved at 5 GHz. 
In such cases (J224827-402515, J224919-400037 and J225436-400531) we can perhaps infer a lower limit 
to the source sizes of 0.15 arcsec. Source J225322-401931, on the other hand, appears as unresolved 
at both low and high frequency. For this source we can infer a tighter upper limit of 0.15 arcsec.       

\begin{table} 
\centering
\caption{Gamma factor statistics at 19 GHz. Source name, gamma factor mean value with error bar,
  gamma factor minimum value, and $\gamma$ lim: 3$\sigma$ below the gamma factor threshold are given in columns 2, 3, and 4.} 
  \label{tab:gfac}
  \begin{tabular}{lccc}
  \hline
Name    &   $\gamma_{\rm mean}$ & $\gamma_{\rm min}$  & $\gamma_{\rm lim}$ \\     
  \hline
2232$-$488 & $0.92\pm0.01$&   0.91   & 0.79       \\ 
src01   &   $0.82\pm0.02$ &   0.68  & 0.76        \\
src03   &   $0.85\pm0.03$ &   0.74  & 0.73        \\
src04   &   $0.78\pm0.03$ &   0.59  & 0.73        \\
src05   &   $0.81\pm0.02$ &   0.58  & 0.76        \\
src06   &   $0.74\pm0.03$ &   0.58  & 0.73        \\
src07   &   $0.84\pm0.02$ &   0.71  & 0.76        \\
src09   &   $0.84\pm0.02$ &   0.73  & 0.76        \\
src10   &   $0.86\pm0.04$ &   0.68  & 0.70        \\
src11   &   $0.88\pm0.03$ &   0.73  & 0.73        \\
src13   &   $0.81\pm0.03$ &   0.64  & 0.73        \\
src14   &   $0.86\pm0.02$ &   0.74  & 0.76        \\
src16   &   $0.79\pm0.03$ &   0.72  & 0.73        \\
src18   &   $0.85\pm0.03$ &   0.60  & 0.73        \\
src20   &   $0.89\pm0.02$ &   0.83  & 0.76        \\
src21   &   $0.87\pm0.02$ &   0.81  & 0.76        \\
src22   &   $0.83\pm0.03$ &   0.79  & 0.73        \\
src25   &   $0.70\pm0.04$ &   0.61  & 0.70        \\
src26   &   $0.84\pm0.03$ &   0.66  & 0.73        \\
  \hline 
\end{tabular}
\end{table}

\begin{figure}[h]
\centering
\includegraphics[width=8cm]{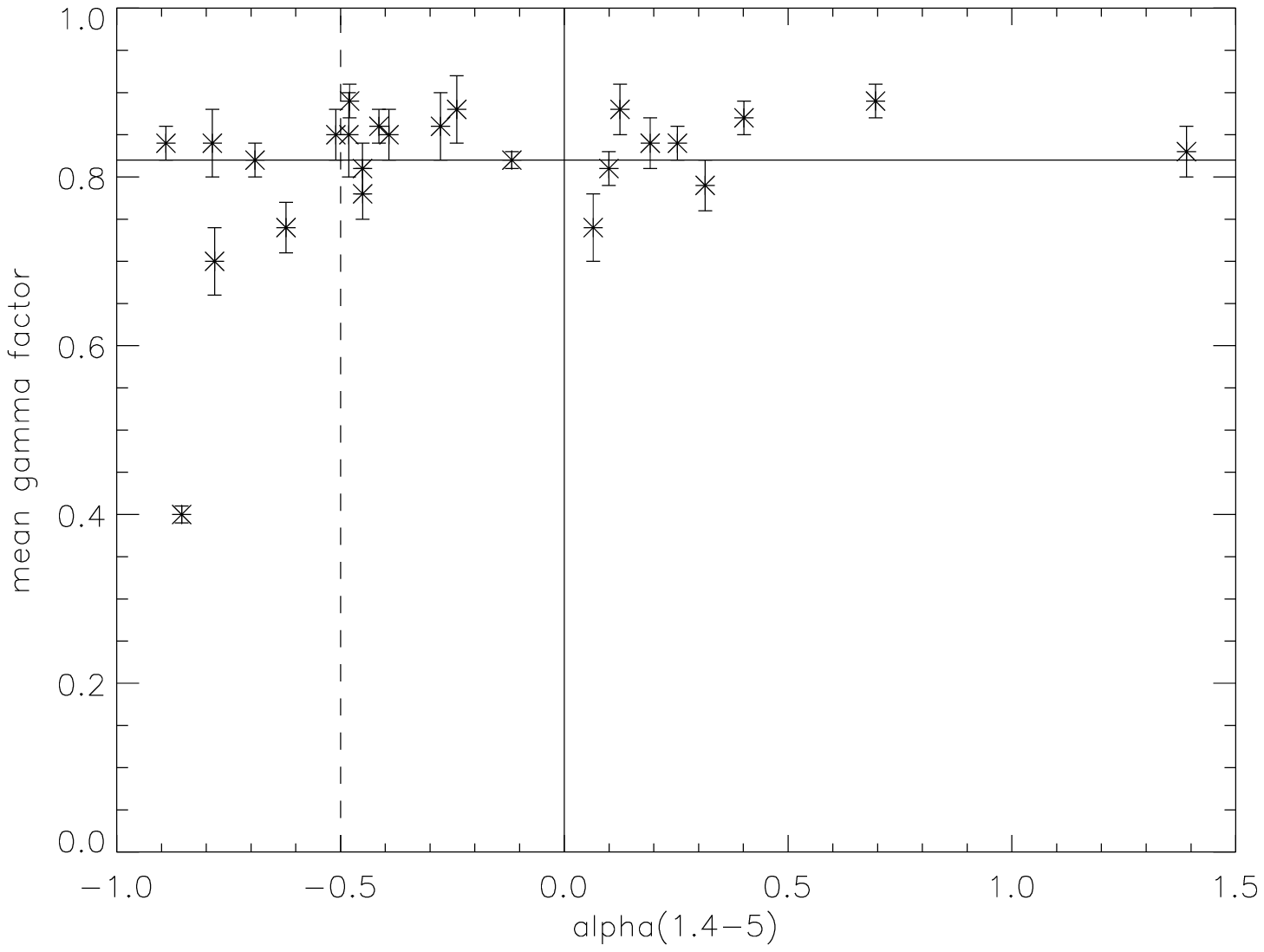}
\caption{Gamma factor vs. $\alpha_{1.4}^{5}$ spectral index.
The vertical dashed and solid lines indicate $\alpha =-0.5$ and 
$\alpha = 0$, respectively, i.e. the canonical values used to separate 
steep from flat and flat from inverted spectrum sources, respectively. The horizontal solid line 
is the gamma factor threshold for compact sources (angular resolution $\theta < 0.15$ 
arcsec).}
\label{fig:gamma}
\end{figure}

For a more meaningful comparison, the 
plot in Fig.~\ref{fig:gamma} showing the mean $\gamma$ factor versus the spectral index
includes all the ATESP sources and not only those unresolved at 38 GHz.
\cite{Chhetri13} found a correlation between flat (steep) and compact (extended) 
sources with extended sources occupying the plot region with $\alpha_{1.4}^{5} < -0.5$ showing
average gamma factor values significantly lower than the threshold. 
Compact sources are instead grouped around the threshold line independently of the $\alpha_{1.4}^5$ 
value. In the ATESP sample most of the sources belong either to the steep-spectrum compact or 
the flat/inverted compact classes, with only one object (the double source J225034-401936) 
clearly in the steep-spectrum/extended part of the plot.  
  
\section{Source variability}

We have multiple 19-GHz flux density measurements of the ATESP sample. We firstly 
used the September 2011 and July 2012 observations to determine the two-epoch
flux density variability. In case a source was not observed in one of these
two epochs or an upper limit was present, the 2007-08 flux densities were used instead.     
We define as probably variable the sources for which the probability that
their flux density is constant over the probed timescale is less than 0.5\% and 
variable the sources for which this probability is less than 0.1\%. 
We obtained this statistic via the variability estimator $VE=|\Delta S/\epsilon_{c}|$ (\citealt{Gregorini86})  
where $\Delta S$ is the difference between the flux 
densities at the two epochs and $\epsilon_{c}$ is the quadratic sum of the 
flux density errors at the two epochs. The probably variable sources 
according to this statistic have a variability estimator value of $VE > 2.81$, 
whereas variable sources have $VE > 3.29$. In cases where a source was tagged as 
variable or probably variable the strength of its flux density variability 
was also measured using the estimator by \cite{Sadler06}:
 
Vrms=$\frac{100}{<S>}\sqrt{\sum_{i}(S_{i}-<S>)^2 - \sum_{i}\sigma_{i}^2}$. 

\begin{figure}[h]
\centering
\includegraphics[width=10cm]{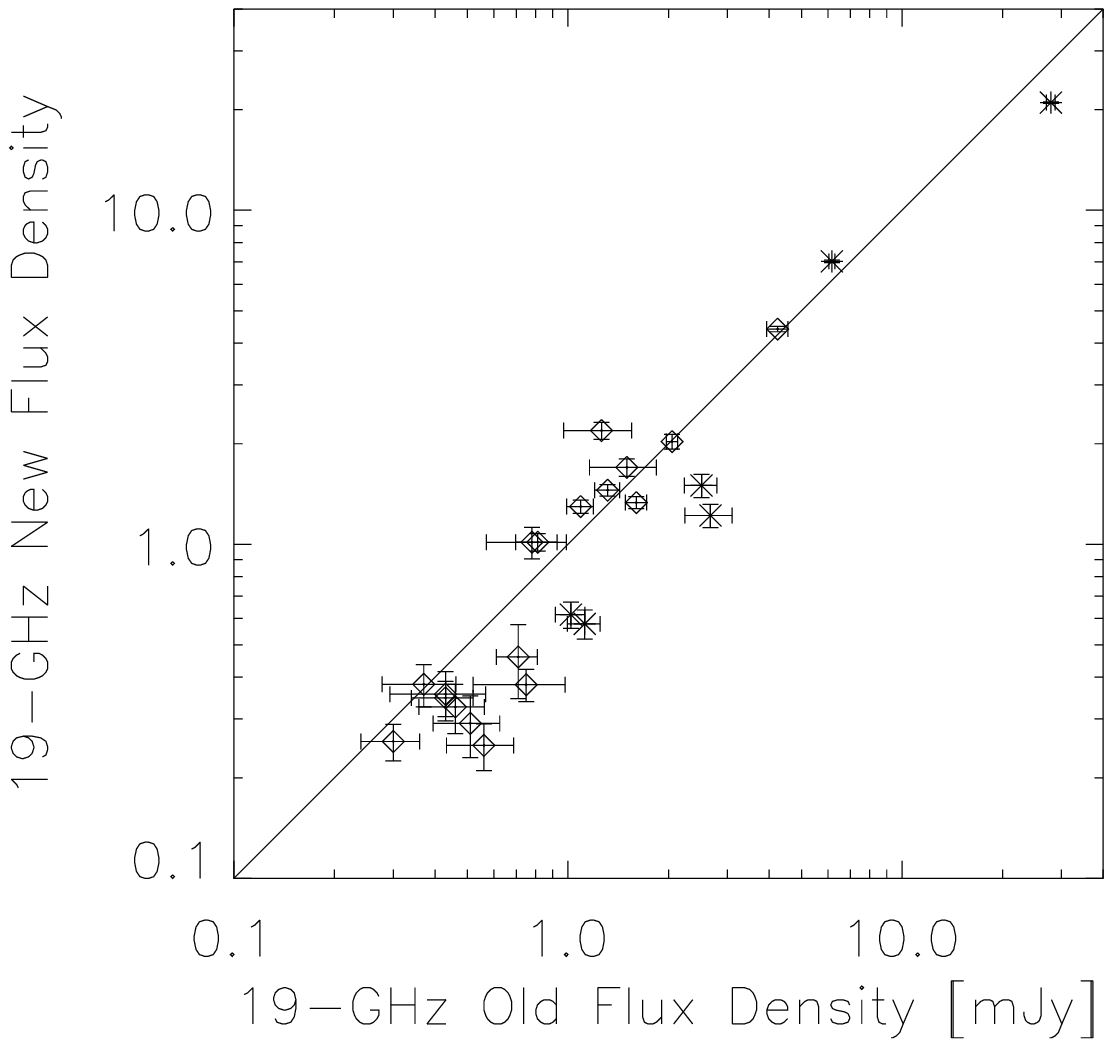}
\caption{Flux-flux scatter plot comparing the 19 GHz flux density 
measurements taken during the 2011-2012 campaign. When one of the two measurements 
was missing or only an upper limit was available, the 2007-2008 flux density 
was used instead, if available.}
\label{fig:flux19}
\end{figure}

These results are shown in Table~\ref{tab:var} (columns 2-4). 
Figure~\ref{fig:flux19} shows the scatter plot comparing 
the 19-GHz flux density measurements at the two epochs. 
We found six variable sources (src02, src09, src13, src15, src16, src25) 
and one probably variable source (src24). 
The probably variable source src02 was also found to be variable in the flux variability analysis presented 
in Paper III (\citealt{Prandoni10}), and src07 and src22 which were tagged as variable 
in Paper III were not flaring in 2011-2012. For four sources, no variability estimates 
were possible: src03, src11, src14 and src28.

\begin{table}[t] 
 \centering
  \caption{19 GHz variability statistics: Columns 1-7: source name; two-epoch
VE value of (III) means that that source was found to be variable 
also in the analysis presented in paper III; variability tag: V for variable, P for probably variable;
variability strength associated to two-epoch VE;  $\chi^2$ test for sources with 
19 GHz flux measurements for more than two epochs; associated probability of random fluctations;
variability strength for sources with P $<$0.1\%.} \label{tab:var}
  \begin{tabular}{lcccccc}
  \hline
  (1)     &    (2)    &    (3)   &   (4)       &  (5)      & (6)     &  (7)           \\
name      &    VE     &   Tag    &  Vrms       &  $\chi^2$ & P       &  Vrms          \\
          &           &          &  (\%)       &           &         &  (\%)          \\
  \hline    
src01    &    1.77   &   $-$    &  $-$        &  $-$    &  $-$       &    $-$         \\
src02    &  8.14(III)&    V     &  13.9       &  $-$    &  $-$       &    $-$         \\
src04    &    2.08   &   $-$    &  $-$        &  5.62   &  0.06      &    $-$         \\
src05    &    1.56   &   $-$    &  $-$        &  $-$    &  $-$       &    $-$         \\  
src06    &    1.09   &   $-$    &  $-$        & 13.86   &  9.8e-4    &    5.0         \\
src07    &  0.63(III)&   $-$    &  $-$        &  0.51   &  0.77      &    $-$         \\
src08    &    0.52   &   $-$    &  $-$        &  $-$    &  $-$       &    $-$         \\
src09    &     3.45   &   V     &  22.5       &  $-$    &  $-$       &    $-$         \\
src10    &     1.16   &  $-$    &  $-$        &  2.43   &  0.30      &    $-$         \\
src12    &     0.56   &  $-$    &  $-$        &  $-$    &  $-$       &    $-$         \\
src13    &     3.90   &   V     &  29.8       & 54.17   & <1e-5      &    26.5        \\  
src15    &     6.10   &   V     &   6.4       & 124.04  & <1e-5      &     6.1        \\
src16    &     3.31   &   V     &  22.8       &  $-$    &  $-$       &    $-$         \\
src17    &     0.50   &  $-$    &  $-$        &   1.11  &  0.57      &    $-$         \\
src18    &     2.33   &  $-$    &  $-$        &  $-$    &  $-$       &    $-$         \\
src19    &     1.64   &  $-$    &  $-$        &  $-$    &  $-$       &    $-$         \\
src20    &     0.84   &  $-$    &  $-$        &  $-$    &  $-$       &    $-$         \\ 
src21    &     1.58   &  $-$    &  $-$        &  $-$    &  $-$       &    $-$         \\
src22    &  0.15(III) &  $-$    &  $-$        &  $-$    &  $-$       &    $-$         \\   
src23    &     0.99   &  $-$    &  $-$        &  $-$    &  $-$       &    $-$         \\
src24    &     2.93   &   P     &  23.6       &  $-$    &  $-$       &    $-$         \\
src25    &     3.29   &   V     &  33.7       &  $-$    &  $-$       &    $-$         \\
src26    &     0.10   &  $-$    &  $-$        &  $-$    &  $-$       &    $-$         \\
src27    &     1.68   &  $-$    &  $-$        &  $-$    &  $-$       &    $-$         \\
  \hline
\end{tabular}
\end{table}

The results of a $\chi^2$ test performed on sources with N $>2$ epochs at 19~GHz between 2007 and 2012
are also presented in Table~\ref{tab:var} (Columns 5-7). Three (src06, src13 and src15) of seven sources with multiple 
19-GHz flux densities are classified as variable. For these sources there is a probability of less than 0.1\% 
of confusing true variability with statistical fluctuations in flux density. 
For those three sources the variability strength Vrms is also reported.
Only two sources (src13 and src25) show $\simeq$30\% variability over a one-year timescale, that is, 7-8 \% of the sources. 
Given the very low sample sizes, this is consistent with the results of \cite{Sadler06}, reporting  5\%  strongly variable sources at K-band. 

Summing together the results obtained from two- and multi-epoch variability analyses we find, 
however, that a significant fraction (about one third)  of our sources show some sign of variability. 
From  the overall spectra shown in Fig.~\ref{fig:spec1}, we see that at least three variable 
sources are doubtful: src6, src15 and src25. Indeed all three are very extended (>100 kpc, see 
Table~\ref{tab:lumlin}) and present relatively smooth steep spectra, 
that can be fitted by a one or two-component power law (see Sect. 6).
        
\section{Model fits of the radio spectra}

Inspection of Fig.~\ref{fig:spec1} shows that there are, broadly
speaking, four categories of radio spectra: i) smooth-steep spectra, sometimes
steepening further at the higher frequencies, ii) flat spectra, sometimes
steepening at the higher frequencies, iii) peaked spectra, that is, inverted at
low frequencies and steep at high frequencies, and iv) spectra with undefined shape, which
may be due to the presence of more than one radio component; in some cases however source
variability may be the most likely cause. Some examples are sources
src04, 10, 13, and 17 where there are clear jumps for flux densities at the 
same frequency but observed in different epochs.

In order to get a more quantitative idea we fitted the spectra with either 
a power-law (PL) or a Jaffe-Perola (JP; \citealt{Jaffe73}) model using the programme Synage (\citealt{Murgia99}). 
We used all detected flux densities and included upper limits only when no detection
was available at a particular frequency. We note that the fitting with a JP model assumes a fixed 
$\alpha_{injection}=0.7$. For a power-law fit, $\alpha_{injection}$ is a free parameter.

When possible we also added flux densities at 843 MHz using the Sydney University Molonglo Sky Survey
(SUMSS, \citealt{Mauch03}) maps. We computed 
the 843-MHz flux densities directly from the images with the help of the AIPS software. This could be done
for 11/28 sources, the remaining being too faint considering the flux limit of SUMSS.

For six sources (src01, src02, src06, src15, src23, src25) there is an entry in the
GaLactic and Extragalactic All-sky Murchison Widefield Array (GLEAM, \citealt{Hurley-Walker17}) catalogue
at 200 MHz. For five of them the 200-MHz point is consistent with the value obtained from the fitting model without the GLEAM point
and the $\chi^2$ to the fit improves by adding the GLEAM point. For src01, the flux density of the GLEAM point is higher than the 
fitting model value and this results in a slightly higher $\chi^2$ value. From visual inspection of the ATESP 1.4 and 5~GHz images
no other confusing source is present in the 2-arcmin beam of the GLEAM survey to justify a spurious flux density increase
and the source itself appears point-like. This probably means that the source has diffuse emission at low frequency 
which is not visible at high frequency.      

Four spectra showed signs of steepening at the highest frequencies. In those cases we applied a
JP model fit. For many other sources a power-law model turned out to provide an appropriate fit.
However, fitting turned out to be impossible for 10 of the 28 sources. For some (three) this is undoubtedly 
due to variability, but other causes need to be invoked for the remainder: either the sources are of the GPS type
(for a definition of GPS, see \citealt{ODea91} and also Section 6), or have multiple components. 
We note that, looking at the entire spectrum, there are in total 
seven GPS-like sources, of which two are also variable: src09, src11, src16 (variable), src21, src22, src24 (variable), src26. 

We summarise the results of the fitting procedure in Table \ref{tab:fit}.
It is worth noting that of the four sources fitted by a JP model, 
three are extended over more than 100 kpc (see Table~\ref{tab:lumlin}), 
and thus represent classical extended radio sources.

The majority of objects that can be fit allow a PL model, but some uncertainty remains; this is indicated by a question
mark in Table \ref{tab:fit}. 

\begin{table}[t] 
 \centering
  \caption{Spectral fitting: $\alpha_{\rm inj}$ is the injection spectral index; $\nu_{\rm b}$ is the model fit break
frequency; $\chi^2_{\rm red}$ is the reduced $\chi^2$ of the model fit; JP stands for the Jaffe-Perola model fit, PL for the 
power-law model fit. Comments: V(PV) stand for variable (probably variable) spectrum (see Table~\ref{tab:var}); Three sources satisfying the variability criteria are not confirmed as such from the inspection of the overall
spectra. These sources are src06, src15, and src25.  
(*) indicates a slightly more uncertain fit, notwithstanding the low $\chi^2$ value (which
is entirely due to the large measurement errors); (**) indicates an acceptable fit but the relatively high
$\chi^2$ value is due to the very small formal errors on the flux density.} \label{tab:fit}
  \begin{tabular}{lcccl}
  \hline
name   &        $\alpha_{\rm inj}$   &      $\nu_{\rm b}$ (GHz)      &  $\chi^2_{\rm red}$     &   Comments             \\
  \hline          
src01 &        0.70               &  169                        & 3.1                   & JP                     \\
src02 &        0.22               &  $-$                        & 5.5                   & PL? V                  \\
src03 &        0.51               &  $-$                        & 1.4                   & PL                     \\
src04 &        0.59               &  $-$                        & 3.4                   & PL                     \\
src05 &        $-$                &  $-$                        & $-$                   & no fit                 \\
src06 &        0.70               &  110                        & 4.0                   & JP                     \\
src07 &        0.75               &  $-$                        & 4.4                   & PL                     \\
src08 &        0.60               &  $-$                        & 1.5                   & PL                     \\
src09 &        $-$                &  $-$                        & $-$                   & no fit V               \\
src10 &        0.34               &  $-$                        & 0.7                   & PL? *                  \\
src11 &        $-$                &  $-$                        & $-$                   & no fit                 \\
src12 &        0.05               &  $-$                        & 3.6                   & PL                     \\
src13 &        $-$                &  $-$                        & $-$                   & no fit; V              \\
src14 &        0.69               &  $-$                        & 6.2                   & PL  **                 \\
src15 &        0.70               &  58                         & 3.6                   & JP                     \\
src16 &        $-$                &  $-$                        & $-$                   & no fit; V              \\
src17 &        0.37               &  $-$                        & 2.0                   & PL? *                  \\
src18 &        $-$                &  $-$                        & $-$                   & no fit                 \\
src19 &        0.30               &  $-$                        & 1.1                   & PL                     \\
src20 &        0.70               &  $-$                        & 5.1                   & PL  **                 \\
src21 &        $-$                &  $-$                        & $-$                   & no fit                 \\
src22 &        $-$                &  $-$                        & $-$                   & no fit                 \\
src23 &        0.70               &  85                         & 1.9                   & JP                     \\
src24 &        $-$                &  $-$                        & $-$                   & no fit; PV             \\
src25 &        1.00               &  $-$                        & 7.0                   & PL  **                 \\ 
src26 &        $-$                &  $-$                        & $-$                   & no fit                 \\
src27 &        0.47               &  $-$                        & 2.7                   & PL?  *                 \\
src28 &        0.26               &  $-$                        & 1.0                   & PL                     \\
  \hline
\end{tabular}
\end{table}

\section{Colour-colour diagrams}

\begin{figure}[h]
\centering
\includegraphics[width=10cm]{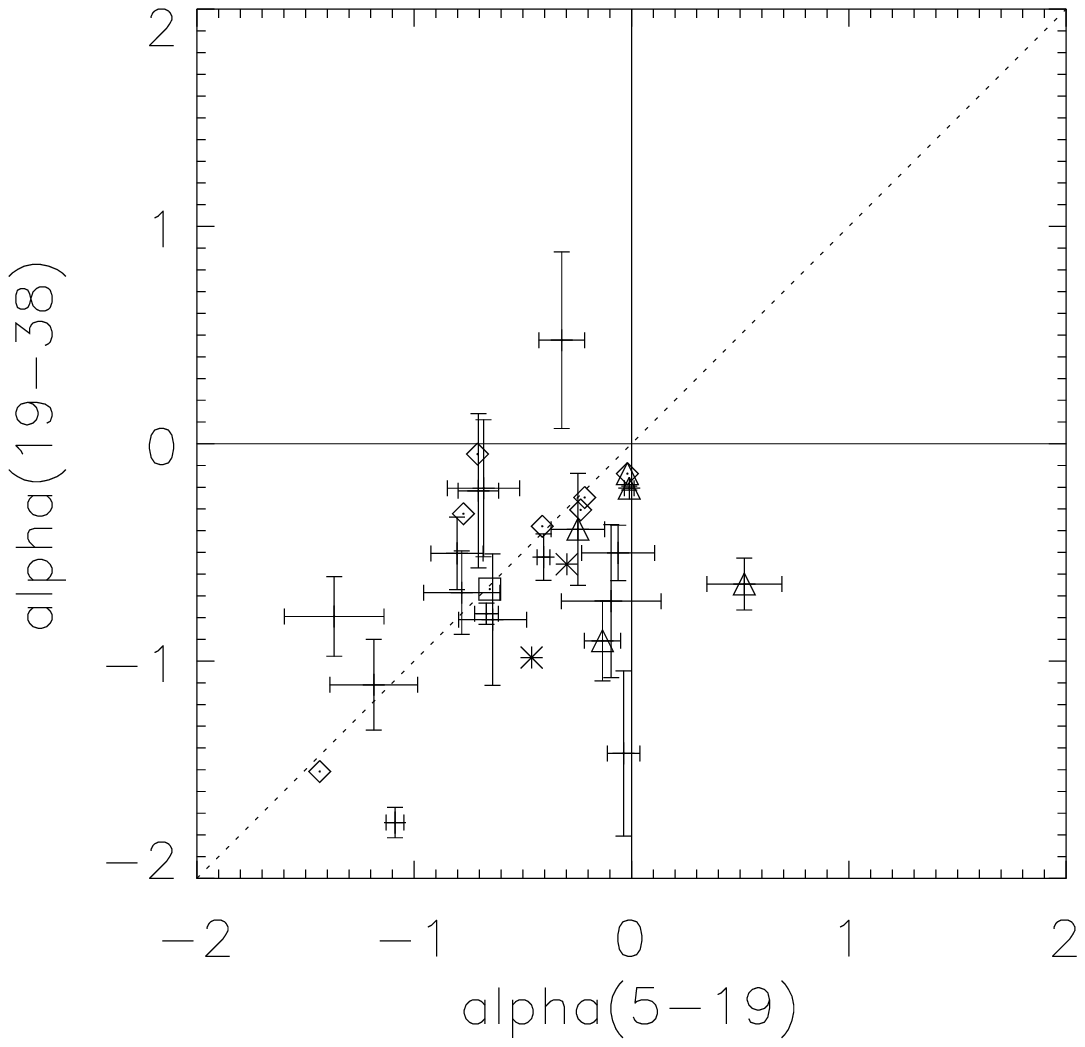}
\caption{Colour-colour plot of the ATESP sample with error bars. The 
triangles are variable sources. Stars are $\alpha_{19}^{38}$ upper limits. 
Diamonds are $\alpha_{5}^{19}$ upper limits. Squares represent upper limits in 
both $\alpha_{5}^{19}$ and $\alpha_{19}^{38}$, while crosses are detections in 
both axes.}
\label{fig:alpalp}
\end{figure}

A very useful tool for further exploring the spectral behaviour of our sample
is offered by colour-colour plots 
(\citealt{Sadler06}) where the spectral index between frequencies in the 
high-frequency part of the radio source spectrum is plotted against a spectral 
index in a low-frequency range. This allows the source sample to be split into 
different populations based on the radio spectral properties: (i) lower-left
quadrant of the plot for steep-spectrum sources; (ii) upper-left quadrant for 
upturn sources, that is,  sources which show a trough in their spectrum 
at a certain frequency; (iii) upper-right quadrant for rising spectrum sources,
and (iv) lower-right quadrant for peaked spectrum sources, that is, sources
with a radio spectrum peaking in the range of frequencies explored. This last 
sub-population is typical of GPS-like (GHz-Peaked-Spectrum) sources (\citealt{ODea98}). 
Figure \ref{fig:alpalp} shows the colour-colour plot of the ATESP early-type source sample with 
error bars. In this analysis we extended the work done in \cite{Prandoni10}
to 38 GHz. We thus plotted the spectral index between 19 and 38 GHz $\alpha_{19}^{38}$ as 
$\alpha_{high}$ versus the spectral index between 5 and 19 GHz $\alpha_{5}^{19}$ as $\alpha_{\rm low}$ 
(which was the $\alpha_{\rm high}$ in the spectral index analysis of \citealt{Prandoni10}).
We found that pure ADAF models are still ruled out by the higher-frequency extension 
of the colour-colour plot, as most of the sources are crowding in the steep-spectrum quadrant,
with just one source (src19) in the upturn quadrant and another one (src24) 
in the peaked-spectrum quadrant. This latter source is GPS-like by its definition
in the colour-colour diagram. Other sources that, based on their global spectra, could 
be classified as GPS-like have their spectral peak at frequencies lower than the ones 
sampled in Fig.\ref{fig:alpalp}, and therefore do not appear in the peaked spectrum source quadrant.        
We also notice that a good number of sources are located above the 1:1 line defining single-power-law spectra. 
Such sources show a flattening going to high frequencies. This flattening is probably associated 
to a core component that shows up and dominates at $>19$~GHz.

\begin{figure}[h]
\centering
\includegraphics[width=8cm]{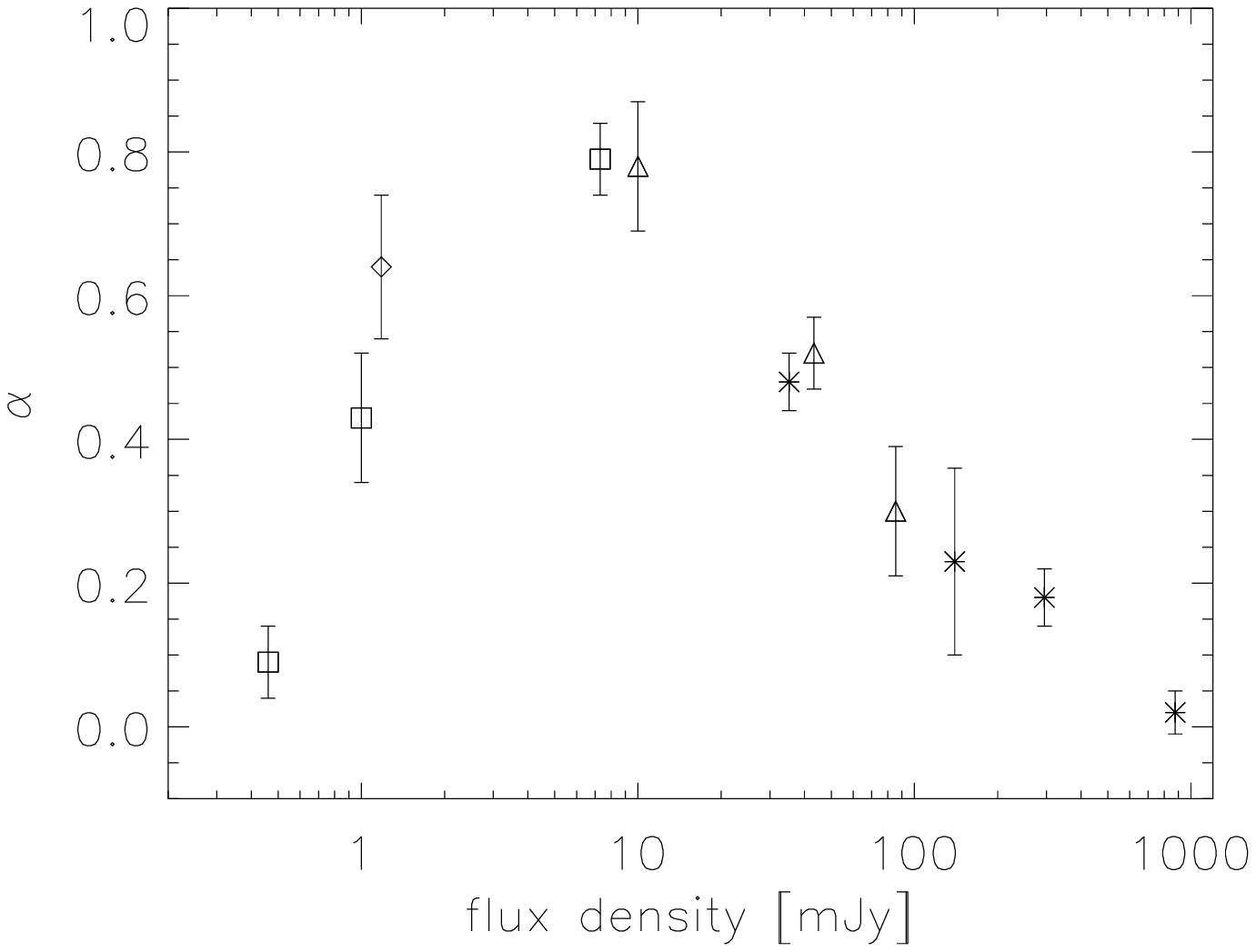}
\caption{The median spectral index as a function of flux density from the ATESP sample ($\alpha_{5}^{19}$, diamond), \citet{Bolton04}
sample ($\alpha_{4.8}^{15}$), the Bright (\citet{Massardi11}) and Faint (\citet{Bonavera11}) PACO samples 
($\alpha_{5.5}^{18}$, asterisks). The 9C (triangles) and 10C (squares) survey points ($\alpha_{1.4}^{15}$) in \citet{Whittam13} 
are also shown for comparison.}
\label{fig:whittam}
\end{figure}

We finally investigated the relationship between median spectral index and flux density.
To compute the median spectral index of the ATESP sample in the presence of upper limits 
we used the Kaplan-Meyer estimator routine inside 
the Survival analysis package  ASURV version 1.2 (\citealt{Feigelson85}). 
We found median (mean) values of $-$0.676 ($-0.619\pm0.099$), $-$0.650 ($-0.651\pm0.099$)
and $-$0.575 ($-0.510\pm0.086$) for the spectral indices $\alpha_{5}^{19}$, 
$\alpha_{19}^{38}$ and $\alpha_{1.4}^{19}$, respectively. In Fig.~\ref{fig:whittam}
we extended the plot of the median spectral index versus flux density, already presented in \cite{Whittam13} 
for the 9C (\citealt{Waldram10}) and 10C (\citealt{Franzen11}; \citealt{Davies11}) samples. 
Here we also include the ATESP sample, the \cite{Bolton04} sample, and the Bright and Faint PACO samples 
(\citealt{Massardi11}; \citealt{Bonavera11}). The latter two samples provide information at the bright end of the plot.
We confirm the trend already found by \cite{Whittam13}: moving from higher to lower flux-density regimes, 
radio source spectra get steeper and steeper until 10-20 mJy when they turn over to become flat 
again at the sub-mJy level. The ATESP point gently fits into the general trend shown in the plot. 
In the ATESP flux density regime (around 1 mJy) the steep-spectrum population is still
dominant, while at fainter fluxes flat-spectrum radio sources take over.    

\begin{figure}[h]
\centering
\includegraphics[width=8cm]{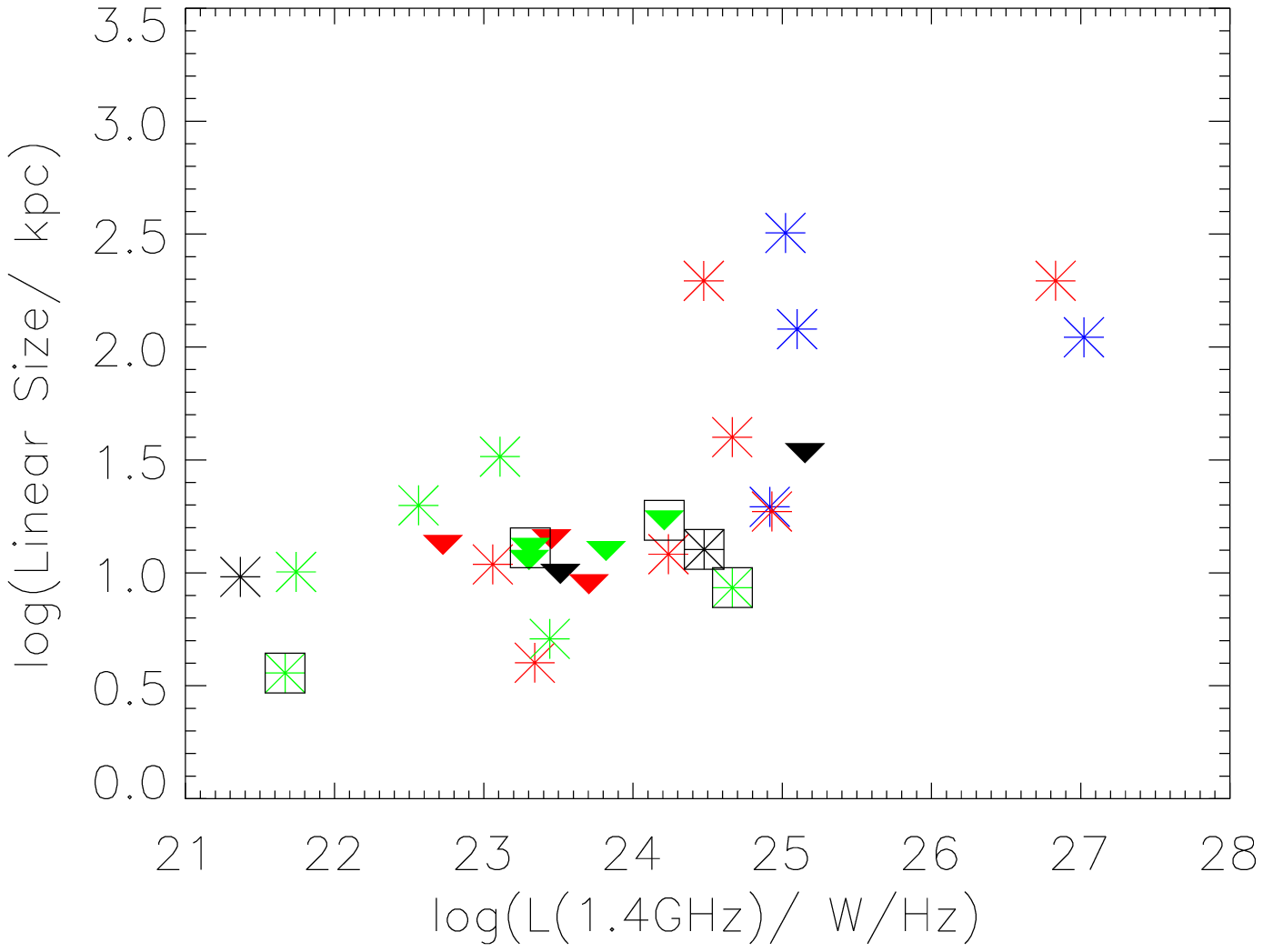}
\caption{Plot of the rest-frame 1.4~GHz luminosity vs. linear size for the ATESP early-type galaxy sample.
Asterisks represent sources with resolved sizes, while downward triangles represent upper limits. 
The quality of the model fit is colour-coded:  blue for JP fits, red for power-law fits, 
black for uncertain fits (all PL marked with a question mark in Table~\ref{tab:fit}) and green 
for sources that did not permit fitting. The 5 reliably variable sources are inside squares.}
\label{fig:lumlin}
\end{figure}

\section{Properties of the ATESP early-type sources}

Our flux-limited (S$>0.6$~mJy) sample of early-type galaxies is fainter than
the sample selected by \cite{Sadler14} from the AT20G survey (\citealt{Murphy10}).
Our sources are distributed over a large redshift range and are also generally more distant 
than those of \cite{Sadler14} (characterised by mean z=0.058).

Intrinsic luminosities at 1.4 and 5~GHz in the rest-frame have been computed making use of the 
$\alpha_{1.4}^5$ spectral index and the redshift information.  
The linear sizes were computed from the source angular sizes obtained from the 2-arcsec-resolution survey maps. 
Sources with an angular size smaller than the angular resolution appear point-like in the survey maps, 
so their linear sizes are upper limits. The numerical values are presented in Table~\ref{tab:lumlin}.
   
How many sources are compact (defined here as those with a linear size smaller than about 30 kpc)?
One should note that sources  $<15$ kpc are called FR0 in \cite{Sadler14}. Because in our sample most sources are 
further away, unresolved sources may have sizes which are typically bigger than those discussed by
Sadler: 11/28 are $<$15 kpc; 10/28 have sizes in the range 15-30 kpc, and 7/28 have larger sizes; 3/28
are very large with sizes $>$ 100 kpc.
In Sadler's sample 136/201 (68 \%) are unresolved and of these 75\% are classified as early-type galaxies.
The rest (32\%) are FRI- and FRII-type radio sources (\citealt{Fanaroff74}). 
We selected only early-type galaxies; nevertheless
the fraction of compact sources is not very different: 75\% of our sources are smaller than 30 kpc.

The 20 cm powers of early-type FR0 sources in Sadler go from $10^{22}$ W/Hz to $10^{26}$ W/Hz. 
This has to be compared with the range spanned by our FR0 sources, $10^{21.37}$ W/Hz to 
$10^{25.15}$ W/Hz with a median value of $10^{23.34}$ W/Hz.

The resolved sources (except one) are relatively powerful, from $10^{24.47}$ W/Hz to $10^{27.02}$ W/Hz.
We calculated the equipartition value of their magnetic field, which ranges from
a few $\mu$G to, for the smallest sources, 10-20 $\mu$G. These values are consistent with the normal 
magnetic field strengths found in early-type radio galaxies.

\begin{table}[t] 
 \centering
  \caption{Intrinsic luminosity and linear size: Columns 1-6: source name; redshift;
 rest-frame luminosity at 5 and 1.4~GHz; spectral index used to compute 
rest-frame luminosity; linear size in kpc. Sources are divided into three groups based on 
the spectral index: (from top to bottom) steep-, flat-, and inverted-spectrum sources.
(*) These sources appear as resolved at 0.6 arcsec resolution. For these sources 
we can infer a lower limit in size equal to~one third  of the quoted values 
(see Sect. 3 for more details).
(**) These sources appear as resolved at 0.15 arcsec resolution. For these sources 
we can infer a lower limit in size equal to one tenth  of the quoted values (see Sect. 3 for more details).
(***) This source appears as unresolved down to 0.15 arcsec. This gives a more stringent 
size upper limit of  <1 kpc  (see Sect. 3 for more details).} \label{tab:lumlin}
  \begin{tabular}{lccccc}
\hline
Name           &    z   &    P$_{\rm 5GHz}$ &  P$_{\rm 1.4 GHz}$ & $\alpha_{1.4}^5$ &  LS \\  
               &        &     W/Hz        &  W/Hz          &                 &  kpc  \\         
\hline
\hline    
src01          &  0.54  &   24.71        &   25.10        & $-$0.71           &  120.0 \\ 
src06          &  0.44  &   24.68        &   25.02        & $-$0.62           &  320.2 \\ 
src07          &  0.61  &   24.17        &   24.66        & $-$0.89           &   39.9 \\ 
src15          &  1.17  &   26.55        &   27.02        & $-$0.86           &  110.5 \\ 
src18          &  0.03  &   21.46        &   21.74        & $-$0.51           &   10.1 \\ 
src23          &  0.43  &   24.48        &   24.92        & $-$0.79           &   19.6 \\ 
src25          &  1.64  &   26.40        &   26.83        & $-$0.78           &  196.4 \\ 
\hline
src02          &  0.19  &   24.41        &   24.48        & $-$0.12           &   12.7 \\ 
src04          &  0.40  &   24.22        &   24.47        & $-$0.45           &  196.3 \\ 
src08          &  0.37  &   24.67        &   24.93        & $-$0.48           &   18.7 \\ 
src10          &  0.35  &   23.36        &   23.51        & $-$0.28           &   <9.9 (**) \\ 
src13          &  0.61  &   24.42        &   24.67        & $-$0.45           &    8.6 \\ 
src14          &  0.13  &   22.83        &   23.06        & $-$0.41           &   10.9 \\ 
src17          &  0.03  &   21.31        &   21.37        & $-$0.11           &    9.6 \\ 
src19          &  0.59  &   24.09        &   24.24        & $-$0.26           &   12.1 \\ 
src20          &  0.30  &   23.44        &   23.70        & $-$0.48           &   <8.9 (***) \\ 
src27          &  1.58  &   24.99        &   25.15        & $-$0.29           &  <33.9 \\ 
src28          &  0.73  &   23.32        &   23.45        & $-$0.24           &  <14.1 (*) \\ 
\hline
src09          &  1.25  &   24.35        &   24.21        &  0.25           &  <17.1 (**) \\ 
src11          &  0.15  &   22.63        &   22.56        &  0.12           &   19.9 \\ 
src12          &  0.25  &   23.38        &   23.34        &  0.06           &    4.0 \\ 
src16          &  0.03  &   21.84        &   21.67        &  0.31           &    3.6 \\ 
src21          &  0.36  &   23.33        &   23.11        &  0.40           &   32.7 \\ 
src22          &  1.40  &   24.21        &   23.44        &  1.39           &    5.1 \\ 
src24          &  0.56  &   23.43        &   23.31        &  0.70           &  <12.9 (*) \\ 
src26          &  0.44  &   23.41        &   23.30        &  0.19           &  <11.4 (**) \\ 
\hline
\end{tabular}
\end{table}

The large majority of the spectra allowing for a model fit can be represented well by a power-law
spectrum, and there is no trace of high-frequency steepening. Therefore such sources will
be relatively young. In the four sources fitted with a JP model the break frequencies obtained from the fit give an
indication of the source age and this turns out to be around a few million years. One should note that the sources
showing steepening are also the most powerful and the biggest sources (of the FRII type), 
but they are still relatively young with respect to typical radio sources with those sizes.

The 1.4~GHz rest-frame luminosities have been plotted against the linear size in Fig.~\ref{fig:lumlin}.
Blue points in the plot represent sources fittable with a JP model, red points are sources following 
a power-law fit, and black points are uncertain fits, while green points are for non-fittable sources.     
Radio sources in the linear size (LS) versus Power plot with larger luminosity (L $> 10^{24.5}$ W/Hz)
have steep or steepening spectra and larger angular sizes comparable with the ones of B2-type
classical radio galaxies. Sources with flat (non-fittable)
radio spectra have instead lower luminosity and smaller linear sizes. There is also a class of compact sources
that can be fit with a power-law spectra at low luminosity that might be the extension at low power of the linear size-luminosity 
relation for the B2-type radio galaxies found by De Ruiter et al. (1990).           

In Fig.~\ref{fig:lumlin} sources with upper limits in their linear size have low luminosity; we thus infer that there 
are three distinct source populations: i) the bright and large sources representing classical radio galaxies (18\%);
ii)  compact (confined within their host galaxies), low-luminosity, power-law (jet-dominated) sources ($\sim
46\%$  of the sample) and iii) compact, more variable, core-dominated sources (36\% of the sample) with linear sizes $<30$~kpc 
(thus confined inside their host galaxy), flat radio spectra and a 1.4-GHz luminosity $\simeq 10^{24}$ W/Hz 
similar to the FR0 sources classified by \cite{Baldi16}.

The $\alpha_{1.4}^5$ spectral index has also been used to separate the sample into spectral classes: 
steep-spectrum sources (with $\alpha_{1.4}^5 < -0.5$, top part of Table~\ref{tab:lumlin}), 
flat-spectrum (with $-0.5 < \alpha_{1.4}^5 \leq 0$, middle part of Table~\ref{tab:lumlin}),
and peaked spectrum (with $\alpha_{1.4}^5 > 0$, bottom part of Table~\ref{tab:lumlin}). 
All steep-spectrum sources are resolved at the 2-arcsec angular resolution. 
Four of eleven flat-spectrum sources and three of the eight peaked spectrum sources 
have upper limits in their linear size. The rest-frame luminosity versus linear size plot is compared with 
the one in Fig. 1 in \cite{Kunert16}. Most of our resolved sources overlap with the FRI sources 
(with LS > a few tens of kiloparsec and L$_{\rm 1.4GHz}$ > 10$^{24.5}$ W/Hz) more than the low-luminosity compact (LLC) sources 
(with LS between 0.1 and 10 kpc and 10$^{24.5}$ W/Hz < L$_{\rm 1.4GHz}$ < 10$^{26.5}$ W/Hz). 

\section{Summary}

Here we present quasi-simultaneous observations at 19, 38, and 94 GHz of a sample of 28 early-type radio galaxies carried out
in two observing campaigns (September 2011 and July 2012). The main results are summarised below.

All 28 sources were detected in 19~GHz low-resolution maps at 3 $\sigma$ level, while 25 sources were detected in 
38~GHz low-resolution maps; for three sources we reported 3-$\sigma$ flux density upper limits. Only one (src02)
of the twelve sources observed was detected at 94~GHz and for the rest of them we report 3-$\sigma$ upper limits.      

Eight sources were detected at the $3-\sigma$ level in 38~GHz maps at an angular resolution of 0.6 arcsec. Seven sources 
appear point-like and one source (src02, the brightest one) shows a tentative NW elongation. When comparing 
low- and high-resolution maps, seven sources have their flux densities resolved out in 38 GHz high-resolution maps.
This sets a 0.6 arcsec lower limit on the angular size of the sources unresolved in 5-GHz survey images (Paper~I). 
To the 18 sources appearing as point-like from this analysis we applied a technique based on the ratio 
of the amplitudes of visibilities (gamma factor) to extract information on source structure down to angular scales of 0.15 arcsec. 
This analysis shows no clear presence of source structures on the sub-arcsec scale. 
For the sources unresolved in the images of the 5-GHz survey, this sets a 0.15 arcsec upper-limit (a factor of ten lower).   
 
Six of 25 sources were found variable at 19~GHz with a confidence level of more than 99.9\% using the 
two-epoch VE test. These sources are variable in a range between 6.4 and 33.7\%. The
percentage of variable sources in the ATESP sample turns out to be higher than the 5\% found by \cite{Sadler06}
on a sub-sample of the AT20G survey. 

From the study of the radio spectra, four categories of sources were identified: (i) smooth steep-spectrum sources; 
(ii) flat-spectrum sources; (iii) peaked-spectrum sources, and (iv) sources having spectra with undefined shape. 
The spectra of the first kind were fitted with a power law or a JP model. The majority of the sources that could be fit
follow a PL model, but for four sources showing signs of spectral steepening at the highest 
frequency a JP model fit was applied.

From the analysis of the colour-colour plot, pure ADAF models are still ruled out, as already found in the 
study at lower frequencies by \cite{Prandoni10}. Twenty-six (93\%) sources populate the steep-spectrum quadrant of the plot, while 
only one source is up-turn and another one has a peaked spectrum. No source is found in the rising spectrum quadrant where
sources with an ADAF model are expected to reside. 

We computed the median spectral index between 5 and 19 GHz ($\alpha_{5}^{19}$) and compared it with data with 
increasing flux density threshold from the 9C and 10C surveys (\citealt{Whittam13}), the \cite{Bolton04} 
sample, and the PACO Faint (\citealt{Bonavera11}) and Bright (\citealt{Massardi11}) samples. All these points follow 
a clear trend which the ATESP point is part of: steep-spectrum sources dominate at around 10 mJy,
and the flat-spectrum sources increase their contribution toward 
fainter flux densities. The ATESP sample with a median flux 
of 1.18~mJy has an intermediate behaviour with a median $\alpha_{5}^{19} \sim -0.64$.  

From the study of the source properties one can conclude that in the ATESP sample three populations are present:
a) bright and large classical radio sources (18\% of the sample) with equipartition magnetic field strengths of 
the order of a few $\mu$G, a luminosity L$_{\rm 1.4 GHz} > 10^{24}$ W Hz$^{-1}$ , and linear sizes > 100 kpc;
b) compact (confined within their host galaxies), low-luminosity, power-law (jet-dominated) sources ($\sim
46\%$  of the sample), and  
c) typically flat-spectrum (i.e. non fittable), variable, and compact (core-dominated) radio sources 
(36\% of the sample) with linear sizes below 30 kpc (confined inside their host galaxies). 
This is in agreement with the result found by \cite{Sadler14} in a sample of local (z=0.058) radio sources 
selected at 20~GHz from the AT20G catalogue.

We conclude that the AGN component of the faint radio population appears as a composite ensemble of source types 
and more accurate studies at higher angular resolution (via VLBI follow-up observations) and higher frequencies 
($>$ 100~GHz via ALMA) are needed to better characterise it.                
                           
\begin{table*}[h]
 \centering
  \caption{Integrated flux densities of the ATESP sources in the 2011 (columns 5-8) and 2012 (columns 9-10)
campaigns, with error bars. Flux densities obtained in the 1.4 and 5-GHz ATESP surveys are also shown for comparison in columns 3-4. 
A $<$ sign in front of a flux density measurement means a 3-$\sigma$ upper limit. LR and HR 
are the short forms for low-angular-resolution and high-angular-resolution (0.6 arcsec) flux densities, respectively. 
Every second line shows the flux-density error bar.} \label{tab:intflux}
  \begin{tabular}{llcc|cccc|cc}
  \hline
  (1) &   (2)         &    (3)  &  (4)  &   (5) &   (6)      &   (7)     &   (8)   &    (9)  &   (10)   \\
      &               &   1.4GHz&   5GHz&  19GHz&    38GHz LR&   38GHz HR&    94GHz&   19GHz &  38 GHz \\
      &               &  (mJy)  &  (mJy)&  (mJy)&  (mJy)     &  (mJy)    &  (mJy)  &    (mJy)& (mJy)   \\  
  \hline 
src01 & J224516-401807  &    9.4     &   3.9    &  1.091  &    $-$  &  <0.198&    $-$  &    1.297 &  0.456  \\  
      &                 &    0.08    &   0.08   & 0.100   &    $-$  &    $-$ &    $-$  & 0.059    &  0.058  \\   
src02 & J224547-400324  &   32.83    &  28.28   &  21.0   & 18.225  &   9.4  &    13.3 &  $-$     & $-$     \\  
      &                 &    0.08    &   0.08   &  0.106  &  0.077  &  0.092 &    0.92 &  $-$     & $-$     \\
src03 & J224628-401207  &    1.35    &   0.82   &   <0.32 &    $-$  &  <0.09 &    $-$  &   0.323  &  0.278  \\  
      &                 &    0.08    &   0.08   &  $-$    &    $-$  &   $-$  &    $-$  &   0.042  &  0.058  \\
src04 & J224654-400107  &         5.59     &   3.15   &  1.602  &    $-$  &  <0.261&   <1.74 &   1.333  &  0.829  \\  
      &                 &   0.08     &   0.07   &  0.117  &    $-$  &  $-$   &    $-$  &   0.055  &  0.105  \\
src05 & J224719-401530  &    0.74    &   0.84   &  0.812  &    $-$  &  <0.183&    $-$  &   1.015  &  0.378  \\  
      &                 &    0.08    &   0.07   &  0.115  &    $-$  &  $-$   &    $-$  &   0.061  &  0.097  \\ 
src06 & J224750-400148  &   13.44    &   6.09   &  1.314  &    $-$  &  <0.297&    $-$  &   1.452  &  0.837  \\  
      &                 &   0.08     &   0.07   &  0.113  &    $-$  &  $-$   &    $-$  &   0.057  &  0.101  \\
src07 & J224753-400455  &    2.08    &   0.67   &  0.3    &    $-$  &  <0.159&    $-$  &   0.257  &  0.223  \\  
      &                 &   0.08     &   0.07   &  0.060  &    $-$  &  $-$   &    $-$  &   0.032  &  0.040  \\
src08 & J224822-401808  &   19.08    &  10.34   &  4.408  &   2.563 &   1.5  &   <1.9  &   $-$    &  $-$    \\  
      &                 &   0.08     &   0.07   &  0.081  &   0.073 &  0.090 &    $-$  &   $-$    &  $-$    \\
src09 & J224827-402515  &   0.58     &   0.80   &  1.020  &    $-$  &  <0.27 &    $-$  &   0.616  &  0.560  \\   
      &                 &   0.08     &   0.07   &  0.103  &    $-$  &  $-$   &    $-$  &   0.056  &  0.093  \\
src10 & J224919-400037  &    0.91    &   0.64   &   0.46  &    $-$  &  <0.192&    $-$  &    0.326 & <0.222  \\  
      &                 &    0.08    &   0.07   &  0.102  &    $-$  &  $-$   &    $-$  &    0.055 &  $-$    \\
src11 & J224935-400816  &    0.7     &   0.82   &   <0.32 &    $-$  &  <0.294&    $-$  &    0.326 &  0.264  \\  
      &                 &    0.08    &   0.06   &    $-$  &    $-$  &  $-$   &    $-$  &     0.041 &  0.032  \\  
src12 & J224958-395855  &    1.52    &   1.65   &   1.7   &    1.2  &  0.87  &    <1.8 &    $-$   &    $-$  \\  
      &                 &    0.09    &   0.07   &  0.103  &  0.077  &  0.092 &    $-$  &    $-$   &    $-$  \\
src13 & J225004-402412  &          3.16    &   1.78   &  1.122  &    $-$  & <0.261 &    <1.3 &   0.578  &  0.440  \\  
      &                 &    0.09    &   0.07   &  0.127  &    $-$  &  $-$   &    $-$  &   0.058  &  0.065  \\
src14 & J225008-400425  &    2.88    &    1.7   &  0.49   &    $-$  & <0.249 &    $-$  &   0.430  &  0.416  \\  
      &                 &    0.09    &   0.07   &  0.141  &    $-$  &  $-$   &    $-$  &   0.055  &  0.073  \\
src15 & J225034-401936  &         76.6     &   25.8   &  6.165  &    $-$  & <0.267 &    <1.0 &   7.031  &  2.100  \\  
      &                 &    0.09    &    0.07  &  0.130  &    $-$  &  $-$   &    $-$  &   0.057  &  0.100  \\
src16 & J225057-401522  &    2.01    &    3.0   &  1.5    &    0.8  &   0.78 &    <2.9 &    $-$   &  $-$    \\  
      &                 &    0.09    &    0.06  &  0.121  &    0.079& 0.084  &    $-$  &    $-$   &  $-$    \\
src17 & J225223-401841  &    0.98    &    0.85  &  0.43   &   <0.228& <0.255 &    $-$  &    0.356 &  <0.180 \\  
      &                 &    0.09    &    0.07  &  0.137  &   $-$   &  $-$   &    $-$  &    0.060 &  $-$    \\
src18 & J225239-401949  &    2.26    &   1.18   &  0.56   &  0.402  & <0.141 &    $-$  &    0.250 &   0.200 \\  
      &                 &    0.09    &   0.07   &  0.127  &  0.046  & $-$          &    $-$  &    0.040 &   0.041 \\
src19 & J225249-401256  &          1.52    &   1.09   &  0.46   &  0.64   & <0.267 &   <0.93 &    $-$   &   $-$   \\  
      &                 &    0.09    &   0.06   &  0.115  &  0.083  &  $-$   &    $-$  &    $-$   &   $-$   \\
src20 & J225322-401931  &    1.86    &   1.01   &  <0.35  &  <0.16  & <0.189 &    $-$  &   0.347  & 0.198   \\  
      &                 &    0.09    &  0.07    &  $-$    &  $-$    & $-$          &    $-$  &   0.042  & 0.034   \\
src21 & J225323-400453  &    0.51    &   0.85   & <0.35   &  <0.25  & <0.261 &    <1.5 &   0.380  & 0.230   \\  
      &                 &    0.09    &   0.06   &  $-$    &  $-$    & $-$    &    $-$  &   0.042  & 0.050   \\
src22 & J225344-401928  &    0.6     &   3.52   & 2.03    & 1.414   & 1.2    &   <3.13 &   $-$    &  $-$    \\  
      &                 &    0.09    &   0.06   & 0.103   & 0.077   & 0.090  &    $-$  &   $-$    &  $-$    \\
src23 & J225404-402226  &   10.34    &   3.8    & 1.014   & 0.47    & 0.22   &    $-$  &   $-$    &  $-$    \\  
      &                 &    0.09    &   0.06   & 0.110   & 0.045   & 0.050  &    $-$  &   $-$    &  $-$    \\
src24 & J225430-400334  &   <0.26    &   0.63   & 2.19    & 1.4     & 0.87   &   <1.4  &   $-$    &  $-$    \\  
      &                 &   $-$      &   0.06   & 0.128   & 0.082   & 0.086  &    $-$  &   $-$    &  $-$    \\
src25 & J225434-401343  & 21.09      &   7.8    & 1.22    & 0.860   & <0.264 &   <3.0  &   $-$    &  $-$    \\  
      &                 &  0.09      &   0.06   & 0.097   & 0.072   & $-$    &    $-$  &   $-$    &  $-$    \\
src26 & J225436-400531  &        0.47      &   0.6    & 0.37    & $-$     & <0.273 &    $-$  &  0.381   &  0.321  \\  
      &                 &  0.09      &   0.06   & 0.092   & $-$     & $-$    &    $-$  &  0.055   &  0.075  \\
src27 & J225449-400918  &  1.24      &   0.86   & 0.51    & $-$     & <0.186 &    $-$  &  0.291   &  <0.183 \\  
      &                 &  0.09      &   0.06   & 0.115   & $-$     &  $-$   &    $-$  &  0.061   &  $-$    \\
src28 & J225529-401101  &  1.48      &   1.09   & 0.761   & 0.585   & 0.279  &    $-$  &   $-$    &  $-$    \\   
      &                 &  0.08      &   0.06   & 0.102   & 0.075   & 0.093  &    $-$  &   $-$    &  $-$    \\
\hline
\end{tabular}
\end{table*}

\begin{acknowledgements}
      The Australia Telescope Compact Array is part of the Australia 
Telescope which is funded by the Commonwealth of Australia for operation
as a National Facility managed by CSIRO.

We thank Matteo Murgia (INAF-Cagliari), who made available to us his
fitting programme SYNAGE.
\end{acknowledgements}


\bibliographystyle{aa} 

\begin{thebibliography}{48}
\expandafter\ifx\csname natexlab\endcsname\relax\def\natexlab#1{#1}\fi

\bibitem[{{AMI Consortium} {et~al.}(2011{\natexlab{a}}){AMI Consortium},
  {Davies}, {Franzen}, {Waldram}, {Grainge}, {Hobson}, {Hurley-Walker},
  {Lasenby}, {Olamaie}, {Pooley}, {Riley}, {Rodr{\'{\i}}guez-Gonz{\'a}lvez},
  {Saunders}, {Scaife}, {Schammel}, {Scott}, {Shimwell}, {Titterington}, \&
  {Zwart}}]{Davies11}
{AMI Consortium}, {Davies}, M.~L., {Franzen}, T.~M.~O., {et~al.}
  2011{\natexlab{a}}, \mnras, 415, 2708

\bibitem[{{AMI Consortium} {et~al.}(2011{\natexlab{b}}){AMI Consortium},
  {Franzen}, {Davies}, {Waldram}, {Grainge}, {Hobson}, {Hurley-Walker},
  {Lasenby}, {Olamaie}, {Pooley}, {Rodr{\'{\i}}guez-Gonz{\'a}lvez}, {Saunders},
  {Scaife}, {Schammel}, {Scott}, {Shimwell}, {Titterington}, \&
  {Zwart}}]{Franzen11}
{AMI Consortium}, {Franzen}, T.~M.~O., {Davies}, M.~L., {et~al.}
  2011{\natexlab{b}}, \mnras, 415, 2699

\bibitem[{{Baldi} {et~al.}(2016){Baldi}, {Capetti}, \& {Giovannini}}]{Baldi16}
{Baldi}, R.~D., {Capetti}, A., \& {Giovannini}, G. 2016, Astronomische
  Nachrichten, 337, 114

\bibitem[{{Blundell} {et~al.}(1999){Blundell}, {Rawlings}, \&
  {Willott}}]{Blundell99}
{Blundell}, K.~M., {Rawlings}, S., \& {Willott}, C.~J. 1999, \aj, 117, 677

\bibitem[{{Bolton} {et~al.}(2004){Bolton}, {Cotter}, {Pooley}, {Riley},
  {Waldram}, {Chandler}, {Mason}, {Pearson}, \& {Readhead}}]{Bolton04}
{Bolton}, R.~C., {Cotter}, G., {Pooley}, G.~G., {et~al.} 2004, \mnras, 354, 485

\bibitem[{{Bonavera} {et~al.}(2011){Bonavera}, {Massardi}, {Bonaldi},
  {Gonz{\'a}lez-Nuevo}, {de Zotti}, \& {Ekers}}]{Bonavera11}
{Bonavera}, L., {Massardi}, M., {Bonaldi}, A., {et~al.} 2011, \mnras, 416, 559

\bibitem[{{Bonzini} {et~al.}(2013){Bonzini}, {Padovani}, {Mainieri},
  {Kellermann}, {Miller}, {Rosati}, {Tozzi}, \& {Vattakunnel}}]{Bonzini13}
{Bonzini}, M., {Padovani}, P., {Mainieri}, V., {et~al.} 2013, \mnras, 436, 3759

\bibitem[{{Chhetri} {et~al.}(2013){Chhetri}, {Ekers}, {Jones}, \&
  {Ricci}}]{Chhetri13}
{Chhetri}, R., {Ekers}, R.~D., {Jones}, P.~A., \& {Ricci}, R. 2013, \mnras,
  434, 956

\bibitem[{{Clark}(1980)}]{Clark80}
{Clark}, B.~G. 1980, \aap, 89, 377

\bibitem[{{Colla} {et~al.}(1975){Colla}, {Fanti}, {Fanti}, {Gioia}, {Lari},
  {Lequeux}, {Lucas}, \& {Ulrich}}]{Colla75}
{Colla}, G., {Fanti}, C., {Fanti}, R., {et~al.} 1975, \aaps, 20, 1

\bibitem[{{Fanaroff} \& {Riley}(1974)}]{Fanaroff74}
{Fanaroff}, B.~L. \& {Riley}, J.~M. 1974, \mnras, 167, 31P

\bibitem[{{Feigelson} \& {Nelson}(1985)}]{Feigelson85}
{Feigelson}, E.~D. \& {Nelson}, P.~I. 1985, \apj, 293, 192

\bibitem[{{Giroletti} {et~al.}(2005){Giroletti}, {Giovannini}, \&
  {Taylor}}]{Giroletti05}
{Giroletti}, M., {Giovannini}, G., \& {Taylor}, G.~B. 2005, \aap, 441, 89

\bibitem[{{Gregorini} {et~al.}(1986){Gregorini}, {Ficarra}, \&
  {Padrielli}}]{Gregorini86}
{Gregorini}, L., {Ficarra}, A., \& {Padrielli}, L. 1986, \aap, 168, 25

\bibitem[{{H{\"o}gbom}(1974)}]{Hogbom74}
{H{\"o}gbom}, J.~A. 1974, \aaps, 15, 417

\bibitem[{{Hurley-Walker} {et~al.}(2017){Hurley-Walker}, {Callingham},
  {Hancock}, {Franzen}, {Hindson}, {Kapi{\'n}ska}, {Morgan}, {Offringa},
  {Wayth}, {Wu}, {Zheng}, {Murphy}, {Bell}, {Dwarakanath}, {For}, {Gaensler},
  {Johnston-Hollitt}, {Lenc}, {Procopio}, {Staveley-Smith}, {Ekers}, {Bowman},
  {Briggs}, {Cappallo}, {Deshpande}, {Greenhill}, {Hazelton}, {Kaplan},
  {Lonsdale}, {McWhirter}, {Mitchell}, {Morales}, {Morgan}, {Oberoi}, {Ord},
  {Prabu}, {Shankar}, {Srivani}, {Subrahmanyan}, {Tingay}, {Webster},
  {Williams}, \& {Williams}}]{Hurley-Walker17}
{Hurley-Walker}, N., {Callingham}, J.~R., {Hancock}, P.~J., {et~al.} 2017,
  \mnras, 464, 1146

\bibitem[{{Jaffe} \& {Perola}(1973)}]{Jaffe73}
{Jaffe}, W.~J. \& {Perola}, G.~C. 1973, \aap, 26, 423

\bibitem[{{Jones} {et~al.}(2009){Jones}, {Read}, {Saunders}, {Colless},
  {Jarrett}, {Parker}, {Fairall}, {Mauch}, {Sadler}, {Watson}, {Burton},
  {Campbell}, {Cass}, {Croom}, {Dawe}, {Fiegert}, {Frankcombe}, {Hartley},
  {Huchra}, {James}, {Kirby}, {Lahav}, {Lucey}, {Mamon}, {Moore}, {Peterson},
  {Prior}, {Proust}, {Russell}, {Safouris}, {Wakamatsu}, {Westra}, \&
  {Williams}}]{Jones09}
{Jones}, D.~H., {Read}, M.~A., {Saunders}, W., {et~al.} 2009, \mnras, 399, 683

\bibitem[{{Kunert-Bajraszewska}(2016)}]{Kunert16}
{Kunert-Bajraszewska}, M. 2016, Astronomische Nachrichten, 337, 27

\bibitem[{{Marriage} {et~al.}(2011){Marriage}, {Baptiste Juin}, {Lin},
  {Marsden}, {Nolta}, {Partridge}, {Ade}, {Aguirre}, {Amiri}, {Appel},
  {Barrientos}, {Battistelli}, {Bond}, {Brown}, {Burger}, {Chervenak}, {Das},
  {Devlin}, {Dicker}, {Bertrand Doriese}, {Dunkley}, {D{\"u}nner},
  {Essinger-Hileman}, {Fisher}, {Fowler}, {Hajian}, {Halpern}, {Hasselfield},
  {Hern{\'a}ndez-Monteagudo}, {Hilton}, {Hilton}, {Hincks}, {Hlozek},
  {Huffenberger}, {Handel Hughes}, {Hughes}, {Infante}, {Irwin}, {Kaul},
  {Klein}, {Kosowsky}, {Lau}, {Limon}, {Lupton}, {Martocci}, {Mauskopf},
  {Menanteau}, {Moodley}, {Moseley}, {Netterfield}, {Niemack}, {Page},
  {Parker}, {Quintana}, {Reid}, {Sehgal}, {Sherwin}, {Sievers}, {Spergel},
  {Staggs}, {Swetz}, {Switzer}, {Thornton}, {Trac}, {Tucker}, {Warne},
  {Wilson}, {Wollack}, \& {Zhao}}]{Marriage11}
{Marriage}, T.~A., {Baptiste Juin}, J., {Lin}, Y.-T., {et~al.} 2011, \apj, 731,
  100

\bibitem[{{Massardi} {et~al.}(2016){Massardi}, {Bonaldi}, {Bonavera}, {De
  Zotti}, {Lopez-Caniego}, \& {Galluzzi}}]{Massardi16}
{Massardi}, M., {Bonaldi}, A., {Bonavera}, L., {et~al.} 2016, \mnras, 455, 3249

\bibitem[{{Massardi} {et~al.}(2011){Massardi}, {Ekers}, {Murphy}, {Mahony},
  {Hancock}, {Chhetri}, {de Zotti}, {Sadler}, {Burke-Spolaor}, {Calabretta},
  {Edwards}, {Ekers}, {Jackson}, {Kesteven}, {Newton-McGee}, {Phillips},
  {Ricci}, {Roberts}, {Sault}, {Staveley-Smith}, {Subrahmanyan}, {Walker}, \&
  {Wilson}}]{Massardi11}
{Massardi}, M., {Ekers}, R.~D., {Murphy}, T., {et~al.} 2011, \mnras, 412, 318

\bibitem[{{Mauch} {et~al.}(2003){Mauch}, {Murphy}, {Buttery}, {Curran},
  {Hunstead}, {Piestrzynski}, {Robertson}, \& {Sadler}}]{Mauch03}
{Mauch}, T., {Murphy}, T., {Buttery}, H.~J., {et~al.} 2003, \mnras, 342, 1117

\bibitem[{{Mignano} {et~al.}(2008){Mignano}, {Prandoni}, {Gregorini}, {Parma},
  {de Ruiter}, {Wieringa}, {Vettolani}, \& {Ekers}}]{Mignano08}
{Mignano}, A., {Prandoni}, I., {Gregorini}, L., {et~al.} 2008, \aap, 477, 459

\bibitem[{{Murgia} {et~al.}(1999){Murgia}, {Fanti}, {Fanti}, {Gregorini},
  {Klein}, {Mack}, \& {Vigotti}}]{Murgia99}
{Murgia}, M., {Fanti}, C., {Fanti}, R., {et~al.} 1999, \aap, 345, 769

\bibitem[{{Murphy} {et~al.}(2010){Murphy}, {Sadler}, {Ekers}, {Massardi},
  {Hancock}, {Mahony}, {Ricci}, {Burke-Spolaor}, {Calabretta}, {Chhetri}, {de
  Zotti}, {Edwards}, {Ekers}, {Jackson}, {Kesteven}, {Lindley}, {Newton-McGee},
  {Phillips}, {Roberts}, {Sault}, {Staveley-Smith}, {Subrahmanyan}, {Walker},
  \& {Wilson}}]{Murphy10}
{Murphy}, T., {Sadler}, E.~M., {Ekers}, R.~D., {et~al.} 2010, \mnras, 402, 2403

\bibitem[{{Nagar} {et~al.}(2001){Nagar}, {Wilson}, \& {Falcke}}]{Nagar01}
{Nagar}, N.~M., {Wilson}, A.~S., \& {Falcke}, H. 2001, \apjl, 559, L87

\bibitem[{{O'Dea}(1998)}]{ODea98}
{O'Dea}, C.~P. 1998, \pasp, 110, 493

\bibitem[{{O'Dea} {et~al.}(1991){O'Dea}, {Baum}, \& {Stanghellini}}]{ODea91}
{O'Dea}, C.~P., {Baum}, S.~A., \& {Stanghellini}, C. 1991, \apj, 380, 66

\bibitem[{{Prandoni} {et~al.}(2010){Prandoni}, {de Ruiter}, {Ricci}, {Parma},
  {Gregorini}, \& {Ekers}}]{Prandoni10}
{Prandoni}, I., {de Ruiter}, H.~R., {Ricci}, R., {et~al.} 2010, \aap, 510, A42

\bibitem[{{Prandoni} {et~al.}(2000{\natexlab{a}}){Prandoni}, {Gregorini},
  {Parma}, {de Ruiter}, {Vettolani}, {Wieringa}, \& {Ekers}}]{Prandoni00a}
{Prandoni}, I., {Gregorini}, L., {Parma}, P., {et~al.} 2000{\natexlab{a}},
  \aaps, 146, 31

\bibitem[{{Prandoni} {et~al.}(2000{\natexlab{b}}){Prandoni}, {Gregorini},
  {Parma}, {de Ruiter}, {Vettolani}, {Wieringa}, \& {Ekers}}]{Prandoni00b}
{Prandoni}, I., {Gregorini}, L., {Parma}, P., {et~al.} 2000{\natexlab{b}},
  \aaps, 146, 41

\bibitem[{{Prandoni} {et~al.}(2006){Prandoni}, {Parma}, {Wieringa}, {de
  Ruiter}, {Gregorini}, {Mignano}, {Vettolani}, \& {Ekers}}]{Prandoni06}
{Prandoni}, I., {Parma}, P., {Wieringa}, M.~H., {et~al.} 2006, \aap, 457, 517

\bibitem[{{Quataert} \& {Narayan}(1999)}]{Quataert99}
{Quataert}, E. \& {Narayan}, R. 1999, \apj, 520, 298

\bibitem[{{Sadler} {et~al.}(2014){Sadler}, {Ekers}, {Mahony}, {Mauch}, \&
  {Murphy}}]{Sadler14}
{Sadler}, E.~M., {Ekers}, R.~D., {Mahony}, E.~K., {Mauch}, T., \& {Murphy}, T.
  2014, \mnras, 438, 796

\bibitem[{{Sadler} {et~al.}(2006){Sadler}, {Ricci}, {Ekers}, {Ekers},
  {Hancock}, {Jackson}, {Kesteven}, {Murphy}, {Phillips}, {Reinfrank},
  {Staveley-Smith}, {Subrahmanyan}, {Walker}, {Wilson}, \& {de
  Zotti}}]{Sadler06}
{Sadler}, E.~M., {Ricci}, R., {Ekers}, R.~D., {et~al.} 2006, \mnras, 371, 898

\bibitem[{{Sajina} {et~al.}(2011){Sajina}, {Partridge}, {Evans}, {Stefl},
  {Vechik}, {Myers}, {Dicker}, \& {Korngut}}]{Sajina11}
{Sajina}, A., {Partridge}, B., {Evans}, T., {et~al.} 2011, \apj, 732, 45

\bibitem[{{Sault} {et~al.}(1995){Sault}, {Teuben}, \& {Wright}}]{Sault95}
{Sault}, R.~J., {Teuben}, P.~J., \& {Wright}, M.~C.~H. 1995, in Astronomical
  Society of the Pacific Conference Series, Vol.~77, Astronomical Data Analysis
  Software and Systems IV, ed. R.~A. {Shaw}, H.~E. {Payne}, \& J.~J.~E.
  {Hayes}, 433

\bibitem[{{Sault} \& {Wieringa}(1994)}]{Sault94}
{Sault}, R.~J. \& {Wieringa}, M.~H. 1994, \aaps, 108, 585

\bibitem[{{Seymour} {et~al.}(2008){Seymour}, {Dwelly}, {Moss}, {McHardy},
  {Zoghbi}, {Rieke}, {Page}, {Hopkins}, \& {Loaring}}]{Seymour08}
{Seymour}, N., {Dwelly}, T., {Moss}, D., {et~al.} 2008, \mnras, 386, 1695

\bibitem[{{Snellen} {et~al.}(2000){Snellen}, {Schilizzi}, {Miley}, {de Bruyn},
  {Bremer}, \& {R{\"o}ttgering}}]{Snellen00}
{Snellen}, I.~A.~G., {Schilizzi}, R.~T., {Miley}, G.~K., {et~al.} 2000, \mnras,
  319, 445

\bibitem[{{Steer} {et~al.}(1984){Steer}, {Dewdney}, \& {Ito}}]{Steer84}
{Steer}, D.~G., {Dewdney}, P.~E., \& {Ito}, M.~R. 1984, \aap, 137, 159

\bibitem[{{Waldram} {et~al.}(2010){Waldram}, {Pooley}, {Davies}, {Grainge}, \&
  {Scott}}]{Waldram10}
{Waldram}, E.~M., {Pooley}, G.~G., {Davies}, M.~L., {Grainge}, K.~J.~B., \&
  {Scott}, P.~F. 2010, \mnras, 404, 1005

\bibitem[{{Whittam} {et~al.}(2017){Whittam}, {Green}, {Jarvis}, \&
  {Riley}}]{Whittam17}
{Whittam}, I.~H., {Green}, D.~A., {Jarvis}, M.~J., \& {Riley}, J.~M. 2017,
  \mnras, 464, 3357

\bibitem[{{Whittam} {et~al.}(2016){Whittam}, {Riley}, {Green}, \&
  {Jarvis}}]{Whittam16}
{Whittam}, I.~H., {Riley}, J.~M., {Green}, D.~A., \& {Jarvis}, M.~J. 2016,
  \mnras, 462, 2122

\bibitem[{{Whittam} {et~al.}(2013){Whittam}, {Riley}, {Green}, {Jarvis},
  {Prandoni}, {Guglielmino}, {Morganti}, {R{\"o}ttgering}, \&
  {Garrett}}]{Whittam13}
{Whittam}, I.~H., {Riley}, J.~M., {Green}, D.~A., {et~al.} 2013, \mnras, 429,
  2080

\bibitem[{{Whittam} {et~al.}(2015){Whittam}, {Riley}, {Green}, {Jarvis}, \&
  {Vaccari}}]{Whittam15}
{Whittam}, I.~H., {Riley}, J.~M., {Green}, D.~A., {Jarvis}, M.~J., \&
  {Vaccari}, M. 2015, \mnras, 453, 4244

\bibitem[{{Windhorst} {et~al.}(1990){Windhorst}, {Mathis}, \&
  {Neuschaefer}}]{Windhorst90}
{Windhorst}, R., {Mathis}, D., \& {Neuschaefer}, L. 1990, in Astronomical
  Society of the Pacific Conference Series, Vol.~10, Evolution of the Universe
  of Galaxies, ed. R.~G. {Kron}, 389--403

\end{thebibliography}

\end{document}